\documentclass[11pt,english,a4paper,twoside]{svjour3}
\usepackage{mathptmx}
\usepackage[T1]{fontenc}
\usepackage[latin9]{inputenc}
\usepackage[a4paper]{geometry}
\usepackage[active]{srcltx}
\usepackage{babel}
\usepackage{mathrsfs}
\usepackage{amsmath}
\usepackage{amssymb}
\usepackage{graphicx}
\usepackage[unicode=true,pdfusetitle,
 bookmarks=true,bookmarksnumbered=false,bookmarksopen=false,
 breaklinks=false,pdfborder={0 0 1},backref=false,colorlinks=false]
 {hyperref}
\usepackage{breakurl}

\makeatletter

\usepackage[a4paper]{geometry}
\usepackage{hyperref}

\hypersetup{
bookmarks=false,
pdfpagemode=UseNone,
pdfstartview=FitH
} 

\makeatother

\begin{document}

\title{Kinetic theory of jet dynamics in the stochastic barotropic and 2D
Navier-Stokes equations}

\author{Freddy Bouchet \and Cesare Nardini \and Tom\'as Tangarife}
\institute{Laboratoire de physique, \'Ecole Normale Sup\'erieure de Lyon
et CNRS, 46 all\'ee d'Italie, 69007 Lyon, France. \email{Freddy.Bouchet@ens-lyon.fr \and cesare.nardini@gmail.com \and tomas.tangarife@ens-lyon.fr}}

\maketitle
\begin{abstract}
We discuss the dynamics of zonal (or unidirectional) jets for barotropic
flows forced by Gaussian stochastic fields with white in time correlation
functions. This problem contains the stochastic dynamics of 2D Navier-Stokes
equation as a special case. We consider the limit of weak forces and
dissipation, when there is a time scale separation between the inertial
time scale (fast) and the spin-up or spin-down time (large) needed
to reach an average energy balance. In this limit, we show that an
adiabatic reduction (or stochastic averaging) of the dynamics can
be performed. We then obtain a kinetic equation that describes the
slow evolution of zonal jets over a very long time scale, where the
effect of non-zonal turbulence has been integrated out. The main theoretical
difficulty, achieved in this work, is to analyze the stationary distribution
of a Lyapunov equation that describes quasi-Gaussian fluctuations
around each zonal jet, in the inertial limit. This is necessary to
prove that there is no ultraviolet divergence at leading order, in
such a way that the asymptotic expansion is self-consistent. We obtain
at leading order a Fokker--Planck equation, associated to a stochastic
kinetic equation, that describes the slow jet dynamics. Its deterministic
part is related to well known phenomenological theories (for instance
Stochastic Structural Stability Theory) and to quasi-linear approximations,
whereas the stochastic part allows to go beyond the computation of
the most probable zonal jet. We argue that the effect of the stochastic
part may be of huge importance when, as for instance in the proximity
of phase transitions, more than one attractor of the dynamics is present.
\end{abstract}
\keywords{}
\PACS{}

\tableofcontents{}

\section{Introduction}

Turbulence in planetary atmospheres leads very often to self organization
and to jet formation (please see for example the special issue of
Journal of Atmospherical Science, named ``Jets and Annular Structures
in Geophysical Fluids'' that contains the paper \cite{Dritschel_McIntyre_2008JAtS}).
Those jet behaviors are at the basis of midlatitude atmosphere dynamics
\cite{VallisBook} and quantifying their statistics is fundamental
for understanding climate dynamics. A similar self-organization into
jets has also been observed in two-dimensional turbulence \cite{Bouchet_Simonnet_2008,Yin_Montgomery_Clercx_2003PhFluids}.

In this paper, we study the jet formation problem in the simplest
possible theoretical framework: the two-dimensional equations for
a barotropic flow with a beta effect $\beta$. These equations, also
called the barotropic quasi-geostrophic equations, are relevant for
the understanding of large scale planetary flows \cite{PedloskyBook}.
When $\beta=0$, they reduce to the two-dimensional Euler or Navier-Stokes
equations. All the formal theoretical framework developed in this
work could be easily extended to the equivalent barotropic quasi-geostrophic
model (also called the Charney--Hasegawa--Mima equation), to the multi-layer
quasi-geostrophic models or to quasi-geostrophic models for continuously
stratified fluids \cite{PedloskyBook}. 

The aim of this work is to consider an approach based on statistical
mechanics. The equilibrium statistical mechanics of two-dimensional
and quasi-two-dimensional turbulence is now well understood \cite{BouchetVenaille-PhysicsReport,Robert:1990_CRAS,Miller:1990_PRL_Meca_Stat,Robert:1991_JSP_Meca_Stat,Majda_Wang_Book_Geophysique_Stat}:
it explains self-organization and why zonal jets (east-west jets)
are natural attractors of the dynamics. However, a drawback of the
equilibrium approach is that the set of equilibrium states is huge,
as it is parametrized by energy, enstrophy and all other inertial
invariant of the dynamics. Moreover, whereas many observed jets are
actually close to equilibrium states, some other jets, for instance
Jupiter's ones, seem to be far from any equilibrium state. It is thus
essential to consider a non-equilibrium statistical mechanics approach,
taking into account forces and dissipation, in order to understand
how real jets are actually selected by a non-equilibrium dynamics.
In this work, we consider the case when the equations are forced by
a stochastic force. We then use classical tools of statistical mechanics
and field theory (stochastic averaging, projection techniques) in
order to develop the kinetic theory from the original dynamics.

Any known relevant kinetic approach is associated with an asymptotic
expansion where a small parameter is clearly identified. Our small
parameter $\alpha$ \cite{Bouchet_Simonnet_2008,BouchetVenaille-PhysicsReport}
is the ratio of an inertial time scale (associated to the jet velocity
and domain size)\textbf{ }divided by the forcing time scale or equivalently
the dissipation time scale (the spin-up or spin-down time scale, needed
to reach a statistically stationary energy balance). As discussed
below, when the force is a white in time stochastic force, the average
energy input rate is known a-priori, and the value of $\alpha$ can
actually be expressed in terms of the control parameters. We call
the limit $\alpha\ll1$ the small force and dissipation limit.

For small $\alpha$, the phenomenology is the following. At any times
the velocity field is close to a zonal jet\textbf{ $\mathbf{v}\simeq U(y,t)\mathbf{e}_{x}$}.
$U(y,t)\mathbf{e}_{x}$ is a steady solution of the inertial dynamics.
This zonal jet then evolves extremely slowly under the effect of the
stochastic forces, of the turbulence (roughly speaking Reynolds stresses),
and of the small dissipation. The non-zonal degrees of freedom have
a turbulent motion which is strongly affected by the dominant zonal
flow $U(y)$. These turbulent fluctuations are effectively damped
by the non-normal linear dynamics close to the zonal jets. This inviscid
damping of the fluctuations is accompanied by a direct transfer of
energy from the fluctuations to the zonal flow. The understanding
and the quantification of these processes is the aim of this work.
The final result will be a kinetic equation that describes the slow
dynamics of the zonal jets, where the effects of the non-zonal turbulence
has been integrated out.

From a technical point of view, we start from the Fokker-Planck equation
describing exactly the evolution of the Probability Density Function
(to be more precise, Functional) (PDF) of the potential vorticity
field (or vorticity field for the two-dimensional Navier-Stokes equations).
We make an asymptotic expansion of this Fokker-Planck equation, by
carefully identifying the slow and fast processes and the order of
magnitude of all fields. At a formal level we follow an extremely
classical route, described for instance in the theory of adiabatic
averaging of stochastic processes \cite{Gardiner_1994_Book_Stochastic},
also called stochastic reduction (see also \cite{RZKhasminskii,brehier2012strong,FW84} for counterpart in mathematics). This leads to a new Fokker-Planck
equation that describes the slow evolution of the zonal jet $U$;
such formal computations are tedious but involve no difficulties.

The main theoretical challenge is to check that the asymptotic expansion
is self-consistent. We have to prove that all quantities appearing
in the kinetic theory remain finite, keeping in such a way the order
of magnitude initially assumed. In field problems, like this one,
this is not granted as ultraviolet divergences often occur. For example,
the slow Fokker-Planck equation involves a non-linear force that is
computed from the Ornstein--Uhlenbeck process, corresponding to the
linearized dynamics close to $U$ stochastically forced. This process
is characterized by the two-points correlation function dynamics,
which is described by a Lyapunov equation. We then need to prove that
this Lyapunov equation has a stationary solution for $\alpha=0$,
in order for the theory to be self-consistent. 

A large part of our work deals with the analysis of this Ornstein--Uhlenbeck
process, and the Lyapunov equation, in the limit $\alpha\rightarrow0$.
The fact that the Lyapunov equation has a stationary solution in the
limit $\alpha\rightarrow0$ is striking, as this is the inertial limit
in which the equation for the non-zonal degrees of freedom contains
a forcing term but no dissipation acts to damp them. The issue is
quite subtle, as we prove that the vorticity-vorticity correlation
function diverges point-wise, as expected based on enstrophy conservation.
However we also prove that any integrated quantity, for instance the
velocity-velocity autocorrelation function, reaches a stationary solution
due to the effect of the zonal shear and a combination of phase mixing
and global effects related to the non-normality of the linearized
operator. As discussed thoroughly in the text, this allows to prove
the convergence in the inertial limit of key physical quantities such
as the Reynolds stress. Most of this analysis strongly relies on the
asymptotic behavior of the linearized deterministic Euler equation,
that we have studied for that purpose in a previous paper \cite{Bouchet_Morita_2010PhyD}.

The linearized equation close to a base flow $U(y)$ has a family
of trivial zero modes that correspond to any function of $y$ only.
The main hypothesis of our work is that the base flows $U$ are linearly
stable (they have no exponentially growing mode), but also that they
have no neutral modes besides the trivial zonal ones. A linear dynamics
with no mode at all may seem strange at first sight, but this is possible
for an infinite dimensional non-normal linear operator. Actually,
the tendency of turbulent flows to produce jets that expel modes has
been recognized long ago \cite{kasahara1980effect}. Moreover, as
discussed in \cite{Bouchet_Morita_2010PhyD}, the absence of non-trivial
modes is a generic situation for zonal flows, even if it is certainly
not always the case.\\

We now describe the physics corresponding to the Fokker-Planck equation
for the slow jets dynamics. It is equivalent to a stochastic dynamics
for the velocity field that we call the kinetic equation. The kinetic
equation has a deterministic drift $F(U)$ and a stochastic force
which is a Gaussian process with white in time correlation function.
$F(U)$ is the Reynolds stress that corresponds to the linearized
dynamics close to the base flow $U$. If we consider the deterministic
drift only, the equation is then related to Stochastic Structural
Stability Theory (SSST) first proposed on a phenomenological basis
by Farrell, Ioannou \cite{Farrel_Ioannou,Farrell_Ioannou_JAS_2007,BakasIoannou2013SSST},
for quasi-geostrophic turbulence. It is also related to a quasi-linear
approximation plus an hypothesis where zonal average and ensemble
average are assumed to be the same, as discussed in details in section
\ref{sub:The-quasi-linear-dynamics}. More recently, an interpretation
in terms of a second order closure (CE2) has also been given \cite{Marston-2010-Chaos,Marston-APS-2011Phy,tobias2013direct}.
All these different forms of quasi-linear approximations have thoroughly
been studied numerically, sometimes with stochastic forces and sometimes
with deterministic ones \cite{delsole1996quasi}. Very interesting
empirical studies (based on numerical simulations) have been performed
recently in order to study the validity of this type of approximation
\cite{Marston_Conover_Schneider_JAS2008,Marston-APS-2011Phy,GormanSchneider-QL-GCM,tobias2013direct},
for the barotropic equations or for more complex dynamics. The SSST
equations have also been used to compare theoretical prediction of
the transition from a turbulence without a coherent structure to a
turbulence with zonal jets \cite{Srinivasan-Young-2011-JAS,BakasIoannou2013SSST,ParkerKrommes2013SSST}.
Our first conclusion is that kinetic theory provides a strong support
to quasi-linear types of approximations, in the limit of weak forces
and dissipation $\alpha\ll1$, in order to compute the attractors
for the slow jet dynamics. 

Beside the deterministic drift $F(U)$, the kinetic theory also predicts
the existence of a small noise. Moreover it predicts the Fokker-Planck
equation describing the full Probability Density Functional of the
zonal jet $U$. This was not described, even phenomenologically, in
any previous approach. This is an essential correction in many respects.
First, it allows to describe the Gaussian fluctuations of the zonal
jet. We also note that the probability of arbitrarily large deviations
from the deterministic attractors can be computed from this Fokker-Planck
equation. For instance, we may implement large deviations theory in
the small noise limit. This is typically the kind of result that cannot
be obtained from a standard cumulant expansion or closure based on
moments. The possibility to compute the full PDF is extremely important
especially in cases where the deterministic part of the dynamics has
more than one attractor. This case actually happens, as noticed by
Ioannou and Farrell (see section \ref{sec:Bistability}). Then, our
approach may predict the actual probability of each attractor and
the transition probabilities from one attractor to the other. We remark
anyway that, from a practical point of view, a further step forward
should be done to obtain explicit results in this direction. \\

Our work has been clearly inspired by the kinetic theory of systems
with long range interactions \cite{Landau_Lifshitz_1996_Book,Nicholson_1991,Binney_Tremaine_1987_Galactic_Dynamics},
described at first order by Vlasov equation and at next order by Lenard--Balescu
equation. We have proven in previous works that this kinetic theory
leads to algebraic relaxations and anomalous diffusion \cite{Bouchet_Dauxois:2005_PRE,Bouchet_Dauxois_2005_AlgebraicCorrelations,Yamaguchi_Bouchet_Dauxois_2007_JSMTE_Anomalous_Diffusion,Campa_Dauxois_Ruffo_Revues_2009_PhR...480...57C}.
The Euler equation is also an example of system with long range interaction,
and there is a strong analogy between the 2D Euler and Vlasov equations.
Quasilinear approximation for the relaxation towards equilibria of
either the 2D Euler equation \cite{Chavanis_Quasilinear_2000PhRvL}
or the point vortex dynamics \cite{Dubin_ONeil_1988_PhysRevLett_Kinetic_Point_Vortex,Chavanis_2001PhRvE_64_PointsVortex,Chavanis_houches_2002}
have actually been proposed and studied in the past.

All the above results started from deterministic dynamics, with no
external forces. In order to prepare this work, and to extend these
kinetic theories to the case with non-equilibrium stochastic forces,
we have first considered the Vlasov equation with stochastic forces
\cite{NardiniGuptaBouchet-2012-JSMTE,Nardini_Gupta_Ruffo_Dauxois_Bouchet_2012_kinetic}.
A kinetic equation was then obtained, similar to the one in this paper,
but with much less technical difficulties. The reason is that the
Landau damping, which is then responsible for the inviscid relaxation,
can be studied analytically through explicit formulas at the linear
level \cite{Nicholson_1991,Landau_Lifshitz_1996_Book}. Non-linear
Landau damping has also been recently established \cite{Mouhot_Villani:2009}.
The kinetic equation for the stochastic dynamics \cite{NardiniGuptaBouchet-2012-JSMTE}
has the very interesting property to exhibit phase transitions and
multistability, leading to a dynamics with random transitions from
one attractor to the other \cite{Nardini_Gupta_Ruffo_Dauxois_Bouchet_2012_kinetic}.
We stress again that beside the formal structure, the main theoretical
difficulty is the analysis of the Lyapunov equation which is a central
object of these kinetic theories. In order to extend this approach
to the barotropic equations, the current work discusses the first
study of the Lyapunov equations for either the 2D Euler, the 2D Navier--Stokes
or the barotropic flow equations in the inertial limit.

The barotropic flow equations include the 2D Stochastic Navier-Stokes
equations as a special case. During last decade, a very interesting
set of mathematical works have proved results related to the existence
and uniqueness of invariant measures, their inertial limit, their
ergodicity, the validity of the law of large number \cite{kuksin2001ergodicity,Kuksin_2004_JStatPhys_EulerianLimit,Kuksin_Penrose_2005_JPhysStat_BalanceRelations,flandoli1995ergodicity,mattingly1999elementary,Weinam_Mattingly_2001_Comm_Pure_Appl_Math_Ergodicity_NS,Hairer_Mattingly_2006ergodicity,Hairer_Mattingly_2008spectral,Bricmont_Kupianen_2001_Comm_Math_Phys_Ergodicity2DNavierStokes,da2003ergodicity,shirikyan2004exponential,ferrario1997ergodic}
and of large deviations principles \cite{gourcy2007large,Sritharana_Sundarb_2006,jaksic2012large}.
Some of the results are summarized in a recent book \cite{kuksin2012mathematics}.
In order to be applied for real physical situations, these works should
be extended in order to deal with a large scale dissipation mechanism,
for instance large scale linear friction. We also note an interesting
work considering the 2D Navier-Stokes equations forced by random vorticity
patches \cite{Majda_Wang_Bombardement_2006_Comm}.
In the case when there exists a scale separation between the forcing scales and the largest scale, our theory is probably very close to appraoch through Rapid Distorsion Theory, or WKB Rapid Distorsion Theory (see for instance \cite{nazarenko1999wkb,nazarenko2000nonlinear} for three dimensionnal flows and \cite{Nazarenko_PhysicsLetterA_2000} for two dimensional near wall turbulence).
The mathematical
literature also contains a lot of interesting studies about stochastic
averaging in partial differential equations \cite{kuksin2008khasminskii},
but we do not know any example dealing with the 2D Navier-Stokes equations
or the barotropic flow equations. \\

In section \ref{sec:section-2}, we discuss the model, the energy
and enstrophy balances, non-dimensional parameters, the quasi-linear
approximation and the Fokker--Planck equations for the potential vorticity
PDF. We develop the formal aspects of the kinetic theory, or stochastic
averaging, in section \ref{sec:Stochastic-Averaging}. This section
ends with the derivation of the kinetic equation: the Fokker--Planck
equations for the slow evolution of the jet (\ref{eq:Fokker-Planck-Zonal}),
and its corresponding stochastic dynamics, Eq. (\ref{eq:Stochastique-Lent})
or (\ref{eq:Stochastique-Lent-U}). Section \ref{sec:Energy-balance}
comes back on energy and enstrophy balances. The analysis of the Lyapunov
equation is performed in section \ref{sec:Lyapunov}. We establish
that it has a stationary solution in the inertial limit, discuss
the divergence of the vorticity-vorticity correlation function, the
nature of its singularity and how they are regularized in a universal
way by a small linear friction or by a small viscosity. Section \ref{sec:Bistability}
discusses the importance of the stochastic part of the kinetic equation,\textbf{
}explains how it predicts zonal jet PDF with jets arbitrarily far
from Gaussian fluctuations, and stress the existence of cases with
multiple attractors, phase transitions and bistability. We discuss
open issues and perspectives in section \ref{sec:Conclusion}. The
paper contains also three appendices reporting the most technical
details of sections \ref{sec:Stochastic-Averaging} and \ref{sec:Lyapunov}.

\section{Quasi-Geostrophic dynamics and Fokker-Planck equation\label{sec:section-2}}

\subsection{2D barotropic flow upon a topography}

We are interested in the non-equilibrium dynamics associated to the
2D motion of a barotropic flow, with topography $h$, on a periodic
domain $\mathscr{D}=[0,2\pi l_{x}L)\times[0,2\pi L)$ with aspect
ratio $1/l_{x}$:
\begin{equation}
\left\lbrace \begin{aligned} & \frac{\partial q}{\partial t}+\mathbf{v}\cdot\mathbf{\nabla}q=-\lambda\omega-\nu_{n,d}\left(-\Delta\right)^{n}\omega+\sqrt{\sigma}\eta,\\
 & \mathbf{v}=\mathbf{e}_{z}\times\mathbf{\nabla}\psi,\quad\omega=\Delta\psi,\quad q=\omega+h(y),
\end{aligned}
\right.\label{eq:barotropic-topography-d}
\end{equation}
where $\omega$, $q$, ${\bf v}$ and $\psi$ are respectively the
vorticity, potential vorticity, the non-divergent velocity, and the
stream function. In the equations above, $\mathbf{r}=(x,y)$ are space
vectors, $x$ is the zonal coordinate and $y$ the meridional one,
$\lambda$ is a Rayleigh (or Ekman) friction coefficient and $\nu_{n,d}$
is the hyper-viscosity coefficient (or viscosity for $n=1$). All
these fields are periodic $f\left(x+2\pi l_{x}L,y\right)=f\left(x,y\right)$
and $f\left(x,y+2\pi L\right)=f\left(x,y\right)$. We have introduced
a forcing term $\eta$, assumed to be a white in time Gaussian noise
with autocorrelation function $\mathbb{E}\left[\eta(\mathbf{r}_{1},t_{1})\eta(\mathbf{r}_{2},t_{2})\right]=C(\mathbf{r}_{1}-\mathbf{r}_{2})\delta(t_{1}-t_{2})$,
where $C$ is an even positive definite function, periodic with respect
to $x$ and $y$. As discussed below, $\sigma$ is the average energy
input rate. When $h=0$, the 2D barotropic flow equations are the
2D-Navier-Stokes equations. Here we assume that the noise autocorrelation
function $C$ is translationally invariant in both direction. The
fact that $C$ is zonally invariant is important for some of the computations
in this paper. However the hypothesis that $C$ is meridionally invariant
is not important and generalization to non-meridionally invariant
forcing would be straightforward.

\paragraph{Dynamical invariants of perfect barotropic flows}

Equations (\ref{eq:barotropic-topography-d}) with $\lambda=\sigma=\nu_{n,d}=0$
describe a perfect barotropic flow. The equations are then Hamiltonian
and they conserve the energy

\begin{equation}
\mathcal{E}[q]=\frac{1}{2}\int_{\mathcal{D}}\mbox{d}\mathbf{r}\:\mathbf{v}^{2}=-\frac{1}{2}\int_{\mathcal{D}}\mbox{d}\mathbf{r}\:\left(q-h\right)\psi,
\end{equation}
 and the Casimir functionals

\begin{equation}
\mathcal{C}_{s}[q]=\int_{\mathcal{D}}\mbox{d}\mathbf{r}\: s(q),
\end{equation}
for any sufficiently regular function $s$. The potential enstrophy
\[
\mathcal{C}_{2}[q]=\frac{1}{2}\int_{\mathcal{D}}\mbox{d}\mathbf{r}\: q^{2}
\]
 is one of the invariants. When $h=0$, the perfect barotropic flow
equations clearly reduce to the 2D Euler equations.

\paragraph{Averaged energy input rate and non-dimensional equations}

Because the force is a white in time Gaussian process, we can compute
a-priori the average, with respect to noise realizations, of the input
rate for quadratic invariants. Without loss of generality, we assume
that 
\[
-2\pi^{2}l_{x}L^{2}\left(\Delta^{-1}C\right)(\mathbf{0})=1,
\]
where $\Delta^{-1}$ denotes the inverse Laplacian operator; indeed,
multiplying $C$ by an arbitrary positive constant amounts at renormalizing
$\sigma$. Then, with the above choice, the average energy input rate
is $\sigma$ and the average energy input rate by unit of mass is
$\epsilon=\sigma/4\pi^{2}l_{x}L^{2}$. Moreover, the average potential
enstrophy input rate is given by 
\[
2\pi^{2}l_{x}L^{2}C(\mathbf{0})\sigma.
\]

We consider the energy balance for equation (\ref{eq:barotropic-topography-d}),
with $E=\mathbb{E}\left[\mathcal{E}[q]\right]$:

\begin{equation}
\frac{dE}{dt}=-2\lambda E-\nu_{n,d}H_{n}+\sigma,\label{eq:Energy-Balance}
\end{equation}
 where $H_{n}=-\mathbb{E}\left[\int_{\mathcal{D}}\psi\left(-\Delta\right)^{n}\omega\right]$.
For most of the turbulent flows we are interested in, the ratio $2\lambda E/\nu_{n,d}H_{n}$
will be extremely large (viscosity is negligible for energy dissipation).
Then, in a statistically stationary regime, the approximate average
energy is $E\simeq\sigma/2\lambda$. We perform a transformation to
non-dimensional variables such that in the new units the domain is
$\mathscr{D}=[0,2\pi l_{x})\times[0,2\pi)$ and the approximate average
energy is 1. This is done introducing a non-dimensional time $t'=t/\tau$
and a non-dimensional spatial variable $\mathbf{r}'=\mathbf{r}/L$
with $\tau=L^{2}\sqrt{2\lambda/\sigma}$. The non-dimensional physical
variables are $q'=\tau q$, $\mathbf{v}'=\tau\mathbf{v}/L$, $h'=\tau h$,
and the non-dimensional parameters are defined by
\[
\alpha=\lambda\tau=L^{2}\sqrt{\frac{2\lambda^{3}}{\sigma}}=\frac{L}{2\pi}\sqrt{\frac{2\lambda^{3}}{\epsilon l_{x}}},
\]
 $\nu_{n}=\nu_{n,d}\tau/L^{2n}=\nu_{n,d}\sqrt{2\lambda/\sigma}/L^{2n-2}$.
We consider a rescaled stochastic Gaussian field $\eta'$ with $\mathbb{E}\left[\eta'(\mathbf{r}'_{1},t'_{1})\eta(\mathbf{r}'_{2},t'_{2})\right]=C'(\mathbf{r}'_{1}-\mathbf{r}'_{2})\delta(t'_{1}-t'_{2})$
with $C'(\mathbf{r}')=L^{4}C(\mathbf{r})$. Performing the non-dimensionalization
procedure explained above, the barotropic equations read 
\begin{equation}
\left\lbrace \begin{aligned} & \frac{\partial q}{\partial t}+\mathbf{v}\cdot\mathbf{\nabla}q=-\alpha\omega-\nu_{n}\left(-\Delta\right)^{n}\omega+\sqrt{2\alpha}\eta,\\
 & \mathbf{v}=\mathbf{e}_{z}\times\mathbf{\nabla}\psi,\quad\omega=\Delta\psi,\quad q=\omega+h(y),
\end{aligned}
\right.\label{eq:barotropic-topography}
\end{equation}
where, for easiness in the notations, we drop here and in the following
the primes. We note that in non-dimensional units, $\alpha$ represents
an inverse Reynolds number based on the large scale dissipation of
energy and $\nu_{n}$ is an inverse Reynolds number based on the viscosity
or hyper-viscosity term that acts predominantly at small scales. The
non dimensional energy balance is

\begin{equation}
\frac{dE}{dt}=-2\alpha E-\nu_{n}H_{n}+2\alpha.\label{eq:Energy-Balance-sans-dimension}
\end{equation}

\subsection{2D barotropic flow with a beta-effect}

For a 2D flow on a rotating sphere, the Coriolis parameter depends
on the latitude. For mid-latitudes, such a dependence can be approached
by a linear function, the so-called beta effect: this corresponds
to consider a linear topography function $h(y)=\beta_{d}y$. In a
periodic geometry, such a function is meaningless, and so is the associated
potential vorticity $q=\omega+h$. However, consistent equations can
be written for the vorticity $\omega$ and the velocity: 
\begin{equation}
\left\lbrace \begin{aligned} & \frac{\partial\omega}{\partial t}+\mathbf{v}\cdot\mathbf{\nabla}\omega+\beta v_{y}=-\alpha\omega-\nu_{n}\left(-\Delta\right)^{n}\omega+\sqrt{2\alpha}\eta,\\
 & \mathbf{v}=\mathbf{e}_{z}\times\mathbf{\nabla}\psi,\quad\omega=\Delta\psi
\end{aligned}
\right.\label{eq:barotropic-beta}
\end{equation}
Observe that, in this case, we have directly written the equations
in a non-dimensional form; the parameter $\beta$ as a function of
dimensional quantities is given by 
\[
\beta=L^{3}\sqrt{\frac{2\lambda}{\sigma}}\beta_{d}=\left(\frac{L}{L_{\beta}}\right)^{2},
\]
where $\beta_{d}$ is the dimensional beta effect, and we have defined
a Rhines scale $L_{\beta}=\sqrt{L/\tau\beta_{d}}$ based on the large
scale velocity $U=L/\tau$.  

 In the case of the barotropic model, we note that an alternative choice for the non-dimensional parameters is often used (see for instance \cite{danilov2004scaling,galperin2010geophysical}), based on a Rhines scales built with an estimate of the RMS velocity  $U_{\textrm{rms}}=\sqrt{\epsilon/\lambda}$ leading to $L_{\textrm{Rhines}}=\left(\epsilon/\beta^2 \lambda\right)^{1/4}$. Such a choice, may be more relevant in regimes where the jet width is actually of the order of magnitude of this Rhine scale. This choice formally leads to same non-dimensional equations, but the expression of the non-dimensional quantities as a function of the dimensional ones are different. As far as the following theory is concerned, the best choice of time unit is the relaxation time of non-zonal perturbations. This time scale may be regime dependent.\\

The perfect barotropic equations ($\alpha=\nu_{n}=0$), in doubly
periodic geometry, with beta effect $(\beta\neq0$), conserve only
two quantities: the kinetic energy

\begin{equation}
\mathcal{E}[\omega]=-\frac{1}{2}\int_{\mathcal{D}}\mathrm{d}\mathbf{r}\,\omega\psi,
\end{equation}
 and the enstrophy

\begin{equation}
\mathcal{Z}[\omega]=\frac{1}{2}\int_{\mathcal{D}}\mathrm{d}\mathbf{r}\,\omega^{2}.
\end{equation}
 For $\beta=0$, the beta-plane equations reduce to the 2D-Navier
Stokes equations discussed in previous section.\\

For sake of simplicity, in the following, we consider the case of
a viscous dissipation $n=1$, and we denote the viscosity $\nu=\nu_{1}$.
One of the consequences of our theoretical work is that, in the limit
$\alpha\ll1$ and $\nu_{n}\ll\alpha$, the results will be independent
on $\nu_{n}$ for generic cases, and at leading order in the small
parameters. 

\subsection{Decomposition into zonal and non-zonal flow}

The physical phenomenon we are interested in is the formation of large
scales structures (jets and vortices). Such large scale features are
slowly dissipated, mainly due to the friction $\alpha$. This dissipation
is balanced by Reynolds stresses due to the transfer of energy from
the forcing scale until the scale of these structures. This phenomenology
is commonly observed in planetary atmospheres (Earth, Jupiter, Saturn)
and in numerical simulations. For the barotropic equations (\ref{eq:barotropic-topography})
and (\ref{eq:barotropic-beta}), the regime corresponding to this
phenomenology is observed when $\alpha\ll1$ (time scale for forcing
and dissipation $1/\lambda$ much larger than time scale for the inertial
dynamics $\tau$) and when $2\lambda E/\nu_{n,d}H_{n}\gg1$ (turbulent
regime). For this reason, we study in the following the limit $\nu_{n}\rightarrow0$,
$\alpha\rightarrow0$ ($\alpha\ll1$ and $\nu_{n}\ll\alpha$).

For sake of simplicity, we consider below the case when the zonal
symmetry (invariance by translation along $x$) is not broken. Then
the large scale structure will be a zonal jet characterized by either
a zonal velocity field $\mathbf{v}_{z}=U(y)\mathbf{e}_{x}$ or its
corresponding zonal potential vorticity $q_{z}(y)=-U'(y)+h(y)$. For
reasons that will become clear in the following discussion (we will
explain that this is a natural hypothesis and prove that it is self-consistent
in the limit $\alpha\ll1$), the perturbation to this zonal velocity
field is of order $\sqrt{\alpha}$.

Defining the zonal projection $\left\langle .\right\rangle $ as 
\[
\left\langle f\right\rangle (y)=\frac{1}{2\pi l_{x}}\int_{0}^{2\pi l_{x}}\mbox{d}x\, f({\bf r}),
\]
 the zonal part of the potential vorticity will be denoted by $q_{z}\equiv\left\langle q\right\rangle $;
the rescaled non-zonal part of the flow $q_{m}=\omega_{m}$ is then
defined through the decomposition 
\[
q({\bf r})=q_{z}(y)+\sqrt{\alpha}\omega_{m}({\bf r}).
\]
 The zonal and non-zonal velocities are then defined through $U'(y)=-q_{z}(y)+h(y)$,
the periodicity condition, and 
\[
\mathbf{v}({\bf r})=U(y)\mathbf{e}_{x}+\sqrt{\alpha}\mathbf{v}_{m}({\bf r}).
\]

We now project the barotropic equation (\ref{eq:barotropic-topography})
into zonal 
\begin{equation}
\frac{\partial q_{z}}{\partial t}=-\alpha\frac{\partial}{\partial y}\left\langle v_{m}^{(y)}\omega_{m}\right\rangle -\alpha\omega_{z}+\nu\frac{\partial^{2}\omega_{z}}{\partial y^{2}}+\sqrt{2\alpha}\eta_{z}\label{eq:Zonal-PV-Evolution}
\end{equation}
 and non-zonal part 
\begin{equation}
\frac{\partial\omega_{m}}{\partial t}+L_{U}\left[\omega_{m}\right]=\sqrt{2}\eta_{m}-\sqrt{\alpha}\mbox{\ensuremath{\mathbf{v}}}_{m}.\nabla\omega_{m}+\sqrt{\alpha}\left\langle \mbox{\ensuremath{\mathbf{v}}}_{m}.\nabla\omega_{m}\right\rangle ,\label{eq:Meridional-PV-Evolution}
\end{equation}
 where $\eta_{z}=\left\langle \eta\right\rangle $ is a Gaussian field
with correlation function $\mathbb{E}\left[\eta_{z}(y_{1},t_{1})\eta_{z}(y_{2},t_{2})\right]=C_{z}(y_{1}-y_{2})\delta(t_{1}-t_{2})$
with $C_{z}=\left\langle C\right\rangle $, $\eta_{m}=\eta-\left\langle \eta\right\rangle $
is a Gaussian field with correlation function $\mathbb{E}\left[\eta_{m}({\bf r}_{1},t_{1})\eta_{m}({\bf r}_{2},t_{2})\right]=C_{m}({\bf r}_{1}-{\bf r}_{2})\delta(t_{1}-t_{2})$
with $C_{m}=C-\left\langle C\right\rangle $, and where 
\begin{equation}
L_{U}\left[\omega_{m}\right]=U(y)\frac{\partial\omega_{m}}{\partial x}+q'_{z}(y)\frac{\partial\psi_{m}}{\partial x}+\alpha\omega_{m}-\nu\Delta\omega_{m}\quad,\quad\omega_{m}=\Delta\psi_{m}.\label{eq:Linearized-Dynamics}
\end{equation}
Observe that the cross correlation between $\eta_z$ and $\eta_m$ is exactly zero, due to the translational invariance along the zonal direction of $C$. Moreover, in the previous equations, $\frac{\partial\omega_{m}}{\partial t}+L_{U}\left[\omega_{m}\right]=0$
is the deterministic dynamics linearized close to the zonal base flow
$U$, whose corresponding potential vorticity is $q_{z}$. In the
following we will also consider the operator 
\begin{equation}
L_{U}^{0}\left[\omega_{m}\right]=U(y)\frac{\partial\omega_{m}}{\partial x}+q'_{z}(y)\frac{\partial\psi_{m}}{\partial x},\label{eq:Linearized-Dynamics-Inertial}
\end{equation}
which is the operator for the linearized inertial dynamics (with no
dissipation) close to the base flow $U$. In all the following, we
assume that the base flow $U(y)$ is linearly stable (the operator
$L_{U}^{0}$ has no unstable mode). We remark that the action of $L_{U}^{0}$
on zonal functions $f(y)$ is trivial: any zonal function is a neutral
mode of $L_{U}^{0}$. We will thus consider the operator $L_{U}^{0}$
acting on non-zonal functions only (functions which have a zero zonal
average). We assume that $L_{U}^{0}$ restricted to non-zonal functions
has no normal mode at all (which is possible as $L_{U}^{0}$ is a
non-normal operator). This hypothesis will be important for the results
presented in section \ref{sec:Lyapunov}. This hypothesis may seem
restrictive, but as explained in \cite{Bouchet_Morita_2010PhyD} this
is a generic case, probably the most common one.\\

The equation for the zonal potential vorticity evolution (\ref{eq:Zonal-PV-Evolution})
can readily be integrated in order to get an equation for the zonal
flow evolution 
\[
\frac{\partial U}{\partial t}=\alpha\left\langle v_{m}^{(y)}\omega_{m}\right\rangle -\alpha U+\nu\frac{\partial^{2}U}{\partial y^{2}}+\sqrt{2\alpha}\eta_{U},
\]
 where $\eta_{U}$ is a Gaussian field with correlation function\textbf{
$\mathbb{E}\left[\eta_{U}(y_{1},t_{1})\eta_{U}(y_{2},t_{2})\right]=C_{U}(y_{1}-y_{2})\delta(t_{1}-t_{2})$
}with $\frac{d^{2}C_{U}}{dy_1^{2}}(y_{1}-y_{2})=-C_{z}(y_{1}-y_{2})$.

We see that the zonal potential vorticity is coupled to the non-zonal
one through the zonal average of the advection term. If our rescaling
of the equations is correct, we clearly see that the natural time
scale for the evolution of the zonal flow is $1/\alpha$. By contrast
the natural time scale for the evolution of the non-zonal perturbation
is one. These remarks show that in the limit $\alpha\ll1$, we have
a time scale separation between the slow zonal evolution and a rapid
non-zonal evolution. Our aim is to use this remark in order to i)
describe precisely the stochastic behavior of the Reynolds stress
in this limit by integrating out the non-zonal turbulence, ii) prove
that our rescaling of the equations and the time scale separation
hypothesis are self-consistent. 

The term $\sqrt{\alpha}\mbox{\ensuremath{\mathbf{v}}}_{m}.\nabla\omega_{m}-\sqrt{\alpha}\left\langle \mbox{\ensuremath{\mathbf{v}}}_{m}.\nabla\omega_{m}\right\rangle $
describes the interactions between non-zonal degrees of freedom (sometimes
called eddy-eddy interactions). If our rescaling is correct, these
terms should be negligible at leading order. Neglecting these terms
leads to the so called quasi-linear dynamics, which is described in
next section.

\subsection{The quasi-linear dynamics\label{sub:The-quasi-linear-dynamics}}

By neglecting the interactions between non-zonal degrees of freedom
in (\ref{eq:Meridional-PV-Evolution}), one obtains the quasi-linear
approximation of the barotropic equations:

\begin{equation}
\left\lbrace \begin{aligned} & \frac{\partial q_{z}}{\partial t}=-\alpha\frac{\partial}{\partial y}\left\langle v_{m}^{(y)}\omega_{m}\right\rangle -\alpha\omega_{z}+\nu\frac{\partial^{2}\omega_{z}}{\partial y^{2}}+\sqrt{2\alpha}\eta_{z},\\
 & \frac{\partial\omega_{m}}{\partial t}+L_{U}\left[\omega_{m}\right]=\sqrt{2}\eta_{m}
\end{aligned}
\right.\label{eq:quasi-lineaire}
\end{equation}
In two-dimensional and geostrophic turbulence, the fluctuations around
a given mean flow are often weak, so this approximation is natural
in this context. If our rescaling is relevant, this corresponds to
the limit $\alpha\ll1$.

The approximation leading to the quasi-linear dynamics amounts at
suppressing some of the triads interactions. As a consequence, the\textbf{
}inertial\textbf{ }quasi-linear dynamics has the same quadratic invariants
as the initial barotropic equations: the energy and potential enstrophy.

As discussed in the previous paragraphs, and as can be seen in (\ref{eq:quasi-lineaire}),
in the regime $\alpha\ll1$, it seems natural to assume that there
is a separation between the time scale for the fluctuation dynamics
(second equation of (\ref{eq:quasi-lineaire})), which is of order
1, and the time scale for the zonal flow (first equation of (\ref{eq:quasi-lineaire})),
which is of order $1/\alpha$. At leading order, the evolution of
$\omega_{m}$ can be considered with the zonal velocity profile held
fixed. The second equation of (\ref{eq:quasi-lineaire}), with a fixed
velocity profile, is then linear. Thus, denoting by $\mathbb{E}$
the average over the realization of the noise $\eta_{m}$ for fixed
$U$, the distribution of $\omega_{m}$ is completely characterized
by the two-points correlation function $g(\mathbf{r}_{1},\mbox{\ensuremath{\mathbf{r}}}_{2},t)=\mathbb{E}\left[\omega_{m}({\bf r}_{1},t)\omega_{m}({\bf r}_{2},t)\right]$.
The evolution of $g$ is given by the so-called Lyapunov equation,
which is obtained by applying the It\={o} formula (with $U$ fixed): 
\[
\frac{\partial g}{\partial t}+L_{U}^{(1)}g+L_{U}^{(2)}g=2C_{m},
\]
 where $ $$L_{U}^{(1)}$ (resp. $L_{U}^{(2)}$) is the linearized
operator $L_{U}$ (\ref{eq:Linearized-Dynamics}) acting on the variable
$ $$\mathbf{r}_{1}$ (resp. $\mathbf{r}_{2}$).

If we assume that this Gaussian process has a stationary distribution,
then considering again the time scale separation $\alpha\ll1$, one
is led to the natural hypothesis that at leading order 
\begin{equation}
\frac{\partial q_{z}}{\partial t}=-\alpha\frac{\partial}{\partial y}\mathbb{E}_{U}\left\langle v_{m}^{(y)}\omega_{m}\right\rangle -\alpha\omega_{z}+\nu\frac{\partial^{2}\omega_{z}}{\partial y^{2}}+\sqrt{2\alpha}\eta_{z},\label{eq:kinetic-equation-determinist}
\end{equation}
 where $\mathbb{E}_{U}\left\langle v_{m}^{(y)}\omega_{m}\right\rangle $
is the average over the stationary Gaussian process corresponding
to the dynamics given by the second equation of (\ref{eq:quasi-lineaire})
at fixed $U$. In our problem, invariant by translation along the
zonal direction, we note that $-\mathbb{E}_{U}\left\langle v_{m}^{(y)}\omega_{m}\right\rangle $
is the divergence of a Reynolds stress \cite{Pope_2000_Livre}. We
call this equation the kinetic equation, because it is the analogous
of the kinetic equation obtained in a deterministic context in the
kinetic theory of plasma physics \cite{Nicholson_1991,Landau_Lifshitz_1996_Book},
systems with long-range interactions \cite{Bouchet:2004_PRE_StochasticProcess,Bouchet_Dauxois:2005_PRE,Bouchet_Dauxois_2005_AlgebraicCorrelations}
and the 2D Euler equations \cite{Chavanis_Quasilinear_2000PhRvL,Chavanis_2001PhRvE_64_PointsVortex}.
The time scale separation hypothesis, leading to consider the asymptotic
solution of the Lyapunov equation is referred to as the Bogolyubov
hypothesis in classical kinetic theory \cite{Nicholson_1991,Landau_Lifshitz_1996_Book}.

Another model related to the quasi-linear dynamics is

\textbf{
\begin{equation}
\left\lbrace \begin{aligned} & \frac{\partial q_{z}}{\partial t}=-\alpha\frac{\partial}{\partial y}\mathbb{E}_{m}\left[\left\langle v_{m}^{(y)}\omega_{m}\right\rangle \right]-\alpha\omega_{z}+\nu\frac{\partial^{2}\omega_{z}}{\partial y^{2}}+\sqrt{2\alpha}\eta_{z}\\
 & \frac{\partial g}{\partial t}+\left(L_{U}^{(1)}+L_{U}^{(2)}\right)g=2C_{m},
\end{aligned}
\right.\label{eq:SSST}
\end{equation}
}where\textbf{ $\mathbb{E}_{m}$ }is the average corresponding to
the Gaussian process with two-point correlation function $g({\bf r},{\bf r}',t)=\mathbb{E}_{m}\left[\omega_{m}({\bf r},t)\omega_{m}({\bf r}',t)\right]$.
By contrast to the kinetic equation (\ref{eq:kinetic-equation-determinist}),
in equation (\ref{eq:SSST}) $g$ and $U$ (or $q_{z}$) evolve simultaneously.
The Reynolds stress divergence $-\mathbb{E}_{m}\left[\left\langle v_{m}^{(y)}\omega_{m}\right\rangle \right]$,
or the gradient of Reynolds stress divergence $-\frac{\partial}{\partial y}\mathbb{E}_{m}\left[\left\langle v_{m}^{(y)}\omega_{m}\right\rangle \right]$,
can be computed as a linear transform of $g$. This model, first proposed
by Farrell and Ioannou \cite{Farrel_Ioannou,Farrell_Ioannou_JAS_2007,BakasIoannou2013SSST},
is referred to as Structural Stochastic Stability Theory (SSST). Farrell
and Ioannou have also used this model in a deterministic context \cite{delsole1996quasi},
where the correlation $C_{m}$ is then added phenomenologically in
order to model the effect of small scale turbulence (the effect of
the neglected non linear terms in the quasi-linear dynamics). This
model, or an analogous one in a deterministic context, is also referred
to as CE2 (Cumulant expansion of order 2) \cite{Marston_Conover_Schneider_JAS2008,Marston-2010-Chaos,Marston-APS-2011Phy}.

As previously observed, the second equation in (\ref{eq:SSST}) can
be deduced from the second equation of (\ref{eq:quasi-lineaire})
as an average over the realizations of the non-zonal part of the noise
$\eta_{m}$, with $U$ held fixed. It can be considered as an approximation of the quasilinear
dynamics, where the instantaneous value of the non-linear term $\left\langle v_{m}^{(y)}\omega_{m}\right\rangle $
is replaced by its ensemble average, with $U$ held fixed. We see
no way how this could be relevant without a good time scale separation
$\alpha\ll1$. With a good time scale separation, both the quasilinear
dynamics (\ref{eq:quasi-lineaire}) and SSST-CE2 (\ref{eq:SSST})
are likely to be described at leading order by the kinetic equation
(\ref{eq:kinetic-equation-determinist}). We see no reason how SSST
dynamics could be relevant to describe the quasilinear dynamics beyond
the limit where it is an approximation of the kinetic equation. However,
from a practical point of view, SSST-CE2 is extremely interesting
as it provides a closed dynamical system which may be extremely easily
computed numerically (by comparison to both direct numerical simulations,
the quasilinear dynamics or the kinetic equation) in order to obtain
the average zonal velocity and its Gaussian corrections. It is probably
the best tool to understand qualitatively the dynamics of barotropic
flows in the limit of small forces and dissipations.

\subsection{Fokker-Planck equation}

We will use the remark that we have a time scale separation between
zonal and non-zonal degrees of freedom in order to average out the
effect of the non-zonal turbulence. This corresponds to treat the
zonal degrees of freedom adiabatically. This kind of problems are
described in the theoretical physics literature as adiabatic elimination
of fast variables \cite{Gardiner_1994_Book_Stochastic} or stochastic
averaging in the mathematical literature. Our aim is to perform the
stochastic averaging of the barotropic flow equation and to find the
equation that describes the slow evolution of zonal flows.

There are several ways to perform this task, for instance using path
integral formalism, working directly with the stochastic equations,
using Mori--Zwanzig or projection formalisms, and so on. A typical
way in turbulence is to write the hierarchy of equations for the moments
of the hydrodynamic variables.\textbf{ }As discussed in the introduction
and in section\textbf{ }\ref{sec:Bistability} this approach may lead
to interesting results sometimes, but it can sometimes be misleading,
and when it works it can describe quasi-gaussian fluctuations at best.\textbf{
}The formalism we find the simplest, the more convenient, and the
more amenable to a clear mathematical study is the one based on the
(functional) Fokker-Planck equations.

At a formal level, we will perform adiabatic reduction of the Fokker-Planck
equation using the classical approach, as described for instance in
Gardiner textbook \cite{Gardiner_1994_Book_Stochastic}. Whereas Gardiner
treats examples with a finite number of degrees of freedom, we are
concerned in this paper with a field problem. At a formal level, this
does not make much difference and the formalism can be directly generalized
to this case. At a deeper level this however hides some mathematical
difficulties,\textbf{ }some of them being related to ultraviolet divergences. We will show that such ultraviolet divergences do not occur, at least
for the quantities arising at leading order.\\

The starting point of our analysis is to write the Fokker-Planck equation
associated with Eq. (\ref{eq:Zonal-PV-Evolution}) and (\ref{eq:Meridional-PV-Evolution}).
We consider the time dependent probability distribution function $P_{t}\left[q_{z},\omega_{m}\right]$
for the potential vorticity field and, for easiness in the notations,
we omit time $t$ in the following. The distribution $P\left[q_{z},\omega_{m}\right]$
is a functional of the two fields $q_{z}$ and $\omega_{m}$ and is
a formal generalization of the probability distribution function for
variables in finite dimensional spaces. From the stochastic dynamics
(\ref{eq:Zonal-PV-Evolution}-\ref{eq:Meridional-PV-Evolution}),
using standard techniques (functional It\={o} calculus), one can prove
that $P$ solves the Fokker-Planck equation 
\begin{equation}
\frac{\partial P}{\partial t}=\mathcal{L}_{0}P+\sqrt{\alpha}\mathcal{L}_{n}P+\alpha\mathcal{L}_{z}P,\label{eq:Evolution-P}
\end{equation}
where the leading order term
is
\begin{equation}
\mathcal{L}_{0}P\equiv\int\mbox{d}\mathbf{r}_{1}\,\frac{\delta}{\delta\omega_{m}(\mathbf{r}_{1})}\left[L_{U}\left[\omega_{m}\right](\mathbf{\mathbf{r}}_{1})P+\int\mbox{d}\mathbf{r}_{2}\, C_{m}(\mathbf{r}_{1}-\mathbf{r}_{2})\frac{\delta P}{\delta\omega_{m}(\mathbf{r}_{2})}\right],\label{eq:L0}
\end{equation}
the zonal part of the perturbation is
\begin{equation}
\mathcal{L}_{z}P\equiv\int\mbox{d}y_{1}\,\frac{\delta}{\delta q_{z}(y_{1})}\left[\left(\frac{\partial}{\partial y}\left\langle v_{m}^{(y)}\omega_{m}\right\rangle (y_{1})+\omega_{z}(y_{1})-\frac{\nu}{\alpha}\Delta\omega_{z}(y_{1})\right)P+\int\mbox{d}y_{2}\, C_{z}(y_{1}-y_{2})\frac{\delta P}{\delta q_{z}(y_{2})}\right],\label{eq:Lz}
\end{equation}
and the nonlinear part of the perturbation is
\begin{equation}
\mathcal{L}_{n}P\equiv\int\mbox{d}\mathbf{r}_{1}\,\frac{\delta}{\delta\omega_{m}(\mathbf{r}_{1})}\left[\left(\mbox{\ensuremath{\mathbf{v}}}_{m}.\nabla\omega_{m}(\mathbf{\mathbf{r}}_{1})-\langle\mbox{\ensuremath{\mathbf{v}}}_{m}.\nabla\omega_{m}(\mathbf{\mathbf{r}}_{1})\rangle\right)P\right].\label{eq:Ln}
\end{equation}
In previous equations, $\frac{\delta}{\delta q(y)}$ are functional derivatives with
respect to the field $q$ at point $y$.  We observe that $\mathcal{L}_{n}$ is a linear operator which describes
the effect of the non-linear terms in the fluid mechanics equations.

The operator $\mathcal{L}_{0}$ is the Fokker-Planck operator that
corresponds to the linearized dynamics close to the zonal flow $U$,
forced by a Gaussian noise, white in time and with spatial correlations
$C_{m}$. This Fokker-Planck operator acts on the non-zonal variables
only and depends parametrically on $U$. This is in accordance with
the fact that on time scales of order $1$, the zonal flow does not
evolve and only the non-zonal degrees of freedom evolve significantly.
It should also be remarked that this term contains dissipation terms
of order $\alpha$ and $\nu$. These dissipation terms could have
equivalently been included in $\mathcal{L}_{z}$, but we keep them
on $\mathcal{L}_{0}$ for later convenience. However it will be an
important part of our work to prove that in the limit $\nu\ll\alpha\ll1$,
at leading order, the operator $\mathcal{L}_{0}$ with or without
these dissipation terms will have the same stationary distributions.
This is a crucial point that will be made clear in section \ref{sec:Lyapunov}.

At order $\sqrt{\alpha}$, the nonlinear part of the perturbation
$\mathcal{L}_{n}$ contains the non-linear interactions between non-zonal
degrees of freedom. We will prove that at leading order $\mathcal{L}_{n}$
has no effect, justifying the quasilinear approach. This is non-trivial
as $\mathcal{L}_{n}$ formally arise at an order in $\sqrt{\alpha}$
lower than $\mathcal{L}_{z}$. At order $\alpha$, the zonal part
of the perturbation $\mathcal{L}_{z}$ contains the terms that describe
the coupling between the zonal and non-zonal flow, the dynamics due
to friction acting on zonal scales and the zonal part of the stochastic
forces.

\section{Stochastic averaging and the Fokker-Planck equation for the slow
evolution of zonal velocity profiles}

\label{sec:Stochastic-Averaging}

In this section, we formally perform the perturbative expansion in
power of $\sqrt{\alpha}$, for the Fokker-Planck equation (\ref{eq:Evolution-P}).
The aim of the computation is to obtain a new equation describing
the slow zonal part only. We will compute explicitly only the leading
order relevant terms. This formal computation will make sense only
if the Gaussian process corresponding to a stochastically forced linearized
equation has a stationary solution. That last point is thus the real
physical issue and the most important mathematical one. We consider
this question in section \ref{sec:Lyapunov}.

\subsection{Stationary distribution of the fast non-zonal variables}

At leading order when $\alpha$ is small, we have 
\begin{equation}
\frac{\partial P}{\partial t}=\mathcal{L}_{0}P.\label{eq:Fokker-Planck-lineaire}
\end{equation}

\noindent Let us first consider the special case when the zonal flow
is a deterministic one: $P\left[q,\omega_{m}\right]=\delta\left(q-q_{z}\right)Q(\omega_{m})$.
Then, (\ref{eq:Fokker-Planck-lineaire}) is a Fokker-Planck equation
corresponding to the dynamics of the non-zonal degrees of freedom
only. It is the Fokker-Planck equation associated to the barotropic
equations linearized around a fixed base flow with zonal velocity
$U$ and forced by a Gaussian noise delta correlated in time and with
spatial correlation function $C_{m}$:
\begin{equation}
\frac{\partial\omega_{m}}{\partial t}+L_{U}[\omega_{m}]=\sqrt{2}\eta_{m}\quad,\quad\mathbb{E}\left[\eta_{m}({\bf r}_{1},t_{1})\eta_{m}({\bf r}_{2},t_{2})\right]=C_{m}({\bf r}_{1}-{\bf r}_{2})\delta(t_{1}-t_{2})\label{eq:dynamique-lineaire}
\end{equation}
 When $U$ is held fixed, this is a linear stochastic Gaussian process,
or Ornstein-Uhlenbeck process. Thus, it is completely characterized
by the two-points correlation function $g(\mathbf{r}_{1},\mbox{\ensuremath{\mathbf{r}}}_{2},t)=\mathbb{E}\left[\omega_{m}({\bf r}_{1},t)\omega_{m}({\bf r}_{2},t)\right]$
(the average $\mathbb{E}[\omega_{m}]$ is equal to zero, here the
average $\mathbb{E}$ refers to an average over the realization of
the noise $\eta_{m}$, for fixed $U$). The evolution of $g$ is given
by the so-called Lyapunov equation, which is obtained by applying
the It\={o} formula to (\ref{eq:dynamique-lineaire}): 
\begin{equation}
\frac{\partial g}{\partial t}+L_{U}^{(1)}g+L_{U}^{(2)}g=2C_{m},\label{eq:equation-Lyapunov}
\end{equation}
 where $L_{U}^{(1)}$ (resp. $L_{U}^{(2)}$) is the linearized operator
$L_{U}$ (\ref{eq:Linearized-Dynamics}) acting on the variable $\mathbf{r}_{1}$
(resp. $\mathbf{r}_{2}$).

For fixed $\alpha$ and $\nu$, the linear operator $L_{U}$ (\ref{eq:Linearized-Dynamics})
is dissipative. If the linearized dynamics is stable, then the Ornstein-Uhlenbeck
will have a stationary distribution. However, as we are computing
an asymptotic expansion for small $\alpha$ and $\nu$, we need to
consider the limit $\nu\rightarrow0$ and $\alpha\rightarrow0$ of
this stationary distribution. In order to do that we will consider
the Lyapunov equation 
\begin{equation}
\frac{\partial g}{\partial t}+L_{U}^{0(1)}g+L_{U}^{0(2)}g=2C_{m},\label{eq:equation-Lyapunov-inertiel}
\end{equation}
where $L_{U}^{0}$ is the inertial linear dynamics (\ref{eq:Linearized-Dynamics-Inertial}).
In this section, we assume that the corresponding Gaussian process
has a stationary distribution. This assumption may seem paradoxical
as there is no dissipation in $L_{U}^{0}$; however, it will be proved
in section \ref{sec:Lyapunov} that this is correct. We denote by
$g^{\infty}\left[q_{z}\right]$ the stationary two-points correlation
function, $g^{\infty}[q_{z}](\mathbf{r}_{1},\mbox{\ensuremath{\mathbf{r}}}_{2})=\lim_{t\rightarrow\infty}g(\mathbf{r}_{1},\mbox{\ensuremath{\mathbf{r}}}_{2},t)$,
and by $\left(g^{\infty}\right)^{-1}\left[q_{z}\right](\mathbf{r}_{1},\mbox{\ensuremath{\mathbf{r}}}_{2})$
its inverse%
\footnote{Here, inverse is understood in the linear operator sense: $\int\mbox{d}{\bf r}'\, g^{\infty}[q_{z}](\mathbf{r}_{1},\mbox{\ensuremath{\mathbf{r}}}')\left(g^{\infty}\right)^{-1}[q_{z}](\mathbf{r}',\mbox{\ensuremath{\mathbf{r}}}_{2})=\delta({\bf r}_{1}-{\bf r}_{2})$.%
}. Then, the stationary distribution of the linear equation (\ref{eq:dynamique-lineaire})
close to the base flow with potential vorticity $q_{z}$ is 
\begin{equation}
G\left[q_{z},\omega_{m}\right]=\frac{1}{\sqrt{\det\left[q_{z}\right]}}\exp\left(-\frac{1}{2}\int\mbox{d}\mathbf{r}_{1}\mbox{d}\mathbf{r}_{2}\,\omega_{m}(\mathbf{r}_{1})\left(g^{\infty}\right)^{-1}\left[q_{z}\right](\mathbf{r}_{1},\mbox{\ensuremath{\mathbf{r}}}_{2})\omega_{m}(\mathbf{r}_{2})\right),\label{eq:distribution-gaussienne}
\end{equation}
 where the normalization constant depends on $\det q_{z}$, the determinant
of $2\pi g^{\infty}\left[q_{z}\right]$. The stationary solution to
(\ref{eq:Fokker-Planck-lineaire}) with any initial condition $P\left[q,\omega_{m}\right]=\delta\left(q-q_{z}\right)Q(\omega_{m})$
is thus given by $\delta\left(q-q_{z}\right)G\left[q_{z},\omega_{m}\right]$.

For any $q_{z}$, the average of any observable $A\left[\omega_{m}\right]$
over the stationary Gaussian distribution will be denoted 
\[
\mathbb{E}_{U}\left[A\right]\equiv\int\mathcal{D}\left[\omega_{m}\right]\, A\left[\omega_{m}\right]G\left[q_{z},\omega_{m}\right]
\]
 and, for any two observables $A\left[\omega_{m}\right]$ and $B\left[\omega_{m}\right]$,
the correlation of $A$ at time $t$ with $B$ et time zero will be
denoted 
\begin{equation}
\mathbb{E}_{U}\left[A(t)B(0)\right]\equiv\int\mathcal{D}\left[\omega_{m}\right]\, A\left[\omega_{m}\right]\mbox{e}^{t\mathcal{L}_{0}}\left[B\left[\omega_{m}\right]G\left[q_{z},\omega_{m}\right]\right].\label{eq:Correlation}
\end{equation}
 The covariance of $A$ at time $t$ with $B$ at time zero is denoted
by 
\begin{equation}
\mathbb{E}_{U}\left[\left[A(t)B(0)\right]\right]\equiv\mathbb{E}_{U}\left[\left(A(t)-\mathbb{E}_{U}\left[A\right]\right)\left(B(0)-\mathbb{E}_{U}\left[B\right]\right)\right].\label{eq:Covariance}
\end{equation}

As already pointed out, we assume that for all $q_{z}$ of interest
the Gaussian process corresponding to the inertial linear dynamics
(\ref{eq:Linearized-Dynamics-Inertial}) has a stationary distribution.
Then, the operator $\exp\left(t\mathcal{L}_{0}\right)$ has a limit
for time going to infinity. We can then define $\mathcal{P}$ the
projector on the asymptotic solutions of this equation 
\[
\mathcal{P}\left[P\right]=\underset{t\rightarrow\infty}{\lim}\exp\left(t\mathcal{L}_{0}\right)P
\]
 for any $P$. We observe that, because $\mathcal{L}_{0}$ does not
act on zonal degrees of freedom, $\mathcal{P}\left[\delta\left(q_{z}-q\right)Q(\omega_{m})\right]=\delta\left(q_{z}-q\right)G\left[q_{z},\omega_{m}\right]$
for any $Q$. Moreover, using the linearity of $\mathcal{L}_{0}$,
we conclude that for any $P$ 
\begin{equation}
\mathcal{P}\left[P\right]=G\left[q_{z},\omega_{m}\right]\int\mathcal{D}\left[\omega_{m}\right]\, P\left[q_{z},\omega_{m}\right].\label{eq:Projecteur}
\end{equation}
The above equation expresses that for any $q_{z}$, the fastly evolving
non-zonal degrees of freedom relax to the stationary distribution
$G$. With this last relation, it is easily checked that $\mathcal{P}$
is a projector: $\mathcal{P}^{2}=\mathcal{P}$. Moreover, because
$\mathcal{P}$ commutes with $\mathcal{L}_{0}$ and projects on its
stationary distribution, we have 
\[
\mathcal{P}\mathcal{L}_{0}=\mathcal{L}_{0}\mathcal{P}=0.
\]

\subsection{Adiabatic elimination of the fast variables}\label{Adiabatic-elimination}

We decompose any PDF $P$ into its slow and fast components: 
\[
P=P_{s}+P_{f}
\]
 with $P_{s}\equiv\mathcal{P}P$ and $P_{f}\equiv(1-\mathcal{P})P$.

Using (\ref{eq:Projecteur}), we obtain 
\[
P_{s}[q_{z},\omega_{m}]=R[q_{z}]G[q_{z},\omega_{m}],
\]
 with $R\left[q_{z}\right]=\int\mathcal{D}\left[\omega_{m}\right]\, P\left[q_{z},\omega_{m}\right]$.
$R$ is the marginal distribution of $P$ on the zonal modes and
describes the statistics of the zonal flow; $P_{s}$ describes the
statistics of the zonal flow assuming that for any $q_{z}$, the fast
non-zonal degrees of freedom instantaneously relax to their stationary
Gaussian distribution (\ref{eq:distribution-gaussienne}). We thus
expect $R$ and $P_{s}$ to evolve slowly, and $P_{f}$ contains the
small corrections to $P_{s}$. The goal of the stochastic reduction
presented here is to get a closed equation for the evolution of $R$,
valid for small $\alpha$.

Applying the projector operator $\mathcal{P}$ to Eq. (\ref{eq:Evolution-P})
and using $\mathcal{P}\mathcal{L}_{0}=0$, we get the evolution equation
for $P_{s}$ 
\[
\frac{\partial P_{s}}{\partial t}=\mathcal{P}\left(\alpha\mathcal{L}_{z}+\sqrt{\alpha}\mathcal{L}_{n}\right)(P_{s}+P_{f}).
\]
 As expected, this equation evolves on a time scale much larger than
$1$, and this is due to the relation $\mathcal{P}\mathcal{L}_{0}=0$,
which is essentially the definition of $\mathcal{P}$.

Using (\ref{eq:Projecteur}) and (\ref{eq:Ln}) we remark that $\mathcal{P}\mathcal{L}_{n}$
involves the integral over $\mathcal{D}\left[\omega_{m}\right]$ of
a divergence with respect to $\omega_{m}$. As a consequence, 
\begin{equation}
\mathcal{P}\mathcal{L}_{n}=0.\label{eq:P-L1-nul}
\end{equation}
 An important consequence of this relation is that the first-order
non-linear correction to the quasi-linear dynamics is exactly zero:
\begin{equation}
\frac{\partial P_{s}}{\partial t}=\alpha\mathcal{P}\mathcal{L}_{z}\left(P_{s}+P_{f}\right).\label{eq:Evolution-Ps}
\end{equation}
 Applying now the operator $(1-\mathcal{P})$ to Eq. (\ref{eq:Evolution-P}),
using $(1-\mathcal{P})\mathcal{L}_{0}=\mathcal{L}_{0}(1-\mathcal{P})$
and (\ref{eq:P-L1-nul}), we obtain 
\begin{equation}
\frac{\partial P_{f}}{\partial t}=\mathcal{L}_{0}P_{f}+\left(\sqrt{\alpha}\mathcal{L}_{n}+\alpha(1-\mathcal{P})\mathcal{L}_{z}\right)(P_{s}+P_{f}).\label{eq:Evolution-Pf}
\end{equation}
 Our goal is to solve (\ref{eq:Evolution-Pf}) and to inject the solution
into (\ref{eq:Evolution-Ps}), which will become a closed equation
for the slowly evolving PDF $P_{s}$. We solve equations (\ref{eq:Evolution-Ps},\ref{eq:Evolution-Pf})
using Laplace transform. The Laplace transform of a function of time
$f$ is defined by 
\[
\tilde{f}(s)=\int_{0}^{\infty}\mbox{d}t\,\mbox{e}^{-st}f(t)\,,
\]
where the real part of $s$ is sufficiently large for the integral
to converge. Using $\tilde{\left(\partial_{t}f\right)}(s)=s\tilde{f}(s)-f(0)$
and the fact that the operators don't depend explicitly on time, equations
(\ref{eq:Evolution-Ps}) and (\ref{eq:Evolution-Pf}) become 
\[
\left\lbrace \begin{aligned} & s\tilde{P}_{s}(s)=\alpha\mathcal{P}\mathcal{L}_{z}\left(\tilde{P}_{s}+\tilde{P}_{f}\right)+P_{s}(0)\\
 & s\tilde{P}_{f}(s)=\mathcal{L}_{0}\tilde{P}_{f}+\left(\alpha(1-\mathcal{P})\mathcal{L}_{z}+\sqrt{\alpha}\mathcal{L}_{n}\right)(\tilde{P}_{s}+\tilde{P}_{f})+P_{f}(0).
\end{aligned}
\right.
\]

For simplicity, we assume that the initial condition satisfies $P_{f}(0)=0$.
In most classical cases, this assumption is not a restriction,
as relaxation towards such a distribution is exponential and fast.
For our case, this may be trickier because as discussed in section
\ref{sec:Lyapunov}, the relaxation will be algebraic for time scales
much smaller than $1/\alpha$. However, we do not discuss this point
in detail before the conclusion (section \ref{sec:Conclusion}). 

Denoting by $\frac{1}{\mathcal{L}}$ the inverse of a generic operator
$\mathcal{L}$, the second equation can be solved as 
\[
\tilde{P}_{f}(s)=\sqrt{\alpha}\frac{1}{s-\mathcal{L}_{0}-\sqrt{\alpha}\mathcal{L}_{n}-\alpha(1-\mathcal{P})\mathcal{L}_{z}}\left[\sqrt{\alpha}(1-\mathcal{P})\mathcal{L}_{z}+\mathcal{L}_{n}\right]\tilde{P}_{s}
%
\]
 and this solution can be injected into the first equation: 
\[
s\tilde{P}_{s}(s)=\left[\alpha\mathcal{P}\mathcal{L}_{z}+\alpha^{3/2}\mathcal{P}\mathcal{L}_{z}\frac{1}{s-\mathcal{L}_{0}-\sqrt{\alpha}\mathcal{L}_{n}-\alpha(1-\mathcal{P})\mathcal{L}_{z}}\left[\sqrt{\alpha}(1-\mathcal{P})\mathcal{L}_{z}+\mathcal{L}_{n}\right]\right]\tilde{P}_{s}+P_{s}(0)\,.
%
%
\]
 At this stage we made no approximation and this last result is exact.

We look at an expansion in powers of $\sqrt{\alpha}$ for the above
equation. At order 4 in $\sqrt{\alpha}$, we get 
\begin{align*}
s\tilde{P}_{s}(s)= & \left\{ \alpha\mathcal{P}\mathcal{L}_{z}+\alpha^{3/2}\mathcal{P}\mathcal{L}_{z}\frac{1}{s-\mathcal{L}_{0}}\mathcal{L}_{n}\right.\\
 & \left.+\alpha^{2}\mathcal{P}\mathcal{L}_{z}\left[\frac{1}{s-\mathcal{L}_{0}}(1-\mathcal{P})\mathcal{L}_{z}+\left(\frac{1}{s-\mathcal{L}_{0}}	\mathcal{L}_{n}\right)^{2}\right]\right\} \tilde{P}_{s}+P_{s}(0)+\mathcal{O}\left(\alpha^{5/2}\right).
\end{align*}
 We use that $\frac{1}{s-\mathcal{L}_{0}}$ is the Laplace transform
of $\exp\left(t\mathcal{L}_{0}\right)$ and that the inverse Laplace transform of a product is a convolution
to conclude that
\begin{align*}
\frac{\partial P_{s}}{\partial t}= & \alpha\mathcal{P}\mathcal{L}_{z}P_{s}+\alpha^{3/2}\mathcal{P}\mathcal{L}_{z}\int_{0}^{\infty}\mbox{d}t'\,\left[\mbox{e}^{t'\mathcal{L}_{0}}\mathcal{L}_{n}P_{s}(t-t')\right]\\
 & +\alpha^{2}\mathcal{P}\mathcal{L}_{z}\int_{0}^{\infty}\mbox{d}t'\,\mbox{e}^{t'\mathcal{L}_{0}}(1-\mathcal{P})\mathcal{L}_{z}P_{s}(t-t')+\alpha^{2}\mathcal{P}\mathcal{L}_{z}\int_{0}^{\infty}\mbox{d}t'\int_{0}^{\infty}\mbox{d}t''\,\mbox{e}^{t'\mathcal{L}_{0}}\mathcal{L}_{n}\mbox{e}^{t''\mathcal{L}_{0}}\mathcal{L}_{n}P_{s}(t-t'-t'')+\mathcal{O}\left(\alpha^{5/2}\right).
\end{align*}
We observe now that the evolution equation for $P_{s}$ contains memory
terms. However, in the limit $\alpha\ll1$, $P_{s}$ evolves very
slowly. Thus, if we make a Markovianization of the time integrals
replacing $P_{s}(t-t')$ by $P_{s}(t)$, we make an error of order
$\frac{\partial P_{s}}{\partial t}(t)$ which is of order $\alpha$.
For example, the first integral in the last equation is 
\[
\int_{0}^{\infty}\mbox{d}t'\,\left[\mbox{e}^{t'\mathcal{L}_{0}}\mathcal{L}_{n}P_{s}(t-t')\right] = \int_{0}^{\infty}\mbox{d}t'\,\mbox{e}^{t'\mathcal{L}_{0}}\mathcal{L}_{n}P_{s}(t)+\mathcal{O}\left(\alpha\right).
\]
 The evolution equation for $P_{s}$ is then 
\begin{equation}
\frac{\partial P_{s}}{\partial t}=\left\{ \alpha\mathcal{P}\mathcal{L}_{z}+\alpha^{3/2}\mathcal{P}\mathcal{L}_{z}\int_{0}^{\infty}\mbox{d}t'\,\mbox{e}^{t'\mathcal{L}_{0}}\mathcal{L}_{n}+\alpha^{2}\mathcal{P}\mathcal{L}_{z}\int_{0}^{\infty}\mbox{d}t'\,\mbox{e}^{t'\mathcal{L}_{0}}\left[(1-\mathcal{P})\mathcal{L}_{z}+\int_{0}^{\infty}\mbox{d}t''\,\mathcal{L}_{n}\mbox{e}^{t''\mathcal{L}_{0}}\mathcal{L}_{n}\right]\right\} P_{s}+\mathcal{O}\left(\alpha^{5/2}\right).\label{eq:Evolution-Ps-1}
\end{equation}
 We now have a closed differential equation for the slow probability
distribution function $P_{s}$.

\subsection{The Fokker-Planck equation for the slow evolution of the zonal flow}

The explicit computation of each term involved in (\ref{eq:Evolution-Ps-1}),
which is reported in the appendix \ref{sec:Explicit-computation-zonal-FP},
leads to the final Fokker-Planck equation for the zonal jets:
\begin{align}
\frac{\partial R}{\partial\tau}= & \int\mbox{d}y_{1}\,\frac{\delta}{\delta q_{z}(y_{1})}\left\{ \left[\frac{\partial F_{1}\left[U\right]}{\partial y_{1}}+\omega_{z}(y_{1})-\frac{\nu}{\alpha}\frac{\partial^{2}\omega_{z}}{\partial y_{1}^{2}}\right]R[q_{z}]+\int\mbox{d}y_{2}\,\frac{\delta}{\delta q_{z}(y_{2})}\left(C_{R}(y_{1},y_{2})\left[q_{z}\right]R\left[q_{z}\right]\right)\right\} ,\label{eq:Fokker-Planck-Zonal}
\end{align}
which evolves over the time scale $\tau=\alpha t$, with the drift
term
\[
F_{1}\left[U\right]=F\left[U\right]+\alpha\int_{0}^{\infty}\mbox{d}t'\int_{0}^{\infty}\mbox{d}t''\,\mathbb{E}_{U}\left[\left\langle v_{m}^{(y)}\omega_{m}\right\rangle (y_{1})M[q_{z},\omega_{m}](t',t'')\right],
\]
with $M$ given in appendix \ref{sec:Explicit-computation-zonal-FP},
with
\[
F\left[U\right]=\mathbb{E}_{U}\left[\left\langle v_{m}^{(y)}\omega_{m}\right\rangle \right],
\]
 and with the diffusion coefficient
\[
C_{R}(y_{1},y_{2})\left[q_{z}\right]=C_{z}(y_{1}-y_{2})+\alpha\int_{0}^{\infty}\mbox{d}t'\,\frac{\partial^{2}}{\partial y_{2}\partial y_{1}}\mathbb{E}_{U}\left[\left[\left\langle v_{m}^{(y)}\omega_{m}\right\rangle (y_{1},t')\left\langle v_{m}^{(y)}\omega_{m}\right\rangle (y_{2},0)\right]\right].
\]
From its definition, we see that $F\left[U\right]$ is the opposite
of the Reynolds stress divergence computed from the quasi-linear approximation.

The Fokker-Planck equation (\ref{eq:Fokker-Planck-Zonal}) is equivalent
to a non-linear stochastic partial differential equation for the potential
vorticity $q_{z}$,
\begin{equation}
\frac{\partial q_{z}}{\partial\tau}=-\frac{\partial F_{1}}{\partial y}[U]-\omega_z+\frac{\nu}{\alpha}\frac{\partial^{2}\omega_z}{\partial y^2}+\zeta,\label{eq:Stochastique-Lent}
\end{equation}
 where $\zeta$ is a white in time Gaussian noise with spatial correlation
$C_{R}$. As $C_{R}$ depends itself on the velocity field $U$, this
is a non-linear noise. 

At first order in $\alpha$, we recover the deterministic kinetic
equation (\ref{eq:kinetic-equation-determinist}) discussed in section
\ref{sub:The-quasi-linear-dynamics}. The main result here is that
the first contribution of the non-linear operator, the order $O(\alpha^{3/2})$
in (\ref{eq:Evolution-Ps-1}), is exactly zero. We could have applied
the same stochastic reduction techniques to the quasi-linear dynamics
(\ref{eq:quasi-lineaire}) and we would have then obtained the same
deterministic kinetic equation at leading order, as expected. We thus
have shown that at order 3 in $\sqrt{\alpha}$, the quasi-linear approximation
and the full non-linear dynamics give the same results for the zonal
flow statistics.

At next order, we see a correction to the drift $F[U]$ due to the
non-linear interactions. At this order, the quasilinear dynamics and
non-linear dynamics differ. We also see the appearance of the noise
term, which has a qualitatively different effect than the drift term.
We note that at this order, we would have obtained the same non-linear
noise $C_{R}$ from the quasilinear dynamics. Pushing our computations
to next orders, we would obtain higher order corrections to the drift
and noise terms, for instance due to eddy-eddy (non-zonal) non-linear
interactions, but also corrections through fourth order operators.
\\

We note that we can also write directly an equation equivalent to
(\ref{eq:Stochastique-Lent}) for the velocity profile
\begin{equation}
\frac{\partial U}{\partial\tau}=F_{1}[U]-U+\frac{\nu}{\alpha}\frac{\partial^{2}U}{\partial y^2}+\xi,\label{eq:Stochastique-Lent-U}
\end{equation}
 where $\xi$ is a white in time Gaussian noise with spatial correlation
\begin{equation}
\mathbb{E}_{\xi}\left[\xi(y_{1},t_{1})\xi(y_{2},t_{2})\right]=\left(C_{U}(y_{1}-y_{2})+\alpha\int_{0}^{\infty}\mbox{d}t'\,\mathbb{E}_{U}\left[\left[\left\langle v_{m}^{(y)}\omega_{m}\right\rangle (y_{1},t')\left\langle v_{m}^{(y)}\omega_{m}\right\rangle (y_{2},0)\right]\right]\right)\delta(t_{1}-t_{2}).\label{eq:minimal-model-correlation-function}
\end{equation}
 Whereas at a formal level, correction to $F$ and the noise term
appear at the same order, their qualitative effect is quite different.
For instance if one is interested in large deviations from the most
probable states, correction of order $\alpha$ to $F$ will still
be vanishingly small, whereas the effect of the noise will be essential.
At leading order, the large fluctuations will be given by
\begin{equation}
\frac{\partial U}{\partial\tau}=F[U]-U+\frac{\nu}{\alpha}\frac{\partial^{2}U}{\partial y^2}+\xi,\label{eq:Minimal-model}
\end{equation}
 Equation (\ref{eq:Minimal-model}) then appears to be the minimal
model in order to describe the evolution of zonal jet in the limit
of weak forcing and dissipation. We will comment further on this issue
in section \ref{sec:Bistability}, dealing with bistability of zonal
jets and phase transitions.

\section{Energy and enstrophy balances\label{sec:Energy-balance}}

We discuss here the energy balance in the inertial limit $\alpha\ll1$
and the consistency of the stochastic reduction at the level of the
energy. It is thus essential to distinguish the different ways to
define the averages, for the original stochastic equation (\ref{eq:barotropic-topography})
or for the zonal Fokker-Planck equation (\ref{eq:Fokker-Planck-Zonal}).
We recall that $\mathbb{E}[\cdot]=\int\mathcal{D}[q_{z}]\mathcal{D}[\omega_{m}]\cdot P[q_{z},\omega_{m}]$
denotes the average with respect to the full PDF $P_{t}$ whose evolution
is given by the Fokker-Planck equation (\ref{eq:Evolution-P}), or
equivalently, the average over realizations of the noise $\eta$ in
equation (\ref{eq:Zonal-PV-Evolution}-\ref{eq:Meridional-PV-Evolution});
$\mathbb{E}_{U}[\cdot]=\int\mathcal{D}[\omega_{m}]\cdot G[q_{z},\omega_{m}]$
denotes the average with respect to the stationary Gaussian distribution
of the non-zonal fluctuations $G$, defined in (\ref{eq:distribution-gaussienne})
and $\mathbb{E}_{R}[\cdot]=\int\mathcal{D}[q_{z}]\cdot R[q_{z}]$
denotes the average with respect to the slowly evolving zonal PDF
$R$ whose evolution is given by the zonal Fokker-Planck equation
(\ref{eq:Fokker-Planck-Zonal}). $\mathbf{\mathbb{E}}_{R}$ is equivalently
an average over the realizations of the noise $\zeta$ in equation
(\ref{eq:Stochastique-Lent}).

As discussed in the presentation of the barotropic equations, we are
interested in the regime where the dissipation of energy is dominated
by the one due to the large-scale linear friction: $2\lambda E/\nu_{n,d}H_{n}\gg1$
in equation (\ref{eq:Energy-Balance}). As a consequence, we will
consider in this section only the case of zero viscosity, $\nu=0$.

\subsection{Energy balance for the barotropic equations}

The total energy balance of the non-dimensional barotropic equations
(\ref{eq:barotropic-topography}) given by Eq. (\ref{eq:Energy-Balance-sans-dimension})
is reported here for convenience: 
\begin{equation}
\frac{dE}{dt}=-2\alpha E+2\alpha\label{eq:energy-balance-total}
\end{equation}
where we recall that $E=\mathbb{E}\left[\mathcal{E}\right]$. With
the orthogonal decomposition into zonal and non-zonal degrees of freedom,
we have a natural decomposition of the energy contained in zonal and
non-zonal degrees of freedom: $E=E_{z}+\alpha E_{m}$, with $E_{z}=\mathbb{E}\left[\frac{1}{2}\int_{\mathcal{D}}U^{2}\right]$
and $E_{m}=\mathbb{E}\left[\frac{1}{2}\int_{\mathcal{D}}{\bf v}_{m}^{2}\right]$.

\paragraph{Zonal energy balance}

From the definition of $E_{z}$, either by direct computation from
the Fokker-Planck equation (\ref{eq:Evolution-P}) or from Eq. (\ref{eq:Zonal-PV-Evolution})
applying the It\={o} formula, we have 
\begin{equation}
\frac{dE_{z}}{dt}=\alpha2\pi l_{x}\int\mbox{d}y\,\mathbb{E}\left[\left\langle v_{m}^{(y)}\omega_{m}\right\rangle (y)U(y)\right]-2\alpha E_{z}+2\alpha\sigma_{z}\label{eq:energy-balance-zonal}
\end{equation}
 with $\sigma_{z}=-2\pi^{2}l_{x}\left(\Delta^{-1}C_{z}\right)(0)$
the rate of energy injected by the forcing directly into the zonal
degrees of freedom. In addition with the expected energy dissipation
and direct energy injection by $\sigma_{z}$, the first term on the
right hand side describes the energy production due to the non-zonal
fluctuations.

\paragraph{Non-zonal energy balance}

Using $E=E_{z}+\alpha E_{m}$ and equations (\ref{eq:energy-balance-total}),
(\ref{eq:energy-balance-zonal}), we obtain 
\begin{equation}
\frac{dE_{m}}{dt}=-2\pi l_{x}\int\mbox{d}y\,\mathbb{E}\left[\left\langle v_{m}^{(y)}\omega_{m}\right\rangle (y)U(y)\right]-2\alpha E_{m}+2\sigma_{m}\,,\label{eq:energy-balance-meridional}
\end{equation}
 where $\sigma_{m}=1-\sigma_{z}$ is the rate of energy injection
by the forcing on non-zonal degrees of freedom. Clearly, Eq. (\ref{eq:energy-balance-zonal})
and (\ref{eq:energy-balance-meridional}) are exact and fully equivalent
to the energy balance (\ref{eq:energy-balance-total}).

\subsection{Energy and enstrophy balance for the kinetic equation}

We now show that the energy balance of the kinetic equation (\ref{eq:Stochastique-Lent})
is consistent with the exact energy balances written above, at leading
order in $\alpha$. We denote by $\tilde{E}_{z}=\mathbb{E}_{R}\left[\frac{1}{2}\int_{\mathcal{D}}U^{2}\right]$
the average zonal energy for the kinetic equation and by $\alpha\tilde{E}_{m}=\alpha\mathbb{E}_{U}\left[\frac{1}{2}\int_{\mathcal{D}}{\bf v}_{m}^{2}\right]$
the energy contained in non-zonal degrees of freedom. We start from
the energy balance for zonal degrees of freedom. Again, working at
the level of the zonal Fokker-Planck (\ref{eq:Fokker-Planck-Zonal})
equation or of the kinetic equation (\ref{eq:Stochastique-Lent})
give the same result: 
\[
\frac{d\tilde{E}_{z}}{d\tau}=2\pi l_{x}\mathbb{E}_{R}\left[\int\mbox{d}y\, F_{1}\left[U\right](y)U(y)\right]-2\tilde{E}_{z}-2\pi l_{x}\mathbb{E}_{R}\left[\int\mbox{d}y\,\left(\Delta^{-1}C_{R}\right)(\mathbf{0})\right].
\]
To approximate the above equation at leading order in $\alpha$, we
observe that $F_{1}\left[U\right]=F[U]+O(\alpha)=\mathbb{E}_{U}\left[\left\langle v_{m}^{(y)}\omega_{m}\right\rangle \right](y_{1})+O(\alpha)$,
and $C_{R}=C_{z}+O(\alpha)$. Then, the stationary energy balance
for zonal degrees of freedom reduces to

\begin{equation}
\frac{d\tilde{E}_{z}}{d\tau}=2\pi l_{x}\mathbb{E}_{R}\left[\int\mbox{d}y\,\mathbb{E}_{U}\left[\left\langle v_{m}^{(y)}\omega_{m}\right\rangle \right]U\right]-2\tilde{E}_{z}+2\sigma_{z}+O(\alpha).\label{eq:energy-balance-zonal-kinetic}
\end{equation}
 In the limit $\alpha\rightarrow0$, the full PDF $P[q_{z},\omega_{m}]$
is given by the slowly evolving part $P_{s}[q_{z},\omega_{m}]=G[q_{z},\omega_{m}]R[q_{z}]$.
Then, the rate of energy transferred from the fluctuations to the
zonal flow becomes 
\[
\mathbb{E}_{R}\left[\int\mbox{d}y\,\mathbb{E}_{U}\left[\left\langle v_{m}^{(y)}\omega_{m}\right\rangle \right]U\right]+O(\alpha)=\int\mbox{d}y\,\mathbb{E}\left[\left\langle v_{m}^{(y)}\omega_{m}\right\rangle U\right]\,;
\]
and equation (\ref{eq:energy-balance-zonal-kinetic}) reduces to the
energy balance for zonal degrees of freedom (\ref{eq:energy-balance-zonal})
at leading order in $\alpha$. This proves the consistency of the
zonal energy balance for the kinetic equation and the barotropic equations:
$E_{z}=\tilde{E}_{z}+O(\alpha)$.

\subsubsection{Non-zonal energy balance and total transfer of energy to the zonal
flow}

Let us now consider the energy balance for $\alpha\tilde{E}_{m}$.
Applying It\={o} formula to (\ref{eq:dynamique-lineaire}), we get, in the stationary state and for any $U$, 
\begin{equation}
0=-2\pi l_{x}\int\mbox{d}y\,\mathbb{E}_{U}\left[\left\langle v_{m}^{(y)}\omega_{m}\right\rangle (y)\right]U(y)-2\alpha\tilde{E}_{m}+2\sigma_{m}\,,\label{eq:energy-balance-meridional-kinetic}
\end{equation}
with $\sigma_{m}=-2\pi^{2}l_{x}\left(\Delta^{-1}C_{m}\right)({\bf 0})$.
Equation (\ref{eq:energy-balance-meridional-kinetic}) does not contain
time evolution which is consistent with the definition of $\tilde{E}_{m}$
in the kinetic equation through a stationary average. In the limit
$\alpha\ll1$, in agreement with the scaling of the variables we expect
the energy of the non-zonal degrees of freedom to be of order $\alpha$,
or $\tilde{E}_{m}=O(1)$. This is essential for the consistency of
the asymptotic expansion and will be proved in section \ref{sec:Lyapunov}. 

From $\tilde{E}_{m}=O(1)$ and equation (\ref{eq:energy-balance-meridional-kinetic}),
the energy dissipated in the non-zonal fluctuations per unit time
is negligible, so the stationary energy balance for the non-zonal
fluctuations gives 
\[
\pi l_{x}\int\mbox{d}y\,\mathbb{E}\left[\left\langle v_{m}^{(y)}\omega_{m}\right\rangle U\right]=\pi l_{x}\int\mbox{d}y\,\mathbb{E}_{U}\left[\left\langle v_{m}^{(y)}\omega_{m}\right\rangle (y)\right]U(y)+O(\alpha)=\sigma_{m}+O(\alpha)\,.
\]
Injecting this relation in the stationary energy balance for the zonal
degrees of freedom (\ref{eq:energy-balance-zonal-kinetic}) gives

\[
E_{z}=\pi l_{x}\int\mbox{d}y\,\mathbb{E}_{R}\left[\left\langle v_{m}^{(y)}\omega_{m}\right\rangle U\right]+\sigma_{z}=\sigma_{z}+\sigma_{m}+O(\alpha)=1+O(\alpha).
\]
The barotropic equations (\ref{eq:barotropic-topography}) are in
units so that $E=1$ in a stationary state: the above relation expresses
the fact that, in the limit $\alpha\ll1$, all the energy is concentrated
in the zonal degrees of freedom: $E_{z}=E+O(\alpha)$.

\subsubsection{Enstrophy balance for the kinetic equation}

We conclude considering the enstrophy balance for the kinetic equation.
As we will see below, the concentration property found for the energy
does not hold for the enstrophy. The zonal and non-zonal enstrophy
balances can be obtained with a very similar reasoning as the one
done for the energy and, at leading order in $\alpha$, one can use
the full Fokker-Planck (\ref{eq:Evolution-P}) or the approximated
one (\ref{eq:Fokker-Planck-Zonal},\ref{eq:dynamique-lineaire}) and
obtain consistent results. Denoting $Z_{z}=\mathbb{E}\left[\frac{1}{2}\int_{\mathcal{D}}q_{z}^{2}\right]$
the enstrophy of zonal degrees of freedom and $\alpha Z_{m}=\alpha\mathbb{E}\left[\frac{1}{2}\int_{\mathcal{D}}\omega_{m}^{2}\right]$
the non-zonal degrees of freedom one, we have 

\begin{equation}
0=-2\pi l_{x}\mathbb{E}\left[\int\mbox{d}y\, U''(y)\left\langle v_{m}^{(y)}\omega_{m}\right\rangle (y)\right]-2Z_{z}+4\pi^{2}l_{x}C_{z}(0)\,,\label{eq:enstrophy-balance-zonal}
\end{equation}
\begin{equation}
0=2\pi l_{x}\mathbb{E}\left[\int\mbox{d}y\, U''(y)\left\langle v_{m}^{(y)}\omega_{m}\right\rangle (y)\right]-2\alpha Z_{m}+4\pi^{2}l_{x}C_{m}({\bf 0})\,,\label{eq:enstrophy-balance-meridional}
\end{equation}
where the first equation is the stationary enstrophy balance for $Z_{z}$
and the second one for $Z_{m}$. The above equations refer explicitly
to the case of the 2D Euler equations, but the generalization to the
barotropic equations is straightforward.

As will be discussed in next section, the enstrophy in the non-zonal
degrees of freedom doesn't converge in the inertial limit: more precisely,
$\alpha Z_{m}=O(1)$. Then, the enstrophy in the zonal degrees of
freedom is 
\[
Z_{z}=-\alpha Z_{m}+2\pi^{2}l_{x}\left(C_{z}(0)+C_{m}({\bf 0})\right)=I_{Z}-\alpha Z_{m}\,,
\]
with the total enstrophy injected by the forcing $I_{Z}$. Then, by
contrast with the zonal energy, the zonal enstrophy is not equal to
the enstrophy injected plus correction of order $\alpha$. There is
no concentration of the enstrophy in the zonal degrees of freedom,
and a non-vanishing part of the enstrophy injected is dissipated in
the non-zonal fluctuations.

Clearly there is no self-similar cascade in the problem considered
here, energy goes directly from the fluctuations at any scales to
the zonal flow through the effect of the advection by the zonal flow.
These results are however in agreement with the phenomenology of a
transfer of the energy to the largest scales, while the excess enstrophy
is transferred to the smallest scales, however with dynamical processes
that are non-local in Fourier space. This non-equilibrium transfer
of energy to the largest scales is also consistent with predictions
from equilibrium statistical mechanics which, roughly speaking, predicts
that the most probable flow concentrates its energy at the largest
possible scales.

\section{The Lyapunov equation in the inertial limit}

\label{sec:Lyapunov}

As discussed in section \ref{sec:Stochastic-Averaging}, it is essential
to make sure that the Gaussian process corresponding to the inertial
linearized evolution of non-zonal degrees of freedom close a base
flow $U$, see Eq. (\ref{eq:Linearized-Dynamics-Inertial}) and (\ref{eq:equation-Lyapunov-inertiel}),
has a stationary distribution. We discuss this issue in this section.
We consider the linear dynamics with stochastic forces Eq. (\ref{eq:dynamique-lineaire}),
that we recall here for convenience 
\begin{equation}
\frac{\partial\omega_{m}}{\partial t}+L_{U}^{0}[\omega_{m}]=\sqrt{2}\eta_{m}\quad,\quad\mathbb{E}\left[\eta_{m}({\bf r}_{1},t)\eta_{m}({\bf r}_{2},t')\right]=C_{m}({\bf r}_{1}-{\bf r}_{2})\delta(t-t'),\label{eq:dynamique-lineaire-1}
\end{equation}
 where 
\begin{equation}
L_{U}^{0}\left[\omega_{m}\right]=U(y)\frac{\partial\omega_{m}}{\partial x}+(-U''(y)+h'(y))\frac{\partial\psi_{m}}{\partial x}\label{eq:Linearized-Dynamics-Inertial-1}
\end{equation}
is the linearized evolution operator close to the zonal flow $U$.

Eq. (\ref{eq:dynamique-lineaire-1}) describes a linear stochastic
Gaussian process, or Ornstein-Uhlenbeck process. Thus, it is completely
characterized by the two-points correlation function $g(\mathbf{r}_{1},\mbox{\ensuremath{\mathbf{r}}}_{2},t)=\mathbb{E}\left[\omega_{m}({\bf r}_{1},t)\omega_{m}({\bf r}_{2},t)\right]$.
The evolution of $g$ is given by the so-called Lyapunov equation,
which is obtained by applying the It\={o} formula to (\ref{eq:dynamique-lineaire-1}):
\begin{equation}
\frac{\partial g}{\partial t}+L_{U}^{0(1)}g+L_{U}^{0(2)}g=2C_{m}.\label{eq:equation-Lyapunov-1}
\end{equation}
We will prove that equation (\ref{eq:equation-Lyapunov-1}) has a
asymptotic limit $g^{\infty}$ for large time. This may seem paradoxical
as we deal with a linearized dynamics with a stochastic force and
no dissipation mechanism. We explain in this section that the Orr mechanism
(the effect of the shear through a non-normal linearized dynamics)
acts as an effective dissipation. However this effects is not uniform
on all observables. We will prove that $g$ has a limit $g^{\infty}$
in the sense of distribution, from which we will be able to prove
that velocity-like observables have a limit. By contrast, observables
involving the vorticity gradient will diverge. Moreover, if the kinetic
energy contained in the non-zonal degrees of freedom converges, the
enstrophy diverges. This non-uniformity for the convergence of the
two-points correlations functions is also related to the fact that
the convergences will be algebraic in time, rather than exponential.
The statement that ``the Gaussian process corresponding to the inertial
linearized evolution close to a base flow $U$ has a stationary distribution''
must thus be understood with care: not all observable converge and
the convergences are algebraic. 

An observable like the Reynolds stress divergence involves both the
velocity and the vorticity gradient. It is thus not obvious that it
has an asymptotic limit. We will also prove that the long time limits
of the Reynolds stress divergence $-\mathbb{E}_{U}\left[\left\langle v_{m}^{(y)}\omega_{m}\right\rangle\right]$
and of its gradient $-\mathbb{E}_{U}\left[\frac{\partial}{\partial y}\left\langle v_{m}^{(y)}\omega_{m}\right\rangle\right]$
are actually well defined. The results in this section ensure that
the asymptotic expansion performed in section \ref{sec:Stochastic-Averaging}
is well posed, at leading order.\\

Because some quantities (such as the enstrophy) diverge in the inertial
limit, it is of interest to understand how the Gaussian process is
regularized by a small viscosity or linear friction. This corresponds
to replace the operator $L_{U}^{0}$ in Eq. (\ref{eq:equation-Lyapunov-1})
with the operator $L_{U}$ defined in Eq. (\ref{eq:Linearized-Dynamics}).
Moreover, to be able to separate the effect of the viscosity and of
the Rayleigh friction, we will introduce in the following the operators
\begin{equation}
L_{U}^{\nu}\left[\omega_{m}\right]=L_{U}^{0}[\omega_{m}]-\nu\Delta\omega_{m}\,,\label{eq:Linearized-Dynamics-Lnu}
\end{equation}
 and 
\begin{equation}
L_{U}^{\alpha}\left[\omega_{m}\right]=L_{U}^{0}[\omega_{m}]+\alpha\omega_{m}\,,\label{eq:Linearized-Dynamics-Lalpha}
\end{equation}
 in which the superscript indicates which of the terms have been retained
from the ordinal operator $L_{U}$. 

For $\alpha=\nu=0$, the two-point correlation function $g$ converges
as a distribution. $g(\mathbf{r},\mathbf{r}',t)$ diverges point-wise
only for value of $y$ and $y'$ such that $U(y)=U(y')$, for instance
$y=y'$. We will prove that for small $\alpha$ this divergence is
regularized on a universal way (independent on $U$) close to points
$(\mathbf{r},\mathbf{r}')$ such that $U(y)=U(y')$, over a scale
$\lambda_{\alpha}=\frac{2\alpha}{kU'(y)}$, where $k$ is the wavenumber
of the meridional perturbation. We stress that only the local behavior
close to the divergence point is universal. All quantities converge
exponentially over a time scale $1/\alpha$. 

We will argue that a small viscosity $\nu\neq0$ also leads to a universal
regularization of the divergence, over a scale $\lambda_{\nu}=\left(\frac{2\nu}{kU'(y)}\right)^{1/3}$.
When both $\alpha$ and $\nu$ are not equal to zero, the way the
singularity is regularized depends on the ratio of the length scales
$\lambda_{\alpha}$ and $\lambda_{\nu}$. More detailed results are
discussed in this section. \\

Our analysis is based on the evaluation of the long time asymptotics
for the deterministic inertial linearized dynamics \cite{Bouchet_Morita_2010PhyD}.
As far as we know, such theoretical results are currently available
only for the linearized Euler equation \cite{Bouchet_Morita_2010PhyD},
$h=\beta=0$. We will discuss the Lyapunov equation in the case $\beta\neq0$
in a forthcoming paper.

\subsection{General discussion\label{sub:Equation-de-Lyapunov}}

\subsubsection{Solution of the Lyapunov equation from the solution of the deterministic
linear equation\label{sub:Equation-de-Lyapunov-sec1}}

In this first section, we show how to obtain a formal solution of
the Lyapunov equation (\ref{eq:equation-Lyapunov-1}) using the solution
of the deterministic dynamics $\frac{\partial}{\partial t}+L_{U}^{0}$
with appropriate initial conditions. This discussion can be trivially
extended to the cases in which the operator appearing in Eq. (\ref{eq:dynamique-lineaire-1})
is $L_{U}^{\nu}$, $L_{U}^{\alpha}$ or $L_{U}$, by simply replacing
the operator $L_{U}^{0}$ with the appropriate one wherever it appears.

We expand the force correlation function $C_{m}$ in Fourier series,
we have 
\[
C_{m}(x,y)=\sum_{k>0\,,l}c_{kl}\cos(kx+ly).
\]
We note that because $C_{m}$ is a correlation, it is a positive definite
function. This explains why $\sin$ contribution are zero in this
expansion. Moreover for all $k\in\mathbb{N}^{*}$ and $l\in\mathbb{Z}$,
we have $c_{kl}\geq0$. The expression $c_{kl}\cos(kx+ly)+c_{k,-l}\cos(kx-ly)$
is the most general positive definite function involving the $ $Fourier
components $\mbox{e}^{ikx}$ and $\mbox{e}^{ikl}$ or their complex
conjugates. Here we have assumed that the correlation function is
homogeneous (it depends only on $x_{1}-x_{2}$ and on $y_{1}-y_{2}$).
Its generalization to the case of an inhomogeneous force, for instance
for the case of a channel would be straightforward. 

Because the Lyapunov equation is linear, the contribution of the effect
of all forcing terms just add up
\begin{equation}
g=\sum_{k>0\,,l}c_{kl}g_{kl}\,,\label{eq:decompositiongk}
\end{equation}
where $g_{kl}$ is the solution of the Lyapunov equation (\ref{eq:equation-Lyapunov-1})
with right hand side $2\cos(kx+ly)$. By direct computation it is easy
to check that 
\begin{equation}
g_{kl}(\mathbf{r}_{1},\mathbf{r}_{2},t)=\int_{0}^{t}\mbox{e}^{-t_{1}L_{U}^{0}}\left[\mbox{e}_{kl}\right](x_{1},y_{1})\mbox{e}^{-t_{1}L_{U}^{0}}\left[\mbox{e}_{kl}^{*}\right](x_{2},y_{2})\,\mathrm{d}t_{1}+\mbox{C.C.}.\label{eq:lyapunov-solution-real-time}
\end{equation}
with $\mbox{e}_{kl}(x,y)=\mbox{e}^{i\left(kx+ly\right)}$, and where
$\mbox{C.C.}$ stands for the complex conjugate %
\footnote{A similar formula can easily been deduced for any stochastic force
of the form $C(\mathbf{r_{1},r_{2}})=f(\mathbf{r}_{1})f(\mathbf{r}_{2})$
from the explicit solution of the Gaussian process from the stochastic
integral $\int_{0}^{t}\mbox{e}^{-t_{1}L_{U}^{0}}\left[f\right]\mbox{d}W_{t_{1}}$.
Here the key point is that the Fourier basis diagonalizes any translationally
invariant correlation function. %
}. We note that $\mbox{e}^{-t_{1}L_{U}^{0}}\left[\mbox{e}_{kl}\right]$
is the solution at time $t_{1}$ of the deterministic linear dynamics
$\partial_{t}+L_{U}^{0}$ with initial condition $\mbox{e}_{kl}$.

Let us observe that from the solution of the Lyapunov equation, we
can also easily obtain the evolution of $\mathbb{E}\left[\omega_{m}(\mathbf{r}_{1},t)\left(S\omega_{m}\right)(\mathbf{r}_{2},t)\right]$
where $S$ is a linear operator. We have\textbf{ 
\[
\mathbb{E}\left[\omega_{m}(\mathbf{r}_{1},t)\left(S\omega_{m}\right)(\mathbf{r}_{2},t)\right]=S^{(2)}\left[g\right]
\]
}where $S^{(2)}$ is the linear operator $S$ acting on the second
variable. The typical operators $S$ that we will use in the following
are those necessary to obtain the stream function and the velocity
from the vorticity.\textbf{ }

\subsubsection{Two explicitly solvable examples\label{sub:Two-trivial-examples}}

Before discussing the long-time behavior of the Gaussian process in
Eq. (\ref{eq:dynamique-lineaire-1}) for a general zonal flow $U$,
we discuss here two simple examples for which the deterministic linear
dynamics $\partial_{t}+L_{U}^{0}$ can be solved analytically. The
two examples are the perfect and the viscous advection by a linear
shear. These two simple examples will put in evidence the mechanisms
that will ensure the convergence of the long-time limit of two-points
correlations in the inertial limit.

For a linear shear this mechanism is the Orr mechanism: the transport
of the vorticity along each streamline leads to a phase mixing in
the computation of all integrated quantities, for instance the velocity.
Then the velocity or the stream function decay algebraically for large
times, the exponent for this decay being related to the singularity
of the Laplacian green function. For more general profile $U$ for
which $U''(y)\neq0$, this shear mechanism still exists, but is also
accompanied by global effects due to the fact that vorticity affects
the velocity field globally. In many cases, those global effects are
the dominant one \cite{Bouchet_Morita_2010PhyD}. They will be taken
into account in section \ref{sub:Stationarity-of-the-lyapunov}.

\paragraph{Perfect advection by a linear shear}

\label{subsection:linear-shear-perfect}

To treat an example that can be worked out explicitly, we consider
the perturbation by a stochastic force of a linear shear $U(y)=sy$,
which corresponds to set in Eq. (\ref{eq:dynamique-lineaire-1}) $L_{U}^{0}=sy\partial_{x}$.
For sake of simplicity, we only treat here the case $s>0$, the corresponding
generalization to $s<0$ being trivial. The following discussion applies
to flows that are periodic in the longitudinal direction $x$, with
period $2\pi l_{x}$, either to the case of the domain $\mathscr{D}=[0,2\pi l_{x})\times(-\infty,\infty)$,
or to flows in a zonal channel with walls at $y=\pm L$.

The Lyapunov equation we have to consider for this problem is 
\[
\frac{\partial g}{\partial t}+L_{U}^{0(1)}g+L_{U}^{0(2)}g=2C_{m},
\]
for the vorticity-vorticity correlation function $g({\bf r}_{1},{\bf r}_{2},t)=\mathbb{E}\left[\omega_{m}({\bf r}_{1},t)\omega_{m}({\bf r}_{2},t)\right]$.
For sake of simplicity, we consider the case where the forcing acts
on a single wave vector 
\begin{equation}
C_{m}=\frac{\epsilon\left(k^{2}+l^{2}\right)}{4}\left[e^{ik(x_{1}-x_{2})+il(y_{1}-y_{2})}+\mbox{C}.\mbox{C}.\right]=\frac{\epsilon\left(k^{2}+l^{2}\right)}{2}\cos\left[k(x_{1}-x_{2})+l(y_{1}-y_{1})\right]\,,\label{eq:Cm}
\end{equation}
where $\mbox{C}.\mbox{C}.$ stands for the complex conjugation of
the first term.\textbf{ }As explained in the previous section, contributions
to the Lyapunov equation from other forcing modes just add up (see
equation (\ref{eq:decompositiongk})). $\epsilon$ is the average
energy input rate per unit of mass (unit $\mbox{m}^{2}\mbox{s}^{-3}$)
(in this section and the following ones dealing with the case of a
linear shear, $\omega_{m}$ is the non-zonal vorticity and has dimensions
$s^{-1}$, whereas in section dealing with the kinetic theory $\sqrt{\alpha}\omega_{m}$
is the non-zonal vorticity).

The deterministic evolution of the linearized dynamics $L_{U}^{0}$
with initial condition $\mbox{e}^{i(kx+ly)}$ is

\begin{equation}
\mbox{e}^{-tL_{U}^{0}}\left[\mbox{e}^{i(kx+ly)}\right]=\mbox{e}^{ikx+ily-iksyt}\,,\label{eq:pure-linear-shear-vorticity}
\end{equation}
as can be easily checked. From (\ref{eq:lyapunov-solution-real-time}),
we have

\begin{equation}
g(\mathbf{r}_{1},\mathbf{r}_{2},t)=\frac{\epsilon\left(k^{2}+l^{2}\right)}{4s}\frac{\mbox{e}^{ik(x_{1}-x_{2})+il\left(y_{1}-y_{2}\right)}}{-ik(y_{1}-y_{2})}\,\left[\mbox{e}^{-iks(y_{1}-y_{2})t}-1\right]+\mbox{C}.\mbox{C}.\,,\label{eq:perfect-shear-gkl}
\end{equation}
 for $y_{1}\neq y_{2}$ and $g(x_{1},y,x_{2},y,t)=\frac{\epsilon\left(k^{2}+l^{2}\right)}{2}\cos\left(k(x_{1}-x_{2})\right)t$.

This result readily shows that the square of the perturbation vorticity
$g(\mathbf{r},\mathbf{r},t)$ diverges proportionally to time $t$.
This is expected as the average enstrophy input rate per unit of area
is $\epsilon\left(k^{2}+l^{2}\right)$, and there is no dissipation
mechanism. However, we also remark that for $y_{1}\neq y_{2}$ the
autocorrelation function is a fast oscillating function. As a consequence
$g$ will have a well defined limit in the sense of distributions:
\begin{align}
g^{\infty}(\mathbf{r}_{1},\mathbf{r}_{2}) & =\lim_{t\to\infty}g(\mathbf{r}_{1},\mathbf{r}_{2},t)\label{eq:linear-shear-g-infty}\\
 & =\frac{\epsilon\left(k^{2}+l^{2}\right)}{4s}\mbox{e}^{ik(x_{1}-x_{2})+il\left(y_{1}-y_{2}\right)}\frac{1}{k}\left[\pi\delta\left(y_{1}-y_{2}\right)-iPV\left(\frac{1}{y_{1}-y_{2}}\right)\right]+\mbox{C}.\mbox{C}.\,,\nonumber 
\end{align}
where $PV$ stands for the Cauchy Principal Value (we have used Plemelj
formula, equation (\ref{eq:Plemelj}) on page \pageref{eq:Plemelj}).
Equivalently, we have 
\begin{equation}
g^{\infty}(\mathbf{r}_{1},\mathbf{r}_{2})=\frac{\epsilon\left(k^{2}+l^{2}\right)}{2s}\left[\frac{\cos k(x_{1}-x_{2})}{k}\pi\delta\left(y_{1}-y_{2}\right)+\frac{\sin\left[k(x_{1}-x_{2})+l(y_{1}-y_{2})\right]}{k}PV\left(\frac{1}{y_{1}-y_{2}}\right)\right].\label{eq:ginfini-reel}
\end{equation}
The fact that the stationary vorticity-vorticity correlation function
has a limit in the sense of distributions is a very important result.
It means that every observable that can be obtained by integration
of a smooth function over $g^{\infty}$ will have a well defined stationary
limit. Actually the formula above will be valid when integrated over
any continuos function. For instance, we can use it to compute the
velocity-vorticity or velocity-velocity correlation functions. Then
all these quantities will have a definite stationary value. This is
a remarkable fact, as we force continuously the perfect flow and no
dissipation is present. Looking at the prefactor $\frac{\epsilon\left(k^{2}+l^{2}\right)}{2s}$,
we remark that it is an injection rate $\epsilon\left(k^{2}+l^{2}\right)$
(the injection rate of enstrophy per unit of area) divided by twice
the shear. By analogy with the equivalent formula in classical Ornstein-Uhlenbeck
process, we see that the shear $s$ acts as an effective damping mechanism.
The effect of the shear, called Orr mechanism, leads to phase mixing
which acts as an effective dissipation for the physical quantities
dominated by the large scales. All integrated quantities, for instance
the kinetic energy, are proportional to $\frac{\epsilon\left(k^{2}+l^{2}\right)}{2s}$
and are independent on the linear friction or viscosity at leading
order. We also note that in the statistically stationary state the
enstrophy is infinite.

As an example, we compute the vorticity-stream function stationary
correlation function
\begin{equation}
h^{\infty}(\mathbf{r}_{1},\mathbf{r}_{2})=\underset{t\rightarrow\infty}{\lim}\,\mathbb{E}[\omega(\mathbf{r}_{1},t)\psi(\mathbf{r}_{2},t)]
\end{equation}
 From Eq. (\ref{eq:ginfini-reel}) we obtain
\begin{equation}
h^{\infty}(\mathbf{r}_{1},\mathbf{r}_{2})=\frac{\epsilon\left(k^{2}+l^{2}\right)}{2s}\left[\frac{\cos k(x_{1}-x_{2})}{k}\pi H_{k}(y_{1}-y_{2})+PV\int\,\frac{\sin\left[k(x_{1}-x_{2})+l(y_{1}-y_{3})\right]}{k}\frac{H_{k}(y_{2}-y_{3})}{y_{1}-y_{3}}\text{d}y_{3}\right],\label{eq:linear-shear-h-k-infty}
\end{equation}
where $H_{k}$ is the Green function for the Laplacian in the $k^{th}$
sector, with the appropriate boundary conditions. From this last expression,
and using $v^{(x)}=-\partial\psi/\partial y$ and $v^{(y)}=\partial\psi/\partial x$
it is easy to obtain the correlation function between the vorticity
and the velocity field. We note that the Green function $H_{k}(y_{1}-y_{2})$
is not a smooth function: its derivative has a jump in $y_{1}=y_{2}$.
This implies that the correlation between the vorticity and the $x$
component of the velocity is defined point-wise only for $y_{1}\neq y_{2}$
or defined globally as a distribution. With analogous computations
one can deduce other stationary two-points correlations: for example,
the Reynolds stress divergence converges to a well defined function.
\\

In this section, we have discussed how the vorticity-vorticity correlation
function has a stationary limit in the sense of distribution. When
looking at this solution point-wise, it is singular for $y_{1}=y_{2}$.
In the following section, we explain how this singularity is regularized
either by a linear friction or by viscosity. The discussion in this
section was relying on the analytic solution of the linear dynamics
close to a linear shear (Eq. (\ref{eq:pure-linear-shear-vorticity})).
Such an analytical solution is not known for generic base flows, so
that we need more refined techniques, as will be explained in section
\ref{sub:Stationarity-of-the-lyapunov}. We will obtain the same conclusion:
the vorticity-vorticity correlation function converges as a distribution,
and is regularized in a universal way.

\paragraph{Advection by a linear shear: regularization by a linear friction }

\label{subsection:linear-shear-perfect-1}

We consider the same problem as in the last paragraph, but adding
a linear friction. We will see how the singularity close to $y_{1}=y_{2}$
is regularized by a linear friction. We solve
\[
\frac{\partial g_{\alpha}}{\partial t}+L_{U}^{\alpha(1)}g_{\alpha}+L_{U}^{\alpha(2)}g_{\alpha}=2C_{m},
\]
with $L_{U}=sy\partial_{x}-\alpha$ and $C_{m}$ given by equation
(\ref{eq:Cm}). It is straightforward to observe that the evolution
under the linearized dynamics $L_{U}^{\alpha}$ is given by 
\begin{equation}
\mbox{e}^{-tL_{U}^{0}}\left[\mbox{e}^{i(kx+ly)}\right]=\mbox{e}^{ikx+ily-iksyt}\mbox{e}^{-\alpha t}\,.\label{eq:rayleigh-linear-shear-vorticity}
\end{equation}
With very similar computations to those of the last section, we can
obtain the stationary value of the vorticity-vorticity correlation
function
\begin{equation}
g_{\alpha}^{\infty}(\mathbf{r}_{1},\mathbf{r}_{2})=\lim_{t\to\infty}g_{\alpha}(\mathbf{r}_{1},\mathbf{r}_{2},t)=\frac{\epsilon\left(k^{2}+l^{2}\right)}{4s}\mbox{e}^{ik(x_{1}-x_{2})+il\left(y_{1}-y_{2}\right)}\frac{1}{k}F_{\frac{2\alpha}{ks}}(y_{1}-y_{2})\,+\mbox{C}.\mbox{C}.\,,\label{eq:perfect-shear-gkl-1-1}
\end{equation}
where
\begin{equation}
F_{\lambda}(y)=\frac{-i}{y-i\lambda}.
\end{equation}
We can thus observe that the function $F_{\lambda}(y)$ is a regularization
of the Plemelj formula (equation (\ref{eq:Plemelj}) on page \pageref{eq:Plemelj})
on the length scale $\lambda$. Indeed we have
\[
F_{\lambda}(y)=\frac{1}{\lambda}F_{1}\left(\frac{y}{\lambda}\right)=\frac{\lambda}{y^{2}+\lambda^{2}}-i\frac{y}{y^{2}+\lambda^{2}}
\]
where the real part of $F_{\lambda}$ is even while the imaginary
part is odd. Moreover the real part of $F_{\lambda}$ is a regularization
of $\pi\delta(y)$,
\[
\lim_{\lambda\rightarrow0^{+}}\mathfrak{R}\left[F_{\lambda}(y)\right]=\lim_{\lambda\rightarrow0^{+}}\frac{\lambda}{y^{2}+\lambda^{2}}=\pi\delta(y),
\]
and the imaginary part of $F_{\lambda}$ is a regularization of $-PV(1/y)$,
\[
\lim_{\lambda\rightarrow0^{+}}\mathfrak{\Im}\left[F_{\lambda}(y)\right]=-\lim_{\alpha\rightarrow0}\frac{y}{y^{2}+\lambda^{2}}=-PV\left(\frac{1}{y}\right)
\]
where $\mathfrak{R}$ and $\mathfrak{I}$ stand, respectively, for
the real and the imaginary parts. We note that for $y\gg\lambda,$
$\lim_{\lambda\rightarrow0^{+}}\mathfrak{\Re}\left[F_{\lambda}(y)\right]=0$,
and $\lim_{\lambda\rightarrow0^{+}}\Im\left[F_{\lambda}(y)\right]=1/y$.

We thus finally obtain
\begin{align}
g_{\alpha}^{\infty}(\mathbf{r}_{1},\mathbf{r}_{2}) & \underset{\frac{\alpha}{ks}\ll1}{\sim}\frac{\epsilon\left(k^{2}+l^{2}\right)}{2s}\biggl\{\cos k(x_{1}-x_{2})\,\frac{1}{k}\mathfrak{R}\left[F_{\frac{2\alpha}{ks}}(y_{1}-y_{2})\right]\label{eq:perfect-shear-gkl-1-1-1}\\
 & \qquad-\sin\left[k(x_{1}-x_{2})+l(y_{1}-y_{2})\right]\frac{1}{k}\mathfrak{I}\left[F_{\frac{2\alpha}{ks}}(y_{1}-y_{2})\right]\biggr\},\nonumber 
\end{align}
where one should observe that, because $\mathfrak{R}\left[F_{\alpha}(ky)\right]$
decays sufficiently fast to zero for $y\gg\alpha$, the factor $\cos\left[k(x_{1}-x_{2})+l(y_{1}-y_{2})\right]$
has been replaced by $\cos\left[k(x_{1}-x_{2})\right]$. Eq. (\ref{eq:perfect-shear-gkl-1-1-1})
is a regularization of the vorticity-vorticity correlation function
found in (\ref{eq:ginfini-reel}), by the effect of a small linear
friction $\alpha>0$.

Quantities which were found to be divergent in the last section are
now regularized by the presence of a small Rayleigh friction. For
example the point-wise rms. non zonal enstrophy density
\begin{equation}
\mathbb{E}_{U}\left[\omega_{m}^{2}\right]=g_{\alpha}^{\infty}(0,0)=\frac{\epsilon\left(k^{2}+l^{2}\right)}{4\alpha}.
\end{equation}

\paragraph{Advection by a linear shear in a viscous fluid}

\label{subsec:viscous-advection}

In this paragraph, we study the regularization of the solution to
the Lyapunov equation by a small viscosity considering the perturbation by a stochastic
force of a linear shear $U(y)=sy$ in a viscous flow. We consider
the domain $\mathscr{D}=[0,2\pi l_{x})\times(-\infty,\infty)$ and
periodic boundary conditions in the $x$ direction.

We solve the Lyapunov equation 
\[
\frac{\partial g}{\partial t}+L_{U}^{\nu(1)}g+L_{U}^{\nu(2)}g=2C_{m},
\]
for the vorticity-vorticity correlation function $g({\bf r}_{1},{\bf r}_{2},t)=\mathbb{E}\left[\omega_{m}({\bf r}_{1},t)\omega_{m}({\bf r}_{2},t)\right]$,
where $L_{U}^{\nu}=sy\partial_{x}-\nu\Delta$, $C_{m}$ is given by
equation (\ref{eq:Cm}) and $\epsilon$ the injection rate of energy
per unit of mass ($\epsilon$ has the dimensions $m^{2}s^{-3}$).

The deterministic evolution of the linear dynamics $\frac{\partial}{\partial t}+L_{U}^{\nu}=0$
with the initial condition $\mbox{e}^{i\left(kx+ly\right)}$ is given
by
\begin{equation}
\mbox{e}^{-tL_{U}^{\nu}}\left[\mbox{e}^{i\left(kx+ly\right)}\right]=\,\mbox{e}^{ikx}\mbox{e}^{i\left(l-skt\right)y}\,\mbox{e}^{-\nu(k^{2}+l^{2})t+\nu sklt^{2}-\frac{1}{3}\nu s^{2}k^{2}t^{3}}.\label{eq:viscous-shear-vorticity}
\end{equation}
This solution, first derived by Lord Kelvin, can be obtained by the
method of characteristics. Alternatively one can directly check a-posteriori
that (\ref{eq:viscous-shear-vorticity}) is a solution to the deterministic
equation. We can see that the second exponential in (\ref{eq:viscous-shear-vorticity})
just gives the inertial evolution of the perturbation and the third
one describes the effect of the viscosity. We thus see that the vorticity
is damped to zero on the time-scale $(1/s^{2}k^{2}\nu)^{1/3}$. This
time scale is the typical time for an initial perturbation with longitudinal
scale $1/k$ to be stretched by the shear until it reaches a scale
where it is damped by viscosity.\\
 We can now calculate the asymptotic solution to the Lyapunov equation.
It is given by 
\begin{align}
g_{\nu}^{\infty}(\mathbf{r}_{1},\mathbf{r}_{2}) & =\frac{\epsilon\left(k^{2}+l^{2}\right)}{4}\int_{0}^{\infty}\,\mbox{e}^{-tL_{U}^{\nu(1)}}\left[\mbox{e}^{ikx_{1}}\mbox{e}^{ily_{1}}\right]\mbox{e}^{-tL_{U}^{\nu(2)}}\left[\mbox{e}^{-ikx_{2}}\mbox{e}^{-ily_{2}}\right]\,\mathrm{d}t+\mbox{C.C.}\nonumber \\
 & =\frac{\epsilon\left(k^{2}+l^{2}\right)}{4s}\mbox{e}^{ik(x_{1}-x_{2})+il(y_{1}-y_{2})}\frac{1}{k}H_{\left(\frac{2\nu}{ks}\right)^{1/3}}(y_{1}-y_{2})+\mbox{C.C.}\label{eq:viscous-shear-gkl-evolution}
\end{align}
where we have used Eq. (\ref{eq:viscous-shear-vorticity}) and the
change of variable $t{}_{1}=\left(2\nu k^{2}s^{2}\right)^{1/3}t$
to write the second equality, and where 
\begin{equation}
H_{\lambda}(y)=\frac{1}{\lambda}\int_{0}^{\infty}\, e^{-\frac{iy}{\lambda}t_{1}-\lambda^{2}(k^{2}+l^{2})t_{1}+\lambda lt_{1}^{2}-\frac{1}{3}t_{1}^{3}}\,\mathrm{d}t_{1}\,.
\end{equation}
In appendix \ref{sec:appendix-regularitazions} we prove that $H_{\lambda}(y)\underset{\lambda\rightarrow0}{=}G_{\lambda}(y)+\mathcal{O}(\lambda)$
with 

\begin{equation}
G_{\lambda}(y)=\frac{1}{\lambda}\int_{0}^{\infty}\, e^{-\frac{iy}{\lambda}t_{1}-\frac{1}{3}t_{1}^{3}}\,\mathrm{d}t_{1}\,.\label{eq:viscous-advection-Glambda}
\end{equation}
We note that $\mbox{Hi}(Y)=G_{\lambda}(i\lambda Y)/\pi$ is one of
the two Scorer's functions, that solve the differential equation $\mbox{d}^{2}\mbox{Hi}/\mbox{dY}^{2}-Y\mbox{Hi}=1/\pi$,
and is related to the family of Airy functions. We stress that the
asymptotic behavior for large $y$ of the real parts of $G_{\lambda}$
and $H_{\lambda}$ are different $\Re\left[H_{\lambda}\left(y\right)\right]\underset{y\gg\sqrt{\frac{2}{k{}^{2}+l{}^{2}}}}{\sim}\left(k{}^{2}+l{}^{2}\right)\frac{\lambda^{3}}{y^{2}}$
whereas $\Re\left[G_{\lambda}\left(y\right)\right]\underset{y\gg\lambda}{\sim}2\frac{\lambda^{3}}{y^{4}}$
but in any cases those are subdominant for small $\lambda$ (please
see appendix \ref{sec:appendix-regularitazions} for more details).

As explained in appendix \ref{sec:appendix-regularitazions}, $G_{\lambda}$
is a regularization of the distributions in Plemelj formula at the
length scale $\lambda$ and has the dimension of the inverse of a
length. The integral over $y$ of the real part of $G_{\lambda}$
is $\pi$, consistently with the fact that it is a regularization
of $\pi\delta(y)$; the imaginary part is a regularization of Cauchy
Principal value of $1/y$, with

\begin{equation}
\mathbb{\Im}\left[G_{\lambda}(y)\right]\,\underset{y\gg\lambda}{\sim}-\frac{1}{y}.\label{eq:viscosity-immag1}
\end{equation}
We thus conclude that
\begin{align}
g_{\nu}^{\infty}(\mathbf{r}_{1},\mathbf{r}_{2}) & \underset{\left(\frac{2\nu}{sk}\right)^{1/3}l\ll1}{\sim}\frac{\epsilon\left(k^{2}+l^{2}\right)}{2s}\biggl\{\frac{\cos\left[k(x_{1}-x_{2})\right]}{k}\mathbb{\Re}\left[G_{\left(\frac{2\nu}{ks}\right)^{1/3}}(y_{1}-y_{2})\right]\label{eq:viscous-advection-g-final}\\
 & \qquad-\frac{\sin\left[k(x_{1}-x_{2})+l(y_{1}-y_{2})\right]}{k}\mathbb{\Im}\left[G_{\left(\frac{2\nu}{ks}\right)^{1/3}}(y_{1}-y_{2})\right]\biggr\}\,.\nonumber 
\end{align}
Observe that, because $\Re\left[G_{\left(\frac{2\nu}{ks}\right)^{1/3}}\left(y\right)\right]$
decays sufficiently fast to zero for $y\gg\left(\frac{2\nu}{ks}\right)^{1/3}$,
the factor $\cos\left[k(x_{1}-x_{2})+l(y_{1}-y_{2})\right]$ has been
replaced by its value for $y_{1}=y_{2}$. The first term of Eq. (\ref{eq:viscous-advection-g-final})
is a local contribution, for values of $y_{1}-y_{2}$ of order $\left(\frac{2\nu}{ks}\right)^{1/3}$,
whereas the second term is a global contribution. Whereas the local
contribution is independent on $l$ and depends on $y_{1}-y_{2}$
through the shape function $G$, the global contribution has a phase
dependance through $l(y_{1}-y_{2})$.

By contrast, the point wise value of the two-point vorticity correlation
function for $(y_{1}-y_{2})\sim\left(\frac{2\nu}{ks}\right)^{1/3}$
diverges for large Reynolds number. For instance 
\[
\mathbb{E}_{U}\left[\omega_{m}^{2}\right]=g_{\nu}^{\infty}(0,0)\underset{\left(\frac{2\nu}{sk}\right)^{1/3}l\ll1}{\sim}\frac{\epsilon\left(k^{2}+l^{2}\right)}{2s}\left(\frac{s}{2k^{2}\nu}\right)^{1/3}\,\int_{0}^{\infty}\,\mbox{e}^{-\frac{1}{3}t_{1}^{3}}\,\mathrm{d}t_{1},
\]
and one can note that $\frac{s}{2\nu k^{2}}$ is a Reynolds number
based on the local shear and the scale of the non-zonal perturbation.
We observe that the enstrophy density $\mathbb{E}_{U}\left[\omega_{m}^{2}\right]$
is regularized by viscosity but diverges for large Reynolds number
to the power $1/3$.

We conclude by stating that physical quantities involving higher order
derivatives will also be regularized by the viscosity and diverge
with the Reynolds number. For instance the palinstrophy density will
diverge as
\[
\mathbb{E}_{U}\left[\left(\nabla\omega_{m}\right){}^{2}\right]\underset{\left(\frac{2\nu}{sk}\right)^{1/3}l\ll1}{\sim}C_{P}\frac{\epsilon\left(k^{2}+l^{2}\right)}{2s}\left(\frac{ks}{2\nu}\right)^{2/3}\left(\frac{s}{2\nu k^{2}}\right)^{1/3},
\]
where $C_{P}$ is a non-dimensional constant.\\

In this section, we have discussed the case of a force at a longitudinal
scale $1/k$ and transverse scale $1/l$. The conclusion for any other
forces can easily be obtained by superposition of the contributions
from all scales. The general conclusion is that in the limit of large
Reynolds number $\frac{s}{2\nu k^{2}}\gg1$, the two-point correlation
function converges as a distribution, and converges point-wise for
values of $y_{1}-y_{2}$ much larger than $\left(\frac{2\nu}{k_{m}s}\right)^{1/3}$
where $1/k_{m}$ is the maximal scale for the forcing. As a consequence,
the velocity-velocity and velocity-vorticity correlation functions
have a limit independent on the Reynolds number, and the kinetic energy
density ($\mbox{m}^{2}$.$\mbox{s}^{-2}$) is roughly proportional
to $\frac{\epsilon}{2s}$ where $\epsilon$ is the average energy
input rate.

We have made explicit computations only in the case of a linear shear
$U(y)=sy$. Explicit computations are not easily done in more complex
situations in the presence of viscosity. However we expect that for
generic $U$, the singularities of the stationary solutions to the
Lyapunov equations are regularized in a universal way.

\paragraph{Advection by a linear shear in a viscous fluid with linear friction}

When both linear friction and viscosity are present, the analysis
above can be easily generalized. The way the two points correlation
function is regularized depends on the relative value of the two length
scales $\lambda_{\alpha}=\frac{2\alpha}{ks}$ and $\lambda_{\nu}=\left(\frac{2\nu}{ks}\right)^{1/3}$.
When $\lambda_{\alpha}\gg\lambda_{\nu}$, the regularization is of
a friction type and formula (\ref{eq:perfect-shear-gkl-1-1-1}) will
be correct. When $\lambda_{\alpha}\ll\lambda_{\nu}$, the regularization
is of a viscous type and formula (\ref{eq:viscous-advection-g-final})
will be correct. 

We stress that, whatever the values of the length scales $\lambda_{\alpha}$
and $\lambda_{\nu}$, the real part of the regularizing function always
decays proportionally to $1/y^{4}$ for small enough values of $y$
and proportionally to $1/y^{2}$ for large enough values of $y$.
The location of the crossover between these two behaviors depends
on the values of the length scales $\lambda_{\alpha}$ and $\lambda_{\nu}$
. A careful discussion of this issue is addressed in appendix \ref{sec:appendix-regularitazions}. 

\noindent Here, we only point out that three cases are possible: $(i)$
when $\lambda_{\alpha}\gg\lambda_{\nu}$ the regularization is of
friction type and the crossover happens for $y\sim\lambda_{\alpha}$;
$(ii)$ when $\lambda_{\alpha}\ll\lambda_{\nu}^{3}(k^{2}+l^{2})$
the regularization is of viscous type and the crossover happens for
$y\sim\sqrt{2/(k^{2}+l^{2})}$; $(iii)$ when $\lambda_{\alpha}\ll\lambda_{\nu}$
but $\lambda_{\alpha}\gg\lambda_{\nu}^{3}(k^{2}+l^{2})$ the singularity
is also of viscous type, but in this case the crossover happens for
$y\sim\sqrt{2\lambda_{\nu}^{3}/\lambda_{\alpha}}$. The emergence
of this last length scale $\sqrt{2\lambda_{\nu}^{3}/\lambda_{\alpha}}$
is due to the different multiplicative factors of the $1/y^{2}$ decay
due to friction and in the $1/y^{4}$ decay due to viscosity (please
see appendix \ref{sec:appendix-regularitazions}).

\subsection{Stationary solutions to the Lyapunov equation in the inertial limit\label{sub:Stationarity-of-the-lyapunov}}

In the previous section, we have considered a base flow with constant
shear, for which analytical solution to the Lyapunov equation can
be computed, and provides a qualitative understanding of its solution.
We have concluded that it has a stationary solution, in the sense
of distributions, even without dissipation. This solution diverges
locally for $y_{1}=y_{2}$, related to an infinite enstrophy. We have
explained how those local divergences are regularized by a small linear
friction or viscosity. The aim of this section is to prove that the
same conclusions remain valid for any generic shear flow $U(y)$,
which is assumed to be linearly stable. We also prove that the Reynolds
stress divergence $-\mathbb{E}_{U}\left[\langle v_{m}^{(y)}\omega_{m}\rangle\right]$
and its gradient $-\mathbb{E}_{U}\left[\frac{\partial}{\partial y}\langle v_{m}^{(y)}\omega_{m}\rangle\right]$
have finite values. Convergence results for other two-point correlations
are also discussed.

At a rough qualitative level, the reason why this is valid is the
same as the one discussed in the previous section: the effect of the
shear (Orr-mechanism). However this explanation can not be considered
as satisfactory. The constant shear flow case has a zero gradient
of vorticity, that's why the equation are so simple and can be solved
analytically. Whenever the vorticity gradient is non-zero, the hydrodynamic
problem becomes drastically different, coupling all fluctuations globally
due to the long range interactions involved through the computation
of the velocity. Any explanation based on local shear only is then
doubtful. Moreover, most of jets in geophysical situations have points
with zero shear $U'(y)=0$. This is also a necessity for jets in doubly
periodic geometries. Then the local shear effect can not be advocated.

As already suggested, explicit analytical results are hopeless in
this case. In order to prove the result, we rely on two main ingredients.
First we use the fact that the stationary solution of the Lyapunov
equation can be computed from solutions of the deterministic linearized
dynamics, as expressed by formulas (\ref{eq:decompositiongk}) and
(\ref{eq:lyapunov-solution-real-time}). Second we prove that these
formulas have limits for large times based on results on the asymptotic
behavior of the linearized 2D Euler equations, discussed in the work
\cite{Bouchet_Morita_2010PhyD}. At a qualitative level, the results
of this paper show that the flow can be divided into areas dominated
by the shear for which the Orr mechanism is responsible for a phase
mixing leading to an effective dissipation. In other flow regions,
for instance close to jet extrema, where no shear is present, a global
mechanism called vorticity depletion at the stationary streamlines
wipes out any fluctuations, extremely rapidly.

We discuss in this section only the case with no beta effect $\beta=0$,
the case with beta effect will be discussed in a forthcoming publication.

\subsubsection{The Orr mechanism and vorticity depletion at the stationary streamlines}

\label{subsection:orr-a-preliminary}

In this section we summarize existing results on the large time asymptotics
of the linearized Euler equations \cite{Bouchet_Morita_2010PhyD}.
We consider the linear deterministic advection with no dissipation.
Because the corresponding linear operator $L_{U}^{0}$ is not normal,
a set of eigenfunctions spanning the whole Hilbert space on which
$L_{U}^{0}$ acts does not necessarily exist. Because $U$ is stable,
$L_{U}^{0}$ has no eigenmodes corresponding to exponential growth.
Moreover it is a very common situation that the Euler operator $L_{U}^{0}$
has no modes at all (neither neutral nor stable nor unstable). A simple
example for which this can be easily checked is the case of the linear
shear treated in Section \ref{subsection:linear-shear-perfect}. We
assume in the following that $L_{U}^{0}$ has no mode and $\beta=0$.

While the vorticity shows filaments at finer and finer scales when
time increases, non-local averages of the vorticity (such as the stream-function
or the velocity) converge to zero in the long-time limit. This relaxation
mechanism with no dissipation is very general for advection equations
and it has an analog in plasma physics in the context of the Vlasov
equation, where it is called Landau damping \cite{Mouhot_Villani:2009}.
This phenomenon was first studied in the context of the Euler equations
in the case of a linear profile $U(y)=sy$ in \cite{Case_1960_Phys_Fluids}.
The work \cite{Bouchet_Morita_2010PhyD} was the first to study the resolvent and to establish results about the asymptotic behavior of the linearized equations in the case in which
the profile $U(y)$ has stationary points $y_{c}$
such that $U'(y_{c})=0$, which is actually the generic case.

We consider the deterministic linear dynamics $\partial_{t}\tilde{\omega}+L_{U}^{0}\tilde{\omega}=0$
with initial condition $\mbox{e}^{ikx}f(y)$. The solution is of the
form $\tilde{\omega}(x,y,t)=\mbox{e}^{ikx}\tilde{\omega}_{k}(y,t)$.
From \cite{Bouchet_Morita_2010PhyD}, we know that %
\footnote{We prefer to use in this section the notation with a tilde to denote
the solutions of the deterministic linear dynamics $\partial_{t}+L_{U}^{0}$.%
} 
\begin{equation}
\tilde{\omega}_{k}(y,t)\underset{t\to\infty}{\sim}\tilde{\omega}_{k}^{\infty}(y)\mbox{e}^{-ikU(y)t}\,.\label{eq:orr-mechanism-voricity}
\end{equation}
 We thus see that the vorticity oscillates on a finer and finer scale
as the time goes on. By contrast to the behavior of the vorticity,
any integral of the vorticity with a differentiable kernel decays
to zero. For instance, the results for the $x$ and $y$ components
of the velocity and for the stream function are: 
\begin{equation}
\tilde{v}_{k}^{(x)}(y,t)\underset{t\to\infty}{\sim}\frac{\tilde{\omega}_{k}^{\infty}(y)}{ikU'(y)}\frac{\mbox{e}^{-ikU(y)t}}{t},\label{eq:orr-mechanism-velocity-x}
\end{equation}
 
\begin{equation}
\tilde{v}_{k}^{(y)}(y,t)\underset{t\to\infty}{\sim}\frac{\tilde{\omega}_{k}^{\infty}(y)}{ik(U'(y))^{2}}\frac{\mbox{e}^{-ikU(y)t}}{t^{2}},\label{eq:orr-mechanism-velocity-y}
\end{equation}
and 
\begin{equation}
\tilde{\psi}_{k}(y,t)\underset{t\to\infty}{\sim}\frac{\tilde{\omega}_{k}^{\infty}(y)}{(ikU'(y))^{2}}\frac{\mbox{e}^{-ikU(y)t}}{t^{2}}\,.\label{eq:orr-mechanism-stream}
\end{equation}
 In all the above formulas, higher order corrections are present and
decay with higher powers in $1/t$. One should also observe that $\tilde{\omega}_{k}^{\infty}(y)$
depends on the initial condition $f(y)$. The asymptotic profile $\tilde{\omega}_{k}^{\infty}(y)$
could be computed numerically, for instance from the resolvent of
the operator $L_{U}^{0}$. An essential point is that $\tilde{\omega}_{k}^{\infty}(y)$
has in general no local approximation, it is not a simple function
of the local shear but depends on the whole profile $U$.

These results have been proven for every shear flow $U$, also in
the presence of stationary points $y_{c}$ such that $U'(y_{c})=0$.
Moreover, it has been proved in \cite{Bouchet_Morita_2010PhyD} that
at the stationary points $\tilde{\omega}_{k}^{\infty}(y_{c})=0$ .
This phenomenon has been called vorticity depletion at the stationary
streamlines. It has been observed numerically that the extend of the
area for which $\tilde{\omega}_{k}^{\infty}(y_{c})\simeq0$ can be
very large, up to half of the total domain, meaning that in a large
part of the domain, the shear is not the explanation for the asymptotic
decay. The formula for the vorticity Eq. (\ref{eq:orr-mechanism-voricity})
is valid for any $y$. The formulas for the velocity and stream functions
are valid for any $y\neq y_{c}$. Exactly at the specific point $y=y_{c}$,
the damping is still algebraic with preliminary explanation given
in \cite{Bouchet_Morita_2010PhyD}, but a complete theoretical prediction
is not yet available.

Using these results, we will study the stationary solutions of the
Lyapunov equation with no dissipation.

\subsubsection{The vorticity auto-correlation function converges to a distribution
in the perfect flow limit}

\label{sub:vorticity-auto-correlation}

We prove now that the Lyapunov equation has a stationary solution,
in the sense of distributions. The hypothesis are the same as in previous
section: $U$ is linearly stable, has no mode, and $\beta=0$. 

In order to understand the long time behavior of the vorticity-vorticity
correlation function, we consider Eq. (\ref{eq:lyapunov-solution-real-time}).
We consider Eq. (\ref{eq:orr-mechanism-voricity}), where $\tilde{\omega}_{k}(y,t)\mbox{e}^{ikx}$
is the solution of the deterministic equation with initial condition
$\mbox{e}^{ikx+ily}$ (we note that $\tilde{\omega}_{k}$ then depends
on $l$ through the initial condition $\mbox{e}^{ily}$). One can
check that the integrand of the r.h.s. of this equation behaves for
long time as 
\begin{equation}
\tilde{\omega}_{k}(y_{1},t_{1})\tilde{\omega}_{k}^{*}(y_{2},t_{1})\underset{t_{1}\to\infty}{\sim}\tilde{\omega}_{k}^{\infty}(y_{1})\tilde{\omega}_{k}^{\infty*}(y_{2})\,\mbox{e}^{-ik\left[U(y_{1})-U(y_{2})\right]t_{1}}\,.\label{eq:lyapunov-general-vorticity-vorticity}
\end{equation}

For $y_{1}$ and $y_{2}$ such that $U(y_{1})\neq U(y_{2})$ a computation
analogous to the one in section \ref{subsection:linear-shear-perfect},
shows that $g_{kl}$, and hence $g^{\infty}$ diverge proportionally
to time $t$. This is related to an infinite value of enstrophy. 

In the following, we write formulas for the case when $U$ is a monotonic
profile. Then each frequency correspond to a single streamline. In
the opposite case, two streamlines may have the same frequency, and
resonances between streamlines should be considered. The formula would
then be more intricate, but the result can be easily obtained from
(\ref{eq:orr-mechanism-voricity}) and the conclusion that the limit
exists is still true%
\footnote{We note that the difficult theoretical result related to those resonances
is to establish (\ref{eq:orr-mechanism-voricity})%
}. From (\ref{eq:lyapunov-general-vorticity-vorticity}), we get 

\begin{align}
g_{kl}(\mathbf{r}_{1},\mathbf{r}_{2},t) & =\int_{0}^{t}\tilde{\omega}_{k}(y_{1},t_{1})\mbox{e}^{ikx_{1}}\tilde{\omega}_{k}^{*}(y_{2},t_{1})\mbox{e}^{-ikx_{2}}\,\mbox{d}t_{1}+\mbox{C.C.}\label{eq:gkl-orr}\\
 & =\left\{ \frac{\tilde{\omega}_{k}^{\infty}(y_{1})\tilde{\omega}_{k}^{\infty*}(y_{2})}{-ik\left[U(y_{1})-U(y_{2})\right]}\,\left[\mbox{e}^{-ik\left[U(y_{1})-U(y_{2})\right]t}-1\right]+g^{r}(y_{1},y_{2},t)\right\} \mbox{e}^{ik(x_{1}-x_{2})}+\mbox{C.C.}\,,\nonumber 
\end{align}
 where $g^{r}$ is a function which remains continuous in the $t\to\infty$
limit. We thus obtain, for the stationary vorticity-vorticity correlation
function
\begin{equation}
g_{kl}^{\infty}(\mathbf{r}_{1},\mathbf{r}_{2})=\,\left\{ \frac{\pi\,|\tilde{\omega}_{k}^{\infty}(y_{1})|^{2}}{|kU'(y_{1})|}\delta(y_{1}-y_{2})-iPV\left(\frac{\tilde{\omega}_{k}^{\infty}(y_{1})\tilde{\omega}_{k}^{\infty,*}(y_{2})}{k\left(U(y_{1})-U(y_{2})\right)}\right)+g^{r}(y_{1},y_{2},\infty)\right\} \mbox{e}^{ik(x_{1}-x_{2})}+\mbox{C.C.}\,,\label{eq:gkl-stationary-orr}
\end{equation}
and conclude that the Lyapunov equation has a stationary solution
understood as a distribution.

Another quantity of interest is the velocity auto-correlation function
and the kinetic energy density. We note that the velocity auto-correlation
function is a quadratic quantity, that can be obtained directly as
a linear transform from the vorticity-vorticity correlation function
$g$, or from the deterministic solution to the linearized operator.
From (\ref{eq:decompositiongk}), we see that the contributions from
each forcing modes add up. Then 
\[
\mathbb{E}_{U}\left[{\bf v}_{m}({\bf r}_{1},t)\cdot{\bf v}_{m}({\bf r}_{2},t)\right]=\sum_{k>0,l}c_{kl}E_{kl}({\bf r}_{1},{\bf r}_{2}),
\]
with 

\begin{equation}
E_{kl}({\bf r}_{1},{\bf r}_{2})=\mbox{e}^{ik(x_{1}-x_{2})}\int_{0}^{\infty}\,\tilde{\mathbf{v}}_{k}(y_{1},u).\tilde{\mathbf{v}}_{k}^{*}(y_{2},u)\,\mbox{d}u+\mbox{C.C.},\label{eq:hkl-1}
\end{equation}
where $\tilde{\mathbf{v}}_{k}\mbox{e}^{ikx}$ is the velocity of the
deterministic solution to the linearized equations $\partial_{t}\tilde{\omega}+L_{U}^{0}\tilde{\omega}=0$,
with initial condition $\mbox{e}^{ikx+ily}$, as in the previous section.
Alternatively, $E_{kl}=\mathbf{V}^{(1)}\mathbf{V}^{(2)*}g_{kl}^{\infty}$,
where $\mathbf{V}=-\nabla\left[\Delta^{-1}(.)\right]\times\mathbf{e}_{z}$
is the linear operator giving the velocity from the vorticity and
$\mathbf{V}^{(1\text{)}}$, resp. $\mathbf{V}^{(2\text{)}}$ are the
operator $\mathbf{V}$ acting on the first, or second variable respectively.
From (\ref{eq:orr-mechanism-velocity-y},\ref{eq:orr-mechanism-velocity-x}),
it is clear that $E_{kl}$ and thus the velocity autocorrelation functions
have finite values in the inertial limit, even if no dissipation is
present in this limit.

We note that $\alpha\mathbb{E}_{U}\left[{\bf v}_{m}({\bf r}_{1},t)\cdot{\bf v}_{m}({\bf r}_{1},t)\right]$
is the non-zonal kinetic energy density (the kinetic energy contained
in the non-zonal degrees of freedom). We thus conclude that in the
limit of very small $\alpha$, the non-zonal kinetic energy is proportional
to $\alpha$, and that its value can be estimated from the Lyapunov
equation with $\alpha=0$. Those results are extremely important,
as they prove that the scaling we have adopted all along this work
in order to make an asymptotic expansion is self-consistent.

\subsubsection{Convergence of the Reynolds stress divergence and of its gradient}

\label{subsec:reynolds-orr} The stationary value of the Reynolds
stress divergence $-\mathbb{E}_{U}\left[\left\langle v_{m}^{(y)}\omega_{m}\right\rangle(y)\right]$
and its gradient $-\mathbb{E}_{U}\left[\frac{\partial}{\partial y}\left\langle v_{m}^{(y)}\omega_{m}\right\rangle(y)\right]$
are central objects in the kinetic theory. Indeed, they enter in the
final kinetic equation when written, respectively, for the evolution
of the zonal velocity (\ref{eq:Stochastique-Lent-U}) or for the evolution
of the zonal vorticity (\ref{eq:Stochastique-Lent}). We have thus
to prove that they are finite.

We note that the Reynolds stress is a quadratic quantity that can
be obtained directly as a linear transform from the vorticity-vorticity
correlation function $g$. From (\ref{eq:decompositiongk}), we see
that the contributions from each forcing modes add up. Then 
\[
F(y)\equiv\mathbb{E}_{U}\left[\left\langle v_{m}^{(y)}\omega_{m}\right\rangle(y)\right]=\sum_{k>0,l}c_{kl}F_{kl}(y),
\]
with 

\begin{equation}
F_{kl}(y)=\underset{t\rightarrow\infty}{\lim}\int_{0}^{t}\,\tilde{\omega}_{k}(y,u)\,\tilde{v}_{k}^{(y)*}(y,u)\,\mbox{d}u+\mbox{C.C.},\label{eq:hkl}
\end{equation}
where $\tilde{\omega}_{k}\mbox{e}^{ikx}$ and $\tilde{v}_{k}^{(y)}\mbox{e}^{ikx}$
are the deterministic solutions to the linearized equations $\partial_{t}\tilde{\omega}+L_{U}^{0}\tilde{\omega}=0$
with initial condition $\mbox{e}^{ikx+ily}$, as in previous section.

Proving that the long-time limit of the integral in Eq. (\ref{eq:hkl})
converges is straightforward. Using Eq. (\ref{eq:orr-mechanism-voricity})
and (\ref{eq:orr-mechanism-velocity-y}), one easily realizes that
the integrands behaves as $1/u^{2}$ for large $u$ and thus the integral
converges. One conclude that the Reynolds stress divergence is finite.

The computation needed to prove that also the gradient of the Reynolds
stress divergence $\sum_{k>0,l}c_{kl}\partial F_{kl}/\partial y$
exists, has to be done with more care. Using (\ref{eq:hkl}), (\ref{eq:orr-mechanism-voricity})
and (\ref{eq:orr-mechanism-velocity-y}), one may be led to the conclusion
that a contribution proportional to $1/u$ exists, that would lead
to a logarithmic divergence once integrated over time. This is actually
not the case because the computation shows that the leading order
contributions to the Reynolds stress cancels out exactly, as can be
checked easily. As a consequence, we conclude that also the divergence
of the Reynolds stress is finite.

In order to reach this conclusion, we have used (\ref{eq:orr-mechanism-voricity})
and (\ref{eq:orr-mechanism-velocity-y}), which have been established
for those values of $y$ that are not stationary for the profile $U(y)$.
For a stationary point $y=y_{c}$ no theoretical results are available
for the asymptotic behavior of the velocity field. However, based
on numerical evidences reported in \cite{Bouchet_Morita_2010PhyD},
we discuss in Appendix \ref{appendix:reynolds-stress-yc} that the
Reynolds stress divergence is finite also for $y=y_{c}$. This will
be also checked by direct numerical computations of the Lyapunov equation
discussed below.

\subsubsection{The effect of the small Rayleigh friction and viscosity on stationary
solutions of the Lyapunov equation \label{sub:Lyapunov-Rayleigh-Viscosity}}

We briefly comment on the effect of the small linear friction, or
of a small viscosity. From the discussions in sections \ref{subsection:orr-a-preliminary}
and \ref{subsec:reynolds-orr}, we have proved that the kinetic energy
density and Reynolds stress have limit values independent on $\alpha$
or the viscosity. By contrast, the vorticity-vorticity correlation
function $g^{\infty}(\mathbf{r}_{1},\mathbf{r}_{2})$ diverges point-wise
for $y_{1}=y_{2}$ (more generally for any two points for which the
streamlines have the same frequency), and that it has a well defined
limit as a distribution. This also implies that the enstrophy contained
in non-zonal degrees of freedom is virtually infinite for $\alpha=0$.
We address how this is regularized for small values of $\alpha$,
or due to a small viscosity. 

We have already discussed this issue for the particular case of a
linear shear $U(y)=sy$, in section \ref{sub:Two-trivial-examples},
both with and without viscosity. For the general case of any stable
base flow $U$ and for a linear friction, we will conclude below that
the same conclusion as for the linear shear holds. More precisely,
we can prove that in the vicinity of $y_{1}=y_{2}$, the singular
behavior is regularized in a universal way (with shape functions independent
on $U$) over a scale $\alpha/kU'(y)$. Moreover we can prove
that $\mathbb{E}_{U}\left[\omega_{m}^{2}(\mathbf{r})\right]$ diverges
proportionally to $1/\alpha$, such that the non-zonal enstrophy density
$\alpha\mathbb{E}_{U}\left[\omega_{m}^{2}(\mathbf{r})\right]$ has
a finite limit (this could have been expected in order to balance
the finite enstrophy input rate provided by the stochastic force). 

When viscous dissipation is present, we think that it is still true
that the divergence of $g$ has a universal regularization, and we
think that the scaling for the divergence and enstrophy found for
the linear shear are also valid for a generic velocity profile $U$.
We do not prove this result, but discuss why this is plausible at
the end of this section.\\

In order to prove those results, we need to consider the operator
$L_{U}^{\alpha}=L_{U}^{0}+\alpha$ instead of the inertial $L_{U}^{0}$
for the analysis of the Lyapunov equation. The effect of replacing
$L_{U}^{0}$ by $L_{U}^{\alpha}$ in section \ref{sub:Equation-de-Lyapunov-sec1}
corresponds simply to add the exponential damping $\mbox{e}^{-2\alpha t_{1}}$
to the integrands of expressions like the one in Eq. (\ref{eq:gkl-orr}).
This observation gives us the possibility of understanding how diverging
quantities are actually regularized by the presence of a small friction
for any zonal flow $U$. Straightforward computation then leads to
a regularization of the inertial result (\ref{eq:gkl-stationary-orr})
as
\begin{equation}
g_{kl}^{\infty}(\mathbf{r}_{1},\mathbf{r}_{2})=\left\{ \frac{\tilde{\omega}_{k}^{\infty}(y_{1})\tilde{\omega}_{k}^{\infty*}(y_{2})}{ik\left(U(y_{1})-U(y_{2})\right)+2\alpha}+g_{\alpha}^{r}(y_{1},y_{2},\infty)\right\} \mbox{e}^{ik(x_{1}-x_{2})}+\mbox{C.C.}\,.\label{eq:gkl-stationary-orr-alpha}
\end{equation}
where, as before, $g_{\alpha}^{r}$ has a finite limit when $\alpha\rightarrow0$.
In the small $\alpha$ limit and for $y_{1}\sim y_{2}$, we have 
\begin{align}
g_{kl}^{\infty}(\mathbf{r}_{1},\mathbf{r}_{2}) & \underset{\frac{\alpha}{\left|kU'(y_{1})\right|}\ll1,\, y_{1}\sim y_{2}}{\sim}2\biggl\{\cos k(x_{1}-x_{2})\,\frac{\left|\tilde{\omega}_{k}^{\infty}(y_{1})\right|^{2}}{\left|kU'(y_{1})\right|}\mathfrak{R}\left[F_{\frac{2\alpha}{\left|kU'(y_{1})\right|}}(y_{1}-y_{2})\right]\label{eq:perfect-shear-gkl-1-1-1-1}\\
 & \qquad+\mathfrak{I}\left[\tilde{\omega}_{k}^{\infty}(y_{1})\tilde{\omega}_{k}^{\infty*}(y_{2})\mbox{e}^{ik(x_{1}-x_{2})}\right]\frac{1}{kU'(y_{1})}\mathfrak{I}\left[F_{\frac{2\alpha}{\left|kU'(y_{1})\right|}}(y_{1}-y_{2})\right]\biggr\}+g_{\alpha}^{r,2}(\mathbf{r}_{1},\mathbf{r}_{2}),\nonumber 
\end{align}
where $g_{\alpha}^{r,2}$ has a finite limit when $\alpha\rightarrow0$.
We also recall that if $U'(y)=0$, then $\tilde{\omega}_{k}^{\infty}(y)=0$.
Comparing (\ref{eq:perfect-shear-gkl-1-1-1}) with (\ref{eq:perfect-shear-gkl-1-1-1-1}),
we conclude that the vorticity-vorticity correlation function is regularized
in a universal way close to $y_{1}=y_{2}$.

The fact that the $\mathbb{E}_{U}\left[\omega_{m}^{2}(\mathbf{r})\right]$
diverges proportionally to $1/\alpha$, is obtained by observing that
Eq. (\ref{eq:lyapunov-general-vorticity-vorticity}) should be replaced
by 
\begin{equation}
\tilde{\omega}_{k}(y_{1},t_{1})\tilde{\omega}_{k}^{*}(y_{2},t_{1})\sim_{t_{1}\to\infty}\tilde{\omega}_{k}^{\infty}(y_{1})\tilde{\omega}_{k}^{\infty*}(y_{2})\,\mbox{e}^{-ik\left[U(y_{1})-U(y_{2})\right]t_{1}-2\alpha t_{1}}\,.
\end{equation}
 We thus have 
\begin{equation}
\mathbb{E}_{U}\left[\omega_{m}^{2}(\mathbf{r})\right]=\sum_{k>0,l}c_{kl}g_{kl}^{\infty}({\bf r},{\bf r})=\frac{\left|\tilde{\omega}_{k}^{\infty}(y)\right|^{2}}{2\alpha}+g_{\alpha}^{r}(y,y,\infty)
\end{equation}

We have not worked out the case with a small viscosity. This is technically
possible. This would require first to describe precisely the large
time asymptotics of the linearized 2D Navier--Stokes equations. Such
results should involve the classical literature on the study of viscous
critical layers \cite{Drazin_Reid_1981}, through matched asymptotics
expansions between what happens close to the critical layers and away
from it. Those classical results exhibit naturally the scale $\left(\frac{2\nu}{ks}\right)^{1/3}$
and the family of Airy functions. It should be possible to extent
those classical deterministic results to the Lyapunov equation, using
the relation between the deterministic solutions and the solutions
to the Lyapunov equation discussed at the beginning of this section.
For this reason, we guess that the regularization obtained for the
linear shear is universal and should explain the regularization for
generic velocity profile $U$. One important technical drawback is
that we guess that the classical results for the deterministic equation
do not yet exist yet for the case where $U$ has stationary streamlines
($U'(y_{c})=0$) and when several streamlines have the same frequency.

\subsection{Numerical solutions of the Lyapunov equation\label{sub:Numerical-solutions-of}}

In the previous section we have proved that the Lyapunov equation
for the linearized Euler dynamics has a stationary solution. In this
section we compute numerically this stationary solution.

For the Lyapunov equation with viscosity, or hyperviscosity, assuming
a stable base flow, we expect the stationary solution to be well behaved.
Then a natural way to solve the Lyapunov equation would be to discretize
the linear operator $L_{U}^{(1)}+L_{U}^{(2)}$ and to directly solve
the approximate dynamics. This is the traditional way and such a technique
has for instance been used in most of previous works using SSST-CE2
\cite{BakasIoannou2013SSST,Farrel_Ioannou,Farrell_Ioannou_JAS_2007,Marston_Conover_Schneider_JAS2008,Srinivasan-Young-2011-JAS,ParkerKrommes2013SSST}.

However, we are here specifically interested in the inertial limit.
Then one could solve the Lyapunov equation for finite values of $\alpha$
and $\nu$ and then study the asymptotic behavior of the results when
these parameters go to zero. While feasible, this route seems extremely
difficult, as the numerical discretization would have to be increased
as $\nu$ goes to zero. We have then chosen to try another route in
the following. We will make a direct numerical computation of $F$,
the divergence of the Reynolds stress which appears in the kinetic
equation (\ref{eq:Stochastique-Lent-U}), which we have proved to
be well behaved. In order to do that, we first establish an integral
equation verified by $h_{kl}$, the vorticity-stream function stationary
correlation function, which will be precisely defined below. As we
will show, this integral equation does not suffer from the instability
in the small viscosity limit that would hinder a partial differential
equation approach, and can be solved even if $\nu=0$.

Using this technique, we illustrate the theoretical results of last
section: $F$ has a finite limit for the zero friction limit $\alpha\to0$,
the energy in the non-zonal degrees of freedom has a finite limit,
and the energy dissipation rate for the non-zonal degrees of freedom
goes to zero when $\alpha$ goes to zero. We also check the analyticity
breaking, and its order, for the correlation function $h_{kl}(y_{1},y_{2})$
at the points $y_{1}=y_{2}$.

\subsubsection{Computing stationary solution of the Lyapunov equation from an integral
equation}

By definition, the stationary vorticity autocorrelation function $g_{kl}^{\infty}$
is the solution of the stationary Lyapunov equation (with linear friction
and no viscosity)
\[
L_{U}^{\alpha(1)}g_{kl}^{\infty}+L_{U}^{\alpha(2)}g_{kl}^{\infty}=2\cos(k(x_{1}-x_{2})+l(y_{1}-y_{2})).
\]
Equivalently, we have $g_{kl}^{\infty}(x_{1},x_{2},y_{1},y_{2})=\mbox{e}^{ik(x_{1}-x_{2})}\tilde{g}_{kl}(y_{1},y_{2})+\mbox{C.C.}$
with
\begin{equation}
L_{U,k}^{\alpha(1)}\tilde{g}_{kl}+L_{U,-k}^{\alpha(2)}\tilde{g}_{kl}=\mbox{e}^{il(y_{1}-y_{2})}\label{eq:gk-alpha}
\end{equation}
where 
\[
L_{U,k}^{\alpha}=ikU-ik\left(U''-\beta\right)\Delta_{k}^{-1}+\alpha.
\]
We define $h_{kl}$ such that $\tilde{g}_{kl}=\Delta_{k}^{(2)}h_{kl}$,\textbf{
}where \textbf{$\Delta_{k}^{(2)}$} is the Laplacian operator with
respect to the second variable. The drift term in Eq. (\ref{eq:Fokker-Planck-Zonal})
can be directly computed from $h_{kl}$, with
\[
F(y)\equiv\mathbb{E}_{U}\left[\langle v_{m}^{(y)}\omega_{m}\rangle(y)\right]=\sum_{k>0,l}c_{kl}F_{kl}(y),
\]
where
\[
F_{kl}(y)=2k\mathfrak{I}\left[h_{kl}(y,y)\right].
\]
In terms of $h_{kl}$, the stationary Lyapunov equation (\ref{eq:gk-alpha})
becomes
\begin{equation}
\Delta_{k}^{(2)}h_{kl}(y_{1},y_{2})=\frac{ik\left(U''(y_{1})-\beta\right)h_{kl}^{*}(y_{2},y_{1})-ik\left(U''(y_{2})-\beta\right)h_{kl}(y_{1},y_{2})+\mbox{e}^{il(y_{1}-y_{2})}}{ik(U(y_{1})-U(y_{2}))+2\alpha}.\label{eq:hk}
\end{equation}
We note that this equation is not a differential equation for $h_{kl}(y_{1},y_{2})$
for each $y_{1}$ fixed, as it involves $h_{kl}^{*}(y_{2},y_{1})$.

In the previous section, we have shown that $g_{kl}^{\infty}$, and
thus $\tilde{g}_{kl}=\Delta_{k}^{(2)}h_{kl}$ diverge point-wise for
$y_{1}\rightarrow y_{2}$ in the limit $\alpha\rightarrow0$. This
is related to the vanishing of the denominator for $\alpha=0$ and
$y_{1}=y_{2}$ in equation (\ref{eq:hk}). On the other hand, it can
be proved, with a very similar reasoning to the one used in section
\ref{subsec:reynolds-orr} for $F_{kl}$, that\textbf{ }$h_{kl}$
is well-defined as a function, even in the limit $\alpha\to0$. Thus,
we chose to turn Eq. (\ref{eq:hk}) to an integral equation. Inverting
the Laplacian operator using the Green function $H_{k}$ (see equation
(\ref{eq:linear-shear-h-k-infty})), we obtain
\begin{align}
h_{kl}(y_{1},y_{2}) & =-\frac{i}{k}\int\frac{H_{k}(y_{2},y'_{2})\mbox{e}^{il(y_{1}-y'_{2})}}{U(y_{1})-U(y'_{2})-\frac{2i\alpha}{k}}\,\mbox{d}y'_{2}\label{eq:equation-integrale}\\
 & +\int\frac{H_{k}(y_{2},y'_{2})}{U(y_{1})-U(y'_{2})-\frac{2i\alpha}{k}}\left[\left(U''(y_{1})-\beta\right)h_{kl}^{*}(y'_{2},y_{1})-\left(U''(y'_{2})-\beta\right)h_{kl}(y_{1},y'_{2})\right]\,\mbox{d}y'_{2}.\nonumber 
\end{align}

The generalization to the 2D barotropic equation upon a topography
is straightforward, replacing $\beta$ by $h'(y)$ in (\ref{eq:equation-integrale}).
The main advantage of this equation is that it involves only well-behaved
functions, even in the limit of no dissipation $\alpha\rightarrow0$.
Indeed, in this limit, the integrals converge to their Cauchy principal
values. Moreover, the fact that it doesn't involve any space derivative
will make it easy to solve numerically, as discussed in next section.

\subsubsection{Numerical implementation}

In order to numerically compute solutions of Eq. (\ref{eq:equation-integrale}),
we chose an iterative scheme. We compute the sequence $\left\{ h_{N}\right\} _{N\geq0}$
with 
\[
h_{N+1}=S+T[h_{N}],
\]
 where $S$ is the first term in the right hand side of Eq. (\ref{eq:equation-integrale})
and $T$ is the integral operator of the second term. If this sequence
converges, then $h_{kl}=\lim_{N\rightarrow\infty}h_{N}$.

We note that we have not been able to establish conditions for which
$T$ is contracting. As a consequence, the convergence of the algorithm
is not guaranteed, and we will establish the convergence empirically
on a case by case basis. More precisely, the convergence of the iterative
algorithm is checked by plotting $\log\parallel h_{N}-h_{N-1}\parallel$
as a function of the number of iterations $N$, where $\left\Vert .\right\Vert $
is the $L^{2}$ norm. 

The computation of integrals of the form $\int\frac{f_{k}(y,y')}{U(y)-U(y')-2i\alpha/k}\,\mbox{d}y'$
requires a particular attention. Indeed, the singularity of the denominator
at the points such that $U(y')=U(y)$ can be the source of important
numerical errors, and we find that the result strongly depends on
the resolution if it is not precise enough. The resolution required
to get robust results is easily understood: in order to avoid numerical
errors, the denominator must satisfy $|U(y)-U(y\pm\Delta y)|\ll\frac{2\alpha}{k}$
for a discretization step $\Delta y$. More precisely, for sufficiently
small $\Delta y$, we have $U(y)-U(y\pm\Delta y)\simeq\pm U'(y)\Delta y$,
so that the condition becomes $\Delta y\ll\frac{2\alpha}{k|U'|}$.
The numerical results confirm this scaling, for base flows with no stationary points. 
This is an important remark,
because the iterative algorithm may converge and yet give a wrong
result if the condition $\Delta y\ll\frac{2\alpha}{k|U'|}$ is not
respected.  For base flows with stationary points, the iterative algorithm often gives problems of convergence in the small $\alpha$ limit.

We have recently devised a different algorithm
that systematically converges and allows to compute stationary solution
of the Lyapunov equation for zero viscosity, with and without stationary points. This much more robust
and versatile algorithm will be presented in a following paper.

\subsubsection{Reynolds stress and stationary correlation functions}

The properties of $F$ and $g_{kl}^{\infty}$ discussed at a theoretical
level in sections \ref{sub:Equation-de-Lyapunov} and \ref{sub:Stationarity-of-the-lyapunov}
are now illustrated in two examples.

\paragraph{Parabolic profile}

We consider the case of the 2D Euler equations ($\beta=h=0$) in the
channel $(x,y)\in[0,2\pi)\times[-1,1]$ with rigid boundary conditions
$\psi_{m}(x,\pm1)=0,\;{\bf v}_{m}(x+2\pi,y)={\bf v}_{m}(x,y)$, and
with a parabolic base flow $U(y)=A(y+2)^{2}-U_{0}$, where the constants
$A$ and $U_{0}$ are chosen so that the total energy is 1 and the
total momentum is 0. This flow has no inflection point, whence its
linear stability by direct application of Rayleigh's inflection point
theorem \cite{Drazin_Reid_1981}. Moreover, this flow has no stationary
points ($y_{0}$ such that $U'(y_{0})=0$). We will consider next
the case of a cosine profile, which has stationary points. We moreover
chose here, for sake of simplicity, to force only one mode $k=l=1$.
This corresponds to the forcing correlation function $C({\bf r})=c_{11}\cos(x+y)$,
with $c_{11}=4.72$.

\begin{figure}
\includegraphics[scale=0.75]{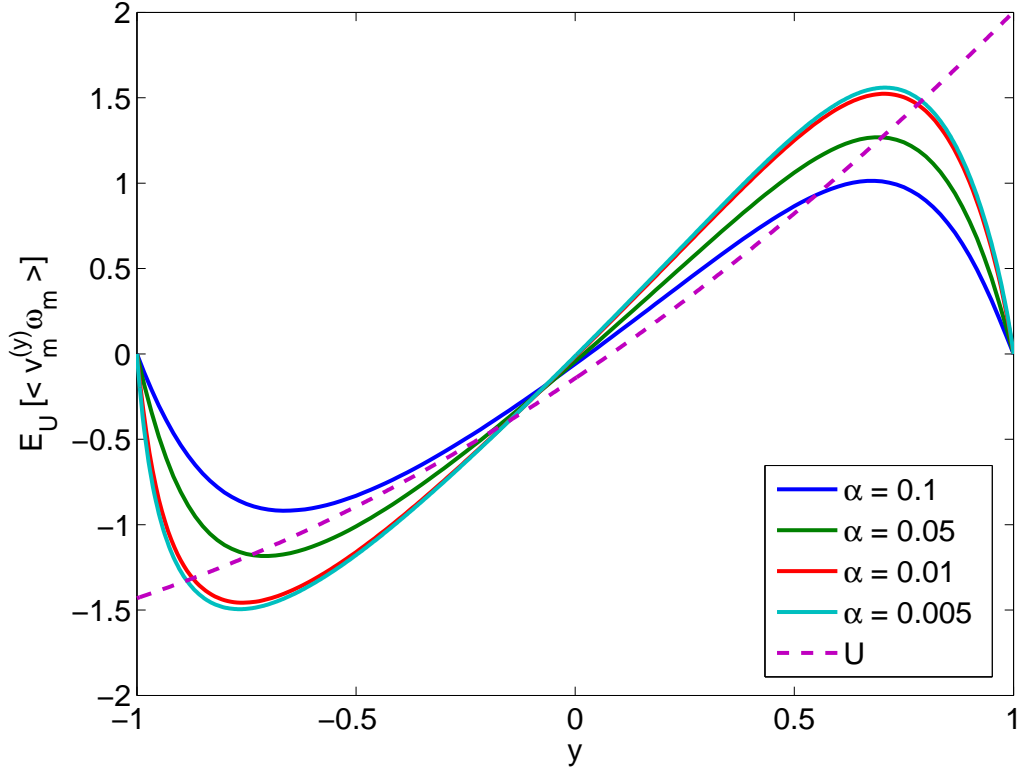}\caption{The Reynolds stress divergence $F(y)=\mathbb{E}_{U}\left[\left\langle v_{m}^{(y)}\omega_{m}\right\rangle \right]=2kc_{kl}\,\mathfrak{I}\left[h_{kl}(y,y)\right]$
in the case of a parabolic base profile in a channel geometry, with
$k=l=1$ and different values of the friction coefficient $\alpha$.
We check the convergence of $F$ to a smooth function in the inertial
limit, as theoretically predicted in section \ref{sec:Lyapunov}.\label{fig:NUM-parabolique-lima0}}
\end{figure}

We have numerically computed the solution of (\ref{eq:equation-integrale})
with the iterative scheme previously explained. As mentioned above,
the necessary resolution to use depends on the value of $\alpha$;
in the present case, it ranges from $\Delta y=1/60$ for the largest
value $\alpha=0.1$ to $\Delta y=1/300$ for the smallest one, $\alpha=0.005$. Figure \ref{fig:NUM-parabolique-lima0} shows the Reynolds
stress divergence \textbf{$F(y)=2kc_{kl}\,\mathfrak{I}\left[h_{kl}(y,y)\right]$}
and it clearly illustrates the convergence of $F$ for $\alpha\rightarrow0$,
as theoretically described in section \ref{sub:Stationarity-of-the-lyapunov}.
We also note that the Reynolds stress divergence and the base flow
profile $U$ have the same sign, except in a small region near $y=0$.
Looking at the evolution equation for $U$ (\ref{eq:Minimal-model})
\begin{equation}
\partial_{t}U=\alpha F-2\alpha U+\alpha\xi,\label{eq:kinetic-equation-h}
\end{equation}
this observation implies that the Reynolds stress is actually forcing
the zonal flow.

\begin{figure}
\includegraphics[scale=0.45]{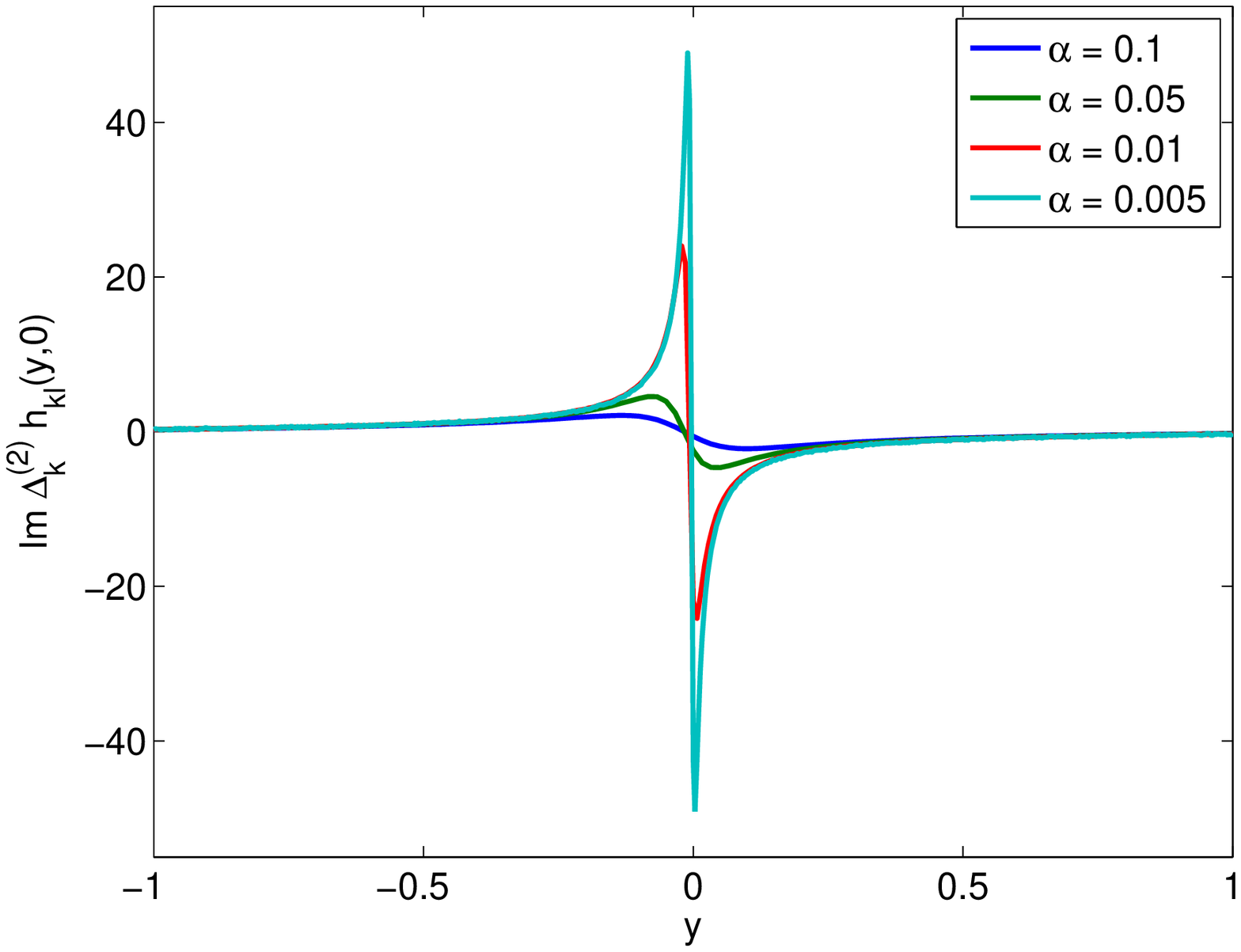}\includegraphics[scale=0.45]{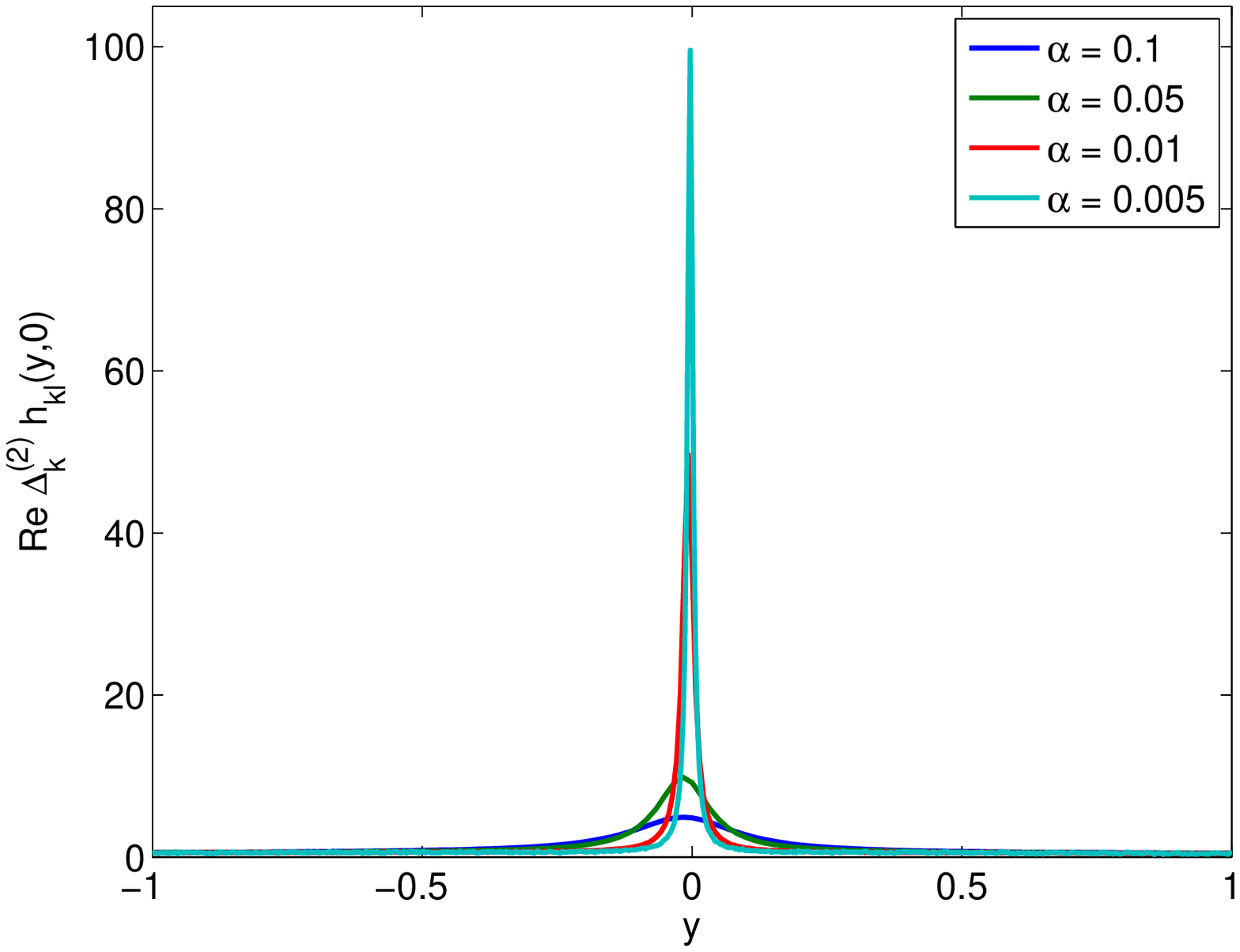}\caption{Real and imaginary parts of the $k$-th Fourier component of the stationary
vorticity auto-correlation function $\Delta_{k}^{(2)}h_{kl}(y,y_{2})|_{y_{2}=0}$
in the case of a parabolic base profile in a channel geometry, with
$k=l=1$ and different values of the friction coefficient $\alpha$.
The plots clearly show the expected divergence at $y=0.$ \label{fig:NUM-parabolique-lima0-g}}
\end{figure}

\begin{figure}
\includegraphics[scale=0.45]{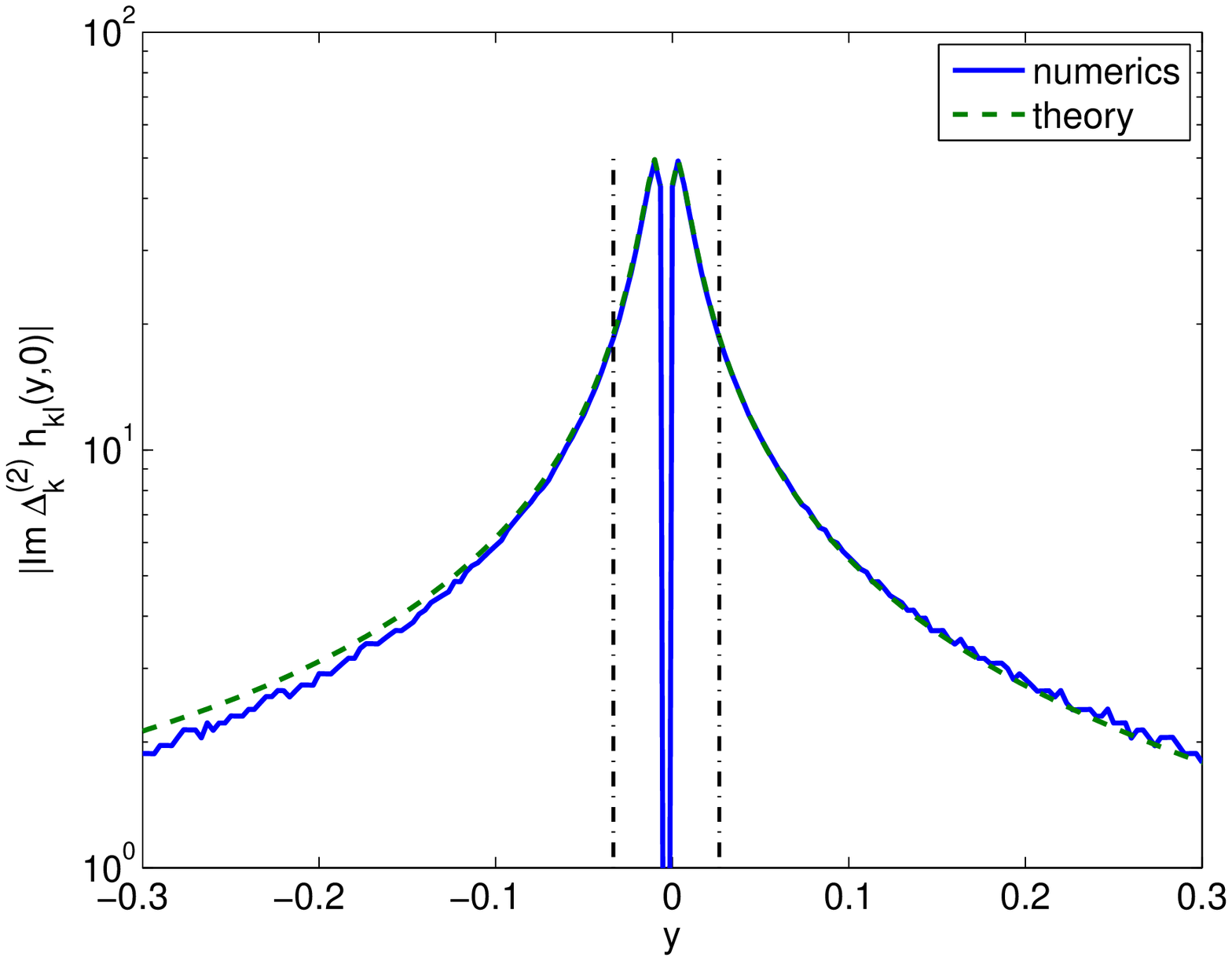}\includegraphics[scale=0.45]{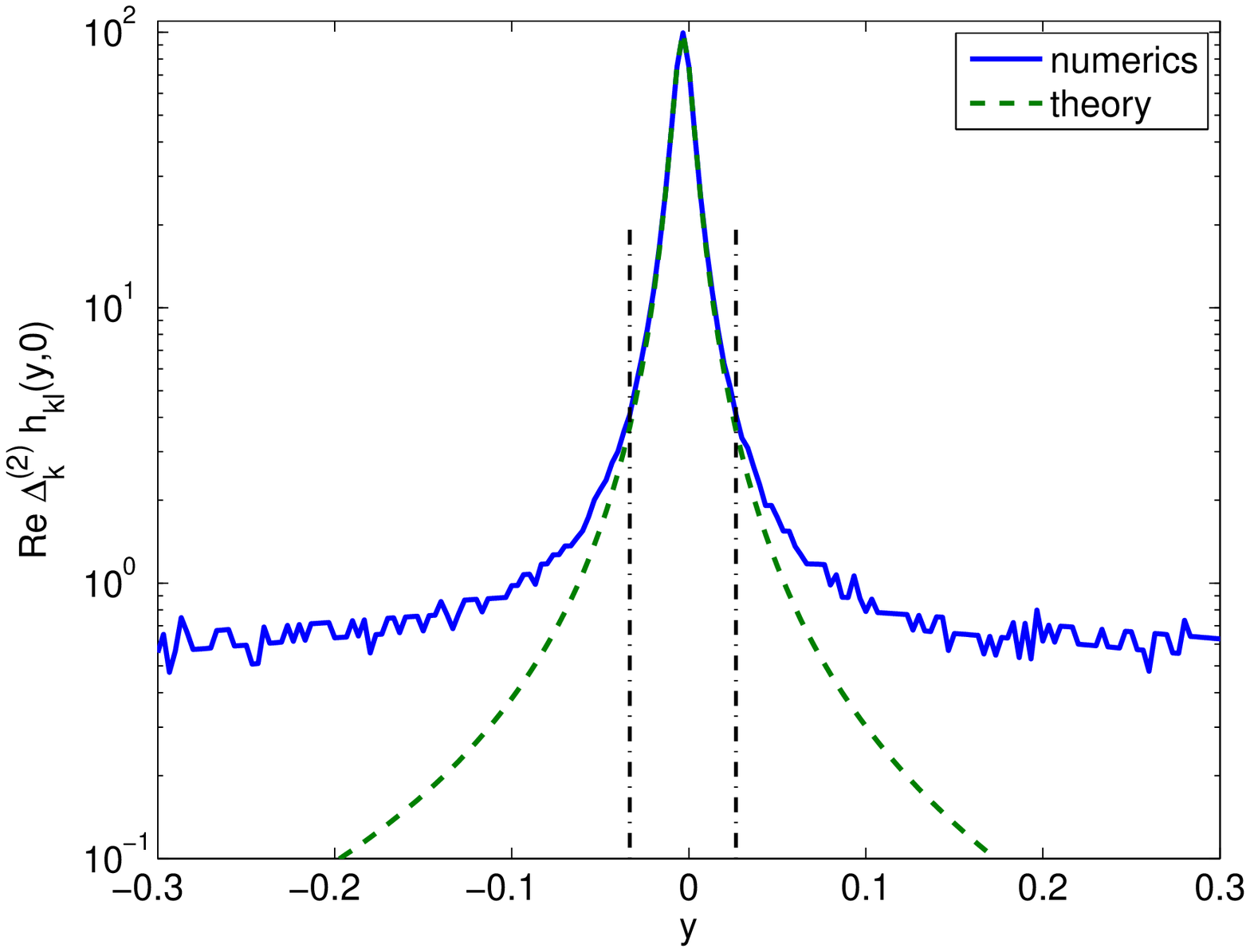}

\caption{Divergence of the stationary vorticity auto-correlation function $\Delta_{k}^{(2)}h_{kl}(y,y_{2})|_{y_{2}=0}$
near $y=0$, in the case of a parabolic base profile in a channel
geometry, with $k=l=1$ and $\alpha=0.005$. As expected, the comparison
between the numerical result and the universal shape (\ref{eq:shape-gkl})
is very good in the range $1\gg y\gg\frac{2\alpha}{ks_{0}}\simeq0.006$
(the area between the vertical lines).\label{fig:NUM-parabolique-lima0-logg}}
\end{figure}

Others correlation functions can be very easily computed from $h_{kl}$
and the the analyticity properties that we have predicted can be directly
checked. For instance, the vorticity auto-correlation function $g_{kl}^{\infty}(x_{1},x_{2},y_{1},y_{2})=\mbox{e}^{ik(x_{1}-x_{2})}\Delta_{k}^{(2)}h_{kl}(y_{1},y_{2})+\mbox{C.C.}\,.$
is represented in Figure \ref{fig:NUM-parabolique-lima0-g}. We clearly
see the divergence at the point $y_{1}=y_{2}=0$ when $\alpha\rightarrow0$.
Moreover, we recover the universal shape of $\Delta_{k}^{(2)}h_{kl}(y_{1},y_{2})$
near this divergence, as expected from equation (\ref{eq:gkl-stationary-orr-alpha}):
for $\frac{2\alpha}{ks_{0}}\ll y\ll1$, where $s_{0}=U'(0)$ is the
local shear at $y=0$, we have
\begin{equation}
\mathfrak{R}\left[\Delta_{k}^{(2)}h_{kl}(y,0)\right]\sim\frac{2\alpha}{k^{2}\left(U(y)-U(0)\right)^{2}+4\alpha^{2}}\quad,\quad\mathfrak{I}\left[\Delta_{k}^{(2)}h_{kl}(y,0)\right]\sim\frac{-k\left(U(y)-U(0)\right)}{k^{2}\left(U(y)-U(0)\right)^{2}+4\alpha^{2}}.\label{eq:shape-gkl}
\end{equation}
The comparison between this theoretical prediction and the numerical
results is shown in Figure \ref{fig:NUM-parabolique-lima0-logg}.

\paragraph{Cosine profile}

\begin{figure}
\includegraphics[scale=0.75]{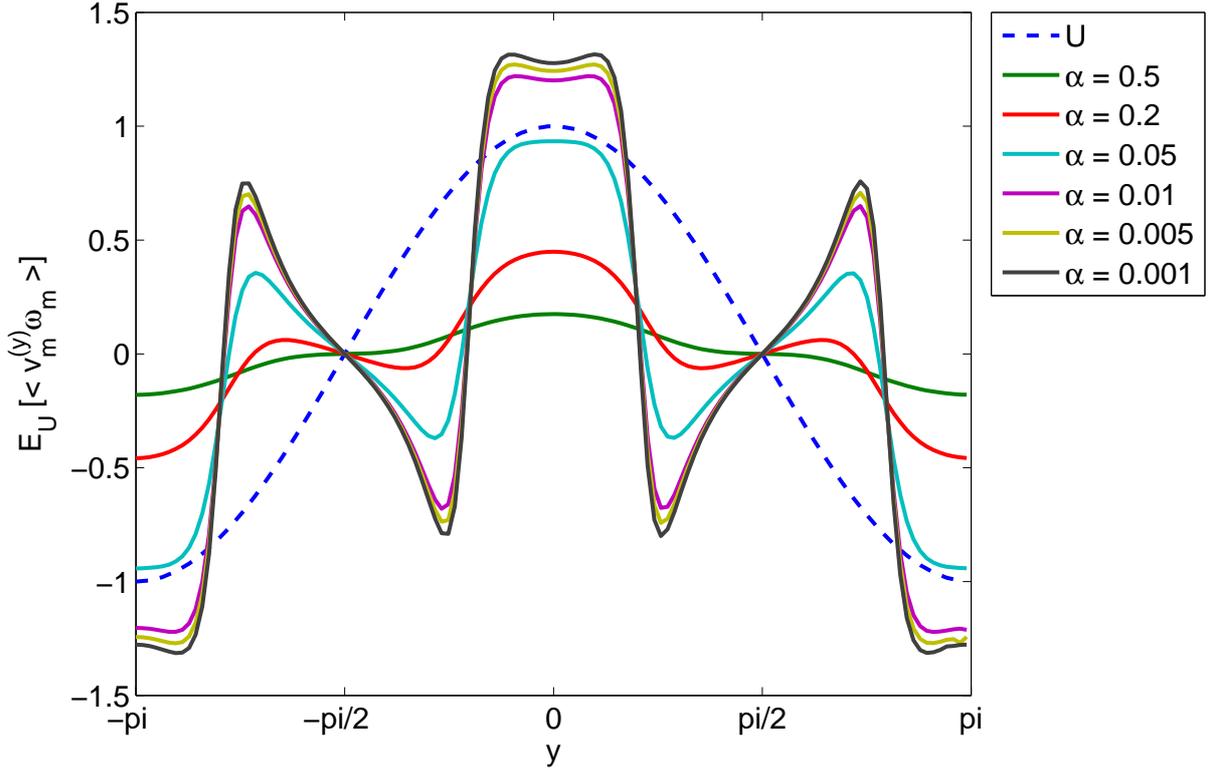}\caption{The Reynolds stress divergence $F(y)=\mathbb{E}_{U}\left[\left\langle v_{m}^{(y)}\omega_{m}\right\rangle \right]=2kc_{kl}\,\mathfrak{I}\left[h_{kl}(y,y)\right]$
in the case of a cosine base profile $U(y)=\cos y$ in a periodic
geometry, with $k=2,\, l=0$ and different values of the friction
coefficient $\alpha$. Again, we observe the convergence of $F$ towards
a smooth function when $\alpha\rightarrow0$, even at the stationary
points $y=0$ and $y=\pi$ where we do not have full theoretical predictions.\label{fig:NUM-cos-lima0-k=00003D00003D2}}
\end{figure}

The second example we consider is the zonal base flow $U(y)=\cos y$
in the domain $(x,y)=[0,2\pi l_{x})\times[-\pi,\pi)$ with periodic
boundary conditions, which is usually referred to as the Kolmogorov
flow \cite{Bouchet_Morita_2010PhyD}. This flow is stable and the
linearized operator associated to this flow has no normal modes for
aspect ratio $l_{x}<1$ \cite{Bouchet_Morita_2010PhyD}. We choose
the parameters $l_{x}=0.5,\, k=2,\, l=0$, corresponding to the forcing
correlation function $C({\bf r})=c_{20}\cos(2x)$, with $c_{20}=1.29$. The Reynolds stress divergence $F$ is
plotted in figure \ref{fig:NUM-cos-lima0-k=00003D00003D2}. It converges
to a smooth function in the inertial limit. For all points such that
$U'(y)\neq0$, this was expected from the theoretical results of section
\ref{sec:Lyapunov}. We note that we have also a convergence of $F(y)$
to a finite limit at the stationary points $y=0,\,\pi$, as discussed
at the end of section \ref{subsec:reynolds-orr}. We observe that
the Reynolds stress is forcing the flow except in some regions around
the zeros of $U$, like in the case of the parabolic zonal flow.

\subsubsection{Conservation laws}

In the channel geometry considered, the linear momentum is always
zero, $P_{x}=\int\mbox{d}y\, v_{x}=0$. This implies the constraint
$\int\mbox{d}y\, U(y)=0$. At the level of the kinetic equation (\ref{eq:kinetic-equation-h}),
this implies that the integral of the divergence of the Reynolds stress
is zero: $\int\mbox{d}y\,\mathfrak{I}\left[h_{kl}(y,y)\right]=0$,
which is also a trivial consequence of the definition of $h_{kl}$.
This constraint is fulfilled by the numerical results in Figure \ref{fig:NUM-parabolique-lima0}
within an error of order $10^{-8}$.\\

\begin{figure}
\includegraphics[scale=0.75]{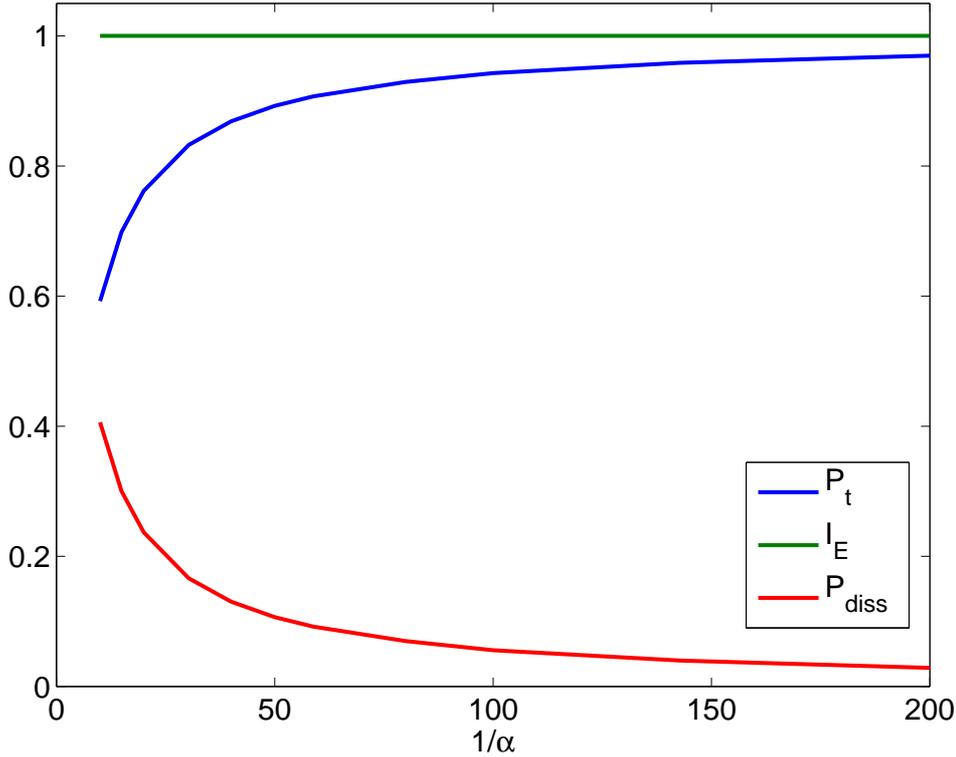}\caption{Stationary energy balance for the fluctuations, in the case of a parabolic
base profile in a channel geometry, with $k=l=1$ and different values
of the friction coefficient $\alpha$. We see that in the inertial
limit $\alpha\ll1$, all the energy injected in the fluctuations is
transferred to the zonal flow, while the energy dissipated in the
fluctuations vanishes.\label{fig:NUM-parabolique-energie}}
\end{figure}

We now discuss the energy and enstrophy balance derived in sections
\ref{sec:Energy-balance} and \ref{sec:Lyapunov}. The theoretical
results predict a vanishing of the non-zonal energy in the inertial
limit, and a divergence of the non-zonal enstrophy as $1/\alpha$.
Figure \ref{fig:NUM-parabolique-energie} shows the three rates of
energy involved in (\ref{eq:energy-balance-meridional-kinetic}) as
a function of $1/\alpha$: the power injected by the forcing (by definition
equal to $1$)
\[
\sigma_{m}=-2\pi^{2}l_{x}\left(\Delta^{-1}C_{m}\right)({\bf 0})=1\,,
\]
the power transferred to the zonal flow
\[
P_{t}=\pi l_{x}\int\mbox{d}y\,\mathbb{E}_{U}\left[\left\langle v_{m}^{(y)}\omega_{m}\right\rangle (y)\right]U(y)=\pi l_{x}\int\mbox{d}y\,2kc_{kl}\,\mathfrak{I}\left[h_{kl}(y,y)\right]U(y)\,,
\]
and the rate of energy dissipated in the non-zonal degrees of freedom
\[
P_{diss}=\alpha E_{m}=-2\alpha\int\mbox{d}y\,\mathfrak{R}\left[h_{kl}(y,y)\right]\,.
\]
It illustrates the fundamental property that in the inertial limit,
all the energy injected in the fluctuations is transferred to the
zonal flow, as expected from the discussion in section \ref{sec:Energy-balance}.
We note that the energy balance (\ref{eq:energy-balance-meridional-kinetic})
is verified in our numerical solutions up to errors of order $10^{-4}$.\\

Enstrophy has a very different behavior. From the general results
of \ref{sec:Lyapunov}, we know that the dissipated enstrophy is not
expected to go to zero as $\alpha\rightarrow0$, so the transferred
enstrophy is not expected to be close to the injection rate. In the
case of a parabolic profile $U(y)=A(y+2)^{2}-U_{0}$, this property
is even more dramatic as the transferred enstrophy vanishes, it is
indeed

\[
Z_{t}=-\pi l_{x}\mathbb{E}\left[\int\mbox{d}y\, U''(y)\left\langle v_{m}^{(y)}\omega_{m}\right\rangle (y)\right]=-\pi l_{x}\int\mbox{d}y\,2A\cdot2kc_{kl}\,\mathfrak{I}\left[h_{kl}(y,y)\right]=0
\]
which is zero because it is proportional to the conservation of the
momentum. Whatever the value of $\alpha$, no enstrophy is transferred
to the zonal flow and all the enstrophy injected by the stochastic
forcing is dissipated in the fluctuations. When it comes to the numerical
results, checking that the enstrophy balance is fulfilled amounts
at checking the conservation of the momentum $\int\mbox{d}y\,\mathfrak{I}\left[h_{kl}(y,y)\right]=0$,
which has already been discussed.

\section{Attractors of the slow dynamics and multistability of atmosphere
jets}

\label{sec:Bistability}

In section \ref{sec:Stochastic-Averaging}, we have obtained that
the jet has a slow evolution whose dynamics is described by equation
(\ref{eq:Minimal-model}), reported here for convenience: 
\begin{equation}
\frac{\partial U}{\partial\tau}=F[U]-U+\frac{\nu}{\alpha}\frac{\partial^{2}U}{\partial y^2}+\xi,\label{eq:Minimal-model-1}
\end{equation}
where $\xi$ is Gaussian process with autocorrelations given by (\ref{eq:minimal-model-correlation-function}).
In most known situations, there is no noise acting on the largest
scales. Then the correlation function of $\xi$, see Eq. (\ref{eq:minimal-model-correlation-function}),
is of an order $\alpha$ smaller than $F\left[U\right]$. In the limit
we consider, we thus expect the noise to be small. Then one may wonder
how important is its effect for practical purposes. We discuss this
issue here, with emphasis on the case where the deterministic equation
$\frac{\partial U}{\partial\tau}=F[U]-U+\nu/\alpha \partial_y^2 U $ has more than one attractor.

First it is useful to note that the deterministic equation $\frac{\partial U}{\partial\tau}=F[U]-U+\nu/\alpha \partial_y^2 U$
is unable to describe the statistics of the small fluctuations close
to a an attractor $U_{0}$. A first very interesting result that can
be derived from (\ref{eq:Minimal-model-1}) is the statistics for
the Gaussian fluctuations of the jet close to its most probable value.

If we now assume that the deterministic equation $\frac{\partial U}{\partial\tau}=F[U]-U+\nu/\alpha \partial_y^2 U$
has more than one attractor, for instance $U_{0}$ and $U_{1}$, the
small noise $\xi$ becomes essential in order to describe the relative
probability of the two attractors and the transition probabilities
between them. These can be computed numerically from (\ref{eq:Minimal-model-1})
by means of large deviation theory or, equivalently using a path integral
representations of the stochastic process (\ref{eq:Minimal-model-1})
and instanton theory. One may wonder if such bistability or multistability
situations actually exist. 

\begin{figure}
\includegraphics[scale=0.6]{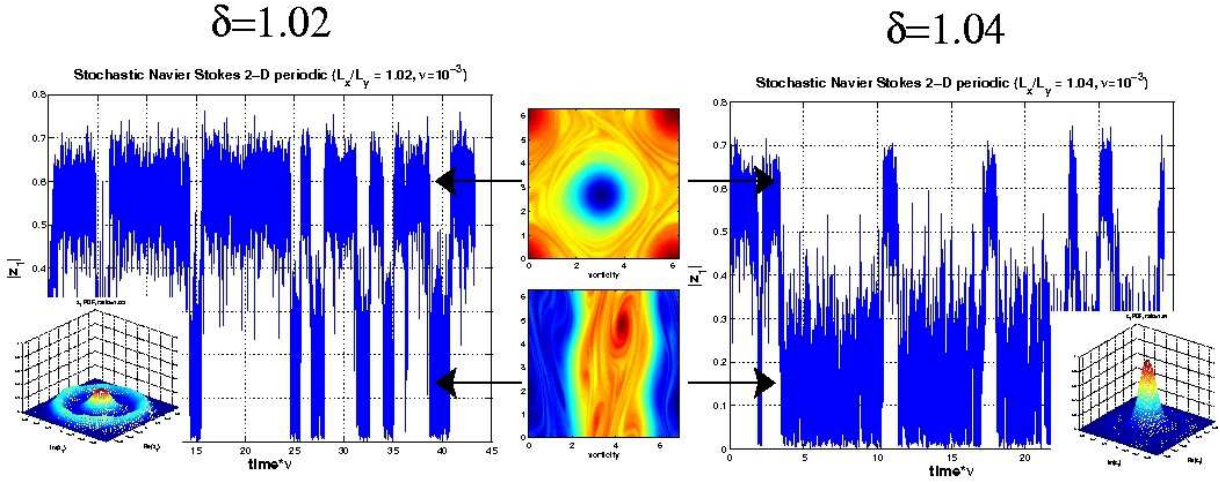}\caption{\label{fig:Transition-Stationnaire}Figure taken from \cite{Bouchet_Simonnet_2008}
showing rare transitions (illustrated by the Fourier component of
the largest $y$ mode) between two large scale attractors of the periodic
2D Navier-Stokes equations. The system spends the majority of its
time close to the vortex dipole and zonal flow configurations.}
\end{figure}

Multistability and phase transitions are actually very common for
turbulent flows (for instance, paths of the Kuroshio current \cite{Schmeits2001},
atmospheric flows \cite{Weeks_Tian_etc_Swinney_Ghil_Science_1997},
Earth's magnetic field reversal and MHD experiments \cite{Berhanu2007},
two--dimensional turbulence simulations and experiments \cite{Sommeria1986,Bouchet_Simonnet_2008,Maassen2003,loxley2013bistability},
and three--dimensional flows \cite{Ravelet_Marie_Chiffaudel_Daviaud_PRL2004}
show this kind of behavior). Figure \ref{fig:Transition-Stationnaire}
shows random transitions in the 2D Navier-Stokes equations \cite{Bouchet_Simonnet_2008}.
These transitions may in principle be described in the kinetic approach
developed in the present paper; however, this would require an extension
of the work to cases where non zonal attractors are also considered.
This is in principle possible, but the Lyapunov equation is then more
tricky. We postpone such an issue for future works.

\begin{figure}
\includegraphics[scale=0.75]{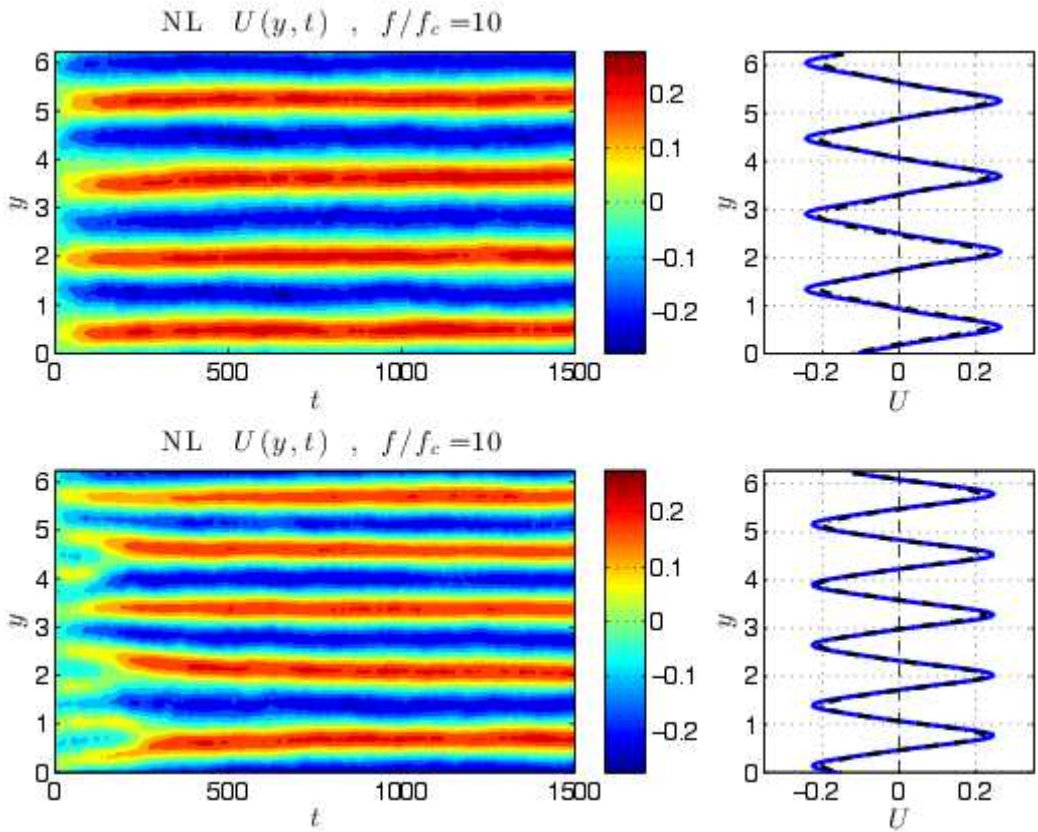}\caption{Evolution of the zonally averaged velocity profile as obtained from
non-linear simulations (left) and comparison of the stationary velocity
profile obtained from non-linear simulations and from the deterministic
part of the kinetic equation (\ref{eq:Minimal-model-1}), or equivalently
SSST \cite{Farrell_Ioannou_JAS_2007,constantinou2012emergence,Farrel_Ioannou}.
The upper and lower pictures are obtained for the same values of the
physical parameters but with different initial conditions. The figure
shows that for a given set of parameters it can converge towards two
attractors, $U_{0}$ and $U_{1}$. Courtesy N. Constantinou.\label{fig:Petros}}
\end{figure}

Remaining in the zonal framework assumed in this paper, we refer to
works by N. Bakas, N. Constantinou, B. Farrell, and P. Ioannou \cite{Farrel_Ioannou,Farrell_Ioannou_JAS_2007,constantinou2012emergence}
for existence of bistability. In those three works the possibility
for multiple attractors for SSST dynamics (the deterministic part
of Eq. (\ref{eq:Minimal-model-1})) is discussed. Figure \ref{fig:Petros},
shows such a case, where the evolution of SSST dynamics has been computed
for some fixed parameters and fixed force spectrum. The two spatiotemporal
diagrams of the zonally averaged velocity profile $U$ obtained from
the barotropic equations, and the two asymptotic profiles $U$ obtained
both from SSST dynamics and the barotropic equations show that for
this set of parameters the deterministic part of the dynamics has
at least two attractors, and that these attractors are observed in
the barotropic equations. In order to compute the relative probability
of the attractors and the transition probabilities from one to the
other, one can rely on the Fokker-Planck equation (\ref{eq:Fokker-Planck-Zonal})
or on the stochastic dynamics (\ref{eq:Minimal-model-1}). These transitions
involve states that are arbitrarily far from the attractors $U_{0}$
and $U_{1}$. 

\section{Conclusion}

\label{sec:Conclusion}

In this paper, we have considered the kinetic theory of the equations
for a barotropic flow with beta effect or topography and with stochastic
forces, in the limit of weak forces and dissipation $\alpha\rightarrow0$.
We have formally computed the leading order dynamics, using the framework
of stochastic averaging for this adiabatic process. At leading order,
we have obtained a Fokker-Planck equation, or equivalently a stochastic
dynamics, that describes the slow jet dynamics (\ref{eq:Minimal-model}).
We have discussed that, at leading order, the deterministic part of
this dynamics are the equations known as Stochastic Structural Stability
Theory or, equivalently what can be obtained from a second order closure
theory. This provides a strong support for those equations if one
wants to compute the attractors for the jet dynamics, when $\alpha$
is sufficiently small. This result also shows that those models do
not describe completely Gaussian fluctuations close to these attractors.
Moreover, in phenomena for which large deviations beyond Gaussian
become essential, a kinetic approach is still relevant, and the full
kinetic equation (\ref{eq:Minimal-model}) should be considered. 

We have also proven that at leading order, quasilinear dynamics will
lead to the same kinetic equation (\ref{eq:Minimal-model}) in the
limit of weak forces and dissipation. We stress however that this
will be wrong at next order. 

Our main hypothesis is that the zonal jets with profile $U(y)$ are
stable and have no non-zonal neutral modes. Whereas the stability
is a crucial hypothesis, the non-neutral mode hypothesis is in principle
not essential. However, if this hypothesis would not be fulfilled,
a different kinetic equation would be obtained. 

For the kinetic theory to be meaningful, the base flows have to be
attractors of the inertial dynamics. This is actually the case for
some of zonal jets. However non-zonal attractors also exist, for instance
dipoles for the 2D Euler equations \cite{Chertkov_Connaughton_andco_2007_PRL_EnergyCondesation,Bouchet_Simonnet_2008}.
The kinetic theory could be extended in principle to non-zonal attractors.
It would be easy to write phenomenologically the kinetic equation,
just as was done for zonal jets. We stress however that proving the
self-consistency of the theory, which is the key point from a theoretical
point of view, would be much more tricky. This requires the analysis
of the asymptotic solutions of the Lyapunov equation, which would
be technically more difficult for non-zonal attractors than for zonal
jets.

We stress that we have proven the consistency of the kinetic theory only for the case of the two-dimensional Navier-Stokes equations. With a beta effect or a topography, the formal structure of the problem will be very similar, so that the formal generalization of the kinetic theory is straightforward. However no theoretical results for the asymptotic decay of the linearized equation are yet available in the literature. As a consequence we cannot draw, for the moment, any conclusion about the convergence of the velocity-vorticity correlation functions and of the Reynolds stress without dissipation in these cases.\\

A number of interesting issues have not been addressed in this work
and should be considered in the future. Whereas the limit $\alpha\rightarrow0$
is the appropriate one in order to set up a clear theoretical framework,
an essentially question arises from a practical point of view: how
small $\alpha$ should be? How does the answer depend on the force
spectrum? These issues could certainly be addressed both numerically
and theoretically. 

We believe that our theoretical approach is the correct way to describe
the statistics of the largest scale of the flow. However, for small
but finite $\alpha$, especially if the forcing has a narrow band
spectrum, it is likely that non-linear effects would become important.
In some regimes, we expect to see together with a quasi-linear dynamics
at the largest scales, either enstrophy cascade, or inverse energy
cascade, or both, at scales much smaller than a typical jet scale.
Those are described by Kraichnan type cascades for the 2D-Navier-Stokes
equations \cite{boffetta2012two}, or by zonostrophic turbulence if
$\beta\neq0$ \cite{galperin2001universal,galperin2010geophysical}.
Those nonlinear effects, basically small scale turbulent mixing, would then be the correct mechanism that would regularize the Lyapunov equations. They would also be responsible for the loss of information about the initial condition, discussed in section \ref{Adiabatic-elimination}.
In such a case, how would our large scale theory relate with smaller
scale cascade pictures for finite $\alpha$? What would be the crossover
scale and how should it depend on $\alpha$? While answer using dimensional
analysis can be given easily, a more detailed theoretical study would
be very interesting. Another related issue is the effects of the small
scale part of the force spectrum. All over the paper, we have assumed
that the spatial correlation $C$ is smooth and that its spectrum
converges sufficiently fast for large wave numbers $\left|\mathbf{k}\right|$,
in order to insure the convergence of the sum of the contributions
of all sectors $k$ to the Lyapunov equation. Could the required hypothesis
be made more explicit?

We have made explicit computations up to leading order only. It would
however be extremely interesting to compute next order corrections.
This may be important in order to have very precise predictions for
small but finite $\alpha$ and also to give a clear indication of
the order of magnitude of $\alpha$ required for the theory to be
trusted. Undoubtedly the answer will depend much on the force spectrum.
We stress that the next order corrections will be different from an
approach using a cumulant expansion and that it should not be hindered
with the usual inconsistency observed in cumulant expansions.

Another important issue is the validity of our approach for more complex
dynamics, for instance for the class of quasi-geostrophic dynamics.
At a formal level, the same approach can be applied. However, the
dynamics is expected to be quite different. For instance, whereas
the fact that there is no non-zonal neutral modes is probably generic
for barotropic flows, it is probably exceptional for equivalent barotropic
quasi-geostrophic dynamics as soon as the Rossby deformation radius
is smaller than the domain size. Also if one considers a deterministic
force in addition to the stochastic one, which is essential for instance
for Earth jet dynamics, then the same formalism can be applied, but
the phenomenology is expected to be quite different. This calls for
new specific studies.

\begin{acknowledgements}
This research has been supported through the ANR program STATOCEAN (ANR-09-SYSC-014) (F. Bouchet), the ANR STOSYMAP (ANR-2011-BS01-015) (F. Bouchet and C. Nardini), the ANR LORIS (ANR-10-CEXC-010-01) (C. Nardini) and the program PEPS-PTI from CNRS. This work has been partially supported by the National Science Foundation under Grant No. PHYS-1066293 and by International Space Science Institute (ISSI). F. Bouchet acknowledges the hospitality of the Aspen Center for Physics. Numerical results have obtained with the PSMN platform in ENS-Lyon.
\end{acknowledgements}

\appendix

\section{Explicit computation of the operators in the zonal Fokker-Planck
equation\label{sec:Explicit-computation-zonal-FP}}

We compute here each term of the equation for the slowly varying part
of the PDF (\ref{eq:Evolution-Ps-1}), which we report:
\begin{equation}
\frac{\partial P_{s}}{\partial t}=\left\{ \alpha\mathcal{P}\mathcal{L}_{z}+\alpha^{3/2}\mathcal{P}\mathcal{L}_{z}\int_{0}^{\infty}\mbox{d}t'\,\mbox{e}^{t'\mathcal{L}_{0}}\mathcal{L}_{n}+\alpha^{2}\mathcal{P}\mathcal{L}_{z}\int_{0}^{\infty}\mbox{d}t'\,\mbox{e}^{t'\mathcal{L}_{0}}\left[(1-\mathcal{P})\mathcal{L}_{z}+\int_{0}^{\infty}\mbox{d}t''\,\mathcal{L}_{n}\mbox{e}^{t''\mathcal{L}_{0}}\mathcal{L}_{n}\right]\right\} P_{s}+\mathcal{O}\left(\alpha^{5/2}\right).\label{eq:Evolution-Ps-1-appendix}
\end{equation}
 This will lead to the final reduced Fokker-Planck equation for the
zonal flow PDF $R$, (\ref{eq:Fokker-Planck-Zonal}). 
\begin{itemize}
\item The first term in equation (\ref{eq:Evolution-Ps-1-appendix}) gives
\begin{align}
 & \mathcal{P}\mathcal{L}_{z}P_{s}=\label{eq:Terme-Deterministe}\\
 & \qquad G\left[q_{z},\omega_{m}\right]\int\mbox{d}y_{1}\,\frac{\delta}{\delta q_{z}(y_{1})}\left[\left(\frac{\partial}{\partial y}\mathbb{E}_{U}\left[\left\langle v_{m}^{(y)}\omega_{m}\right\rangle \right](y_{1})+\omega_{z}(y_{1})-\frac{\nu}{\alpha}\Delta\omega_{z}(y_{1})\right)R+\int\mbox{d}y_{2}\, C_{z}(y_{1},y_{2})\frac{\delta R}{\delta q_{z}(y_{2})}\right].\nonumber 
\end{align}

\item The next term is exactly zero: we have 
\[
\mathcal{L}_{n}P_{s}=\int\mbox{d}\mathbf{r}_{1}\,\frac{\delta}{\delta\omega_{m}(\mathbf{r}_{1})}\left[\left(\mbox{\ensuremath{\mathbf{v}}}_{m}.\nabla\omega_{m}(\mathbf{\mathbf{r}}_{1})-\langle\mbox{\ensuremath{\mathbf{v}}}_{m}.\nabla\omega_{m}(\mathbf{\mathbf{r}}_{1})\rangle\right)G\left[q_{z},\omega_{m}\right]R\left[q_{z}\right]\right].
\]
We remark that the action of the functional derivative $\frac{\delta}{\delta\omega_{m}(\mathbf{r}_{1})}$
will produce factors which are either linear or cubic in the variable
$\omega_{m}$. When applying $\mathcal{L}_{z}$, those terms will
be multiplied by constant or quadratic factors. The projection $\mathcal{P}$
will then lead to the computation of odd moments of the centered Gaussian
distribution $G$, which are all null.
As a consequence, we conclude that 
\begin{equation}
\mathcal{P}\mathcal{L}_{z}\mbox{e}^{t'\mathcal{L}_{0}}\mathcal{L}_{n}P_{s}=0.\label{eq:Diffusion-Non-Lineaire-Nulle}
\end{equation}
This is a very important result: at first-order, the corrections to
the quasi-linear dynamics due to the full nonlinearity exactly vanish.
This is a partial explanation to the success of the quasi-linear dynamics:
the non-linear interactions between the non-zonal degrees of freedom
(eddy-eddy interaction) don't have any influence on the statistics
of the zonal flow until the order $\alpha^{2}$. 
\item Terms of order $O(\alpha^{2})$ in equation (\ref{eq:Evolution-Ps-1-appendix})
give non-trivial contributions. We write
\[
\mbox{e}^{t'\mathcal{L}_{0}}\mathcal{L}_{n}\mbox{e}^{t''\mathcal{L}_{0}}\mathcal{L}_{n}P_{s}=M[q_{z},\omega_{m}](t',t'')G[q_{z},\omega_{m}]R[q_{z}].
\]
Then,
\begin{align*}
\mathcal{P}\mathcal{L}_{z}\mbox{e}^{t'\mathcal{L}_{0}}\mathcal{L}_{n}\mbox{e}^{t''\mathcal{L}_{0}}\mathcal{L}_{n}P_{s}& =G[q_{z},\omega_{m}]\text{\ensuremath{\int}}\mbox{d}y_{1}\frac{\delta}{\delta q_{z}(y_{1})}\left[\frac{\partial}{\partial y_{1}}\mathbb{E}_{U}\left[\langle v_{m}^{(y)}\omega_{m}\rangle(y_{1})M[q_{z},\omega_{m}](t',t'')\right]R[q_{z}]\right]\\
 & +G[q_{z},\omega_{m}]\text{\ensuremath{\int}}\mbox{d}y_{1}\frac{\delta}{\delta q_{z}(y_{1})}\left[\left(\omega_{z}(y_{1})-\frac{\nu}{\alpha}\Delta \omega_{z}(y_{1})\right)\mathbb{E}_{U}\left[M[q_{z},\omega_{m}](t',t'')\right]R[q_{z}]\right]\\
 & +G[q_{z},\omega_{m}]\int\mbox{d}y_{1}\mbox{d}y_{2}\, C_{z}(y_{1},y_{2})\frac{\delta^2}{\delta q_{z}(y_{1})\delta q_{z}(y_{2})}\left[\mathbb{E}_{U}\left[M[q_{z},\omega_{m}](t',t'')\right]R[q_{z}]\right].
\end{align*}

We first consider the average value of $M$. As $\mathbb{E}_{U}$
is an average over the stationnary measure $G$, we have
\[
\mathbb{E}_{U}\left[M[q_{z},\omega_{m}](t',t'')\right]=\mathbb{E}_{U}\left[M_{1}[q_{z},\omega_{m}](t'-t'')\right]
\]
with $M_{1}$ defined by
\[
\mathcal{L}_{n}\mbox{e}^{t'-t''\mathcal{L}_{0}}\mathcal{L}_{n}G[q_{z},\omega_{m}]=M_{1}[q_{z},\omega_{m}](t'-t'')G[q_{z},\omega_{m}].
\]
Using the expression of $\mathcal{L}_{n}$ as a functional derivative
with respect to $\omega_{m}$ (Eq. (\ref{eq:Ln})), we then easily show that $\mathbb{E}_{U}\left[M[q_{z},\omega_{m}](t',t'')\right]=0$, so that the interesting quantity simplifies to
\begin{align*}
\mathcal{P}\mathcal{L}_{z}\mbox{e}^{t'\mathcal{L}_{0}}\mathcal{L}_{n}\mbox{e}^{t''\mathcal{L}_{0}}\mathcal{L}_{n}P_{s} & =G[q_{z},\omega_{m}]\text{\ensuremath{\int}}\mbox{d}y_{1}\frac{\delta}{\delta q_{z}(y_{1})}\left(\frac{\partial}{\partial y_{1}}\mathbb{E}_{U}\left[\langle v_{m}^{(y)}\omega_{m}\rangle(y_{1})M[q_{z},\omega_{m}](t',t'')\right]R[q_{z}]\right).
\end{align*}
\item In the last term, we have
\[
\left(1-\mathcal{P}\right)\mathcal{L}_{z}P_{s}=G\left[q_{z},\omega_{m}\right]\int\mbox{d}y_{1}\,\frac{\delta}{\delta q_{z}(y_{1})}\left[\frac{\partial}{\partial y}\left(\left\langle v_{m}^{(y)}\omega_{m}\right\rangle -\mathbb{E}_{U}\left[\left\langle v_{m}^{(y)}\omega_{m}\right\rangle \right]\right)(y_{1})R\right],
\]
 thus
\begin{align*}
 & \mathcal{L}_{z}\mbox{e}^{t'\mathcal{L}_{0}}\left(1-\mathcal{P}\right)\mathcal{L}_{z}P_{s}=\\
 & \qquad\int\mbox{d}y_{2}\,\frac{\delta}{\delta q_{z}(y_{2})}\left\{ \left(\frac{\partial}{\partial y_{2}}\left\langle v_{m}^{(y)}\omega_{m}\right\rangle (y_{2})+\omega_{z}(y_{2})-\frac{\nu}{\alpha}\Delta\omega_{z}(y_{2})+\int\mbox{d}y_{3}\, C_{z}(y_{2},y_{3})\frac{\delta}{\delta q_{z}(y_{3})}\right)...\right.\\
 & \left.\qquad\mbox{e}^{t'\mathcal{L}_{0}}\left[G\left[q_{z},\omega_{m}\right]\int\mbox{d}y_{1}\,\frac{\delta}{\delta q_{z}(y_{1})}\left[\frac{\partial}{\partial y_{1}}\left(\left\langle v_{m}^{(y)}\omega_{m}\right\rangle -\mathbb{E}_{U}\left[\left\langle v_{m}^{(y)}\omega_{m}\right\rangle \right]\right)(y_{1})R\right]\right]\right\} .
\end{align*}
 Using the definition of the correlation and covariance (\ref{eq:Correlation})
and (\ref{eq:Covariance}) we conclude that
\begin{align}
 & \mathcal{P}\mathcal{L}_{z}\mbox{e}^{t'\mathcal{L}_{0}}\left(1-\mathcal{P}\right)\mathcal{L}_{z}P_{s}=\label{eq:Calcul-Diffusion}\\
 & \qquad G\left[q_{z},\omega_{m}\right]\int\mbox{d}y_{2}\mbox{d}y_{1}\,\frac{\delta^{2}}{\delta q_{z}(y_{1})\delta q_{z}(y_{2})}\frac{\partial^{2}}{\partial y_{2}\partial y_{1}}\left\{ \mathbb{E}_{U}\left[\left[\left\langle v_{m}^{(y)}\omega_{m}\right\rangle (y_{1},t')\left\langle v_{m}^{(y)}\omega_{m}\right\rangle (y_{2},0)\right]\right]R\left[q_{z}\right]\right\} .\nonumber 
\end{align}

\end{itemize}

\section{Linear friction and viscous regularization of Sokhotskyi--Plemelj
formula\label{sec:appendix-regularitazions}}

The Sokhotskyi--Plemelj formula adapted to our notations is

\begin{equation}
\int_{0}^{\infty}e^{-ity}\,\mbox{d}t=\lim_{\alpha\rightarrow0^{+}}\frac{-i}{y-i\alpha}=\pi\delta\left(y\right)-iPV\left(\frac{1}{y}\right).\label{eq:Plemelj}
\end{equation}
In this Appendix, we consider different regularizations of the above
equation that correspond to a viscous regularization or to the combination
of the regularizations given by viscosity and by linear friction.
We thus deduce few useful technical results. 

In the following, we have to deal with the large $y$ behavior of
oscillatory integrals of the form

\begin{equation}
\int_{0}^{\infty}e^{-iyt}f(t)\,\mbox{d}t\,.
\end{equation}
In this respect, we will use the well known result that 

\begin{equation}
\mathfrak{R}\left[\int_{0}^{\infty}e^{-iyt}f(t)\,\mbox{d}t\right]=\sum_{n=1}^{N}\frac{f^{(2n-1)}(0)}{i^{2n}y^{2n}}+\textrm{O}\left(\frac{1}{y^{2N+2}}\right)\label{eq:appendix-regularizations-oscillatory-1}
\end{equation}
and

\begin{equation}
\mathfrak{I}\left[\int_{0}^{\infty}e^{-iyt}f(t)\,\mbox{d}t\right]=-\sum_{n=0}^{N}\frac{f^{(2n)}(0)}{i^{2n}y^{2n+1}}+\textrm{O}\left(\frac{1}{y^{2N+3}}\right)\,.\label{eq:appendix-regularizations-oscillatory-2}
\end{equation}
The above formula holds for any function $f$ which is $(2N+1)$ times
differentiable, such that $\underset{t\to\infty}{\lim}\left[f^{(n)}(t)\right]=0$
for any $n\leq2N$ and such that $\int_{0}^{\infty}\textrm{d}te^{iyt}f^{(n)}(t)$
is finite for any $n\leq2N+1$. They can be found by successive part
integrations.

\paragraph{Viscous regularization.}

We consider the regularization of Plemelj formula that correspond
to the effect of viscosity. We consider

\begin{equation}
H_{\lambda}(y)=\frac{1}{\lambda}\int_{0}^{\infty}\, e^{-i\frac{y}{\lambda}u-\lambda^{2}(k^{2}+l^{2})u+\lambda lu^{2}-\frac{1}{3}u^{3}}\,\mathrm{d}u\,.\label{eq:appendix-regularizations-h-lambda}
\end{equation}
We note that the real part of $H_{\lambda}$ is an even function and
the imaginary part of $H_{\lambda}$ is an odd function.

\noindent We can also compute the integral of $H_{\lambda}$ using
the fact that

\begin{equation}
\int_{-\infty}^{\infty}H_{\lambda}(y)\textrm{d}y=\frac{1}{2\lambda}\int_{-\infty}^{\infty}\mathrm{d}y'\int_{-\infty}^{\infty}\, e^{-i\frac{y'}{\lambda}t-\lambda^{2}(k^{2}+l^{2})t+\lambda lt^{2}-\frac{1}{3}t^{3}}\,\mathrm{d}t\,,
\end{equation}
which can be proved by integrating Eq. (\ref{eq:appendix-regularizations-h-lambda})
and with the two changes of variable $y'=-y$ and $t=-u$. Using
in the last expression the integral representation of the Dirac delta,
we obtain the desired result

\begin{equation}
\int_{-\infty}^{+\infty}H_{\lambda}(y)\,\mbox{d}y=\pi.
\end{equation}

We want now to prove that $H_{\lambda}(y)\underset{\lambda\to0}{=}G_{\lambda}(y)+\mathcal{O}(\lambda)$
for every value of $y$. Let us consider $\tilde{H}_{\lambda}(Y)=H_{\lambda}(\lambda Y)$ and
study the large $Y$ asymptotic behavior of $\tilde{H}_{\lambda}$.
Supposing small viscosity, $\lambda l\ll1$ and $\lambda^{2}(k^{2}+l^{2})\ll1$,
and by applying Eq. (\ref{eq:appendix-regularizations-oscillatory-1})
and (\ref{eq:appendix-regularizations-oscillatory-2}) to $\tilde{H}_{\lambda}$,
we obtain for the real part

\begin{equation}
\Re\left[\tilde{H}_{\lambda}(Y)\right]\underset{1\ll Y\ll\frac{1}{\lambda}\sqrt{\frac{2}{k{}^{2}+l{}^{2}}}}{\sim}-\frac{2}{\lambda Y^{4}}\,\qquad\textrm{and}\qquad\Re\left[\tilde{H}_{\lambda}(Y)\right]\underset{Y\gg\frac{1}{\lambda}\sqrt{\frac{2}{k{}^{2}+l{}^{2}}}}{\sim}\frac{\lambda\left(k{}^{2}+l{}^{2}\right)}{Y^{2}}\,,
\end{equation}
while the behavior of the imaginary part is 

\begin{equation}
\mathbb{\Im}\left[\tilde{H}_{\lambda}(Y)\right]\,\underset{Y\gg1}{\sim}-\frac{1}{\lambda Y}.
\end{equation}
Observe that the crossover in the asymptotic behavior of $\Re\left[\tilde{H}_{\lambda}(Y)\right]$
happens for $Y\gg1$, so that the asymptotic expansion performed is
justified.

By performing the change of variables $Y=y/\lambda$, we thus obtain
the asymptotic behavior of $H_{\lambda}$

\begin{equation}
\Re\left[H_{\lambda}(y)\right]\underset{\lambda\ll y\ll\sqrt{\frac{2}{k{}^{2}+l{}^{2}}}}{\sim}-\frac{2\lambda^{3}}{y^{4}}\qquad\textrm{and }\qquad\Re\left[H_{\lambda}(y)\right]\underset{y\gg\sqrt{\frac{2}{k{}^{2}+l{}^{2}}}}{\sim}\frac{\lambda^{3}\left(k{}^{2}+l{}^{2}\right)}{y^{2}}\label{eq:appendix-h-lambda-asymptotic}
\end{equation}
for the real part, and 

\begin{equation}
\mathbb{\Im}\left[H_{\lambda}(y)\right]\,\underset{y\gg\lambda}{\sim}-\frac{1}{y}\,
\end{equation}
for the imaginary part.

A very similar reasoning applied to the function $G_{\lambda}$ defined
in Eq. (\ref{eq:viscous-advection-Glambda}) leads us to the conclusion
that 

\begin{equation}
\Re\left[G_{\lambda}(y)\right]\underset{y\gg\lambda}{\sim}-\frac{2\lambda^{3}}{y^{4}}\,,
\end{equation}
while the leading order behavior of the imaginary part remains the
one found for $H_{\lambda}$. We can thus conclude that $H_{\lambda}(y)\underset{\lambda\to0}{=}G_{\lambda}(y)+\mathcal{O}(\lambda)$
for every value of $y$.

\paragraph{Combination of a viscous and a linear friction regularization.}

We can now also discuss the regularization of Plemelj formula that
corresponds to the combined effect of viscosity and linear friction,
by a simple generalization of the argument given above. We start observing
that the stationary vorticity-vorticity correlations is
\begin{equation}
g^{\infty}(x_{1}-x_{2},y_{1}-y_{2})=\epsilon\left(k^{2}+l^{2}\right)e^{ik(x_{1}-x_{2})}e^{il(y_{1}-y_{2})}\,\int_{0}^{\infty}\, e^{-iks(y_{1}-y_{2})t}\, e^{-2\nu(k^{2}+l^{2})t+2\nu sklt^{2}-\frac{2}{3}\nu s^{2}k^{2}t^{3}}e^{-2\alpha t}\,\mathrm{d}t+\mbox{C.C.}\,;
\end{equation}
we have thus to study the following oscillating integral (obtained
from the one contained in the above formula by a change of variable)

\begin{equation}
H_{\lambda_{\nu},\lambda_{\alpha}}(y)=\frac{1}{\lambda_{\nu}}\int_{0}^{\infty}\, e^{-i\frac{y}{\lambda_{\nu}}t_{1}-\left[\lambda_{\nu}^{2}(k^{2}+l^{2})+\frac{\lambda_{\alpha}}{\lambda_{\nu}}\right]t_{1}+\lambda_{\nu}lt_{1}^{2}-\frac{1}{3}t_{1}^{3}}\,\mathrm{d}t_{1}\,,
\end{equation}
where $\lambda_{\nu}=\left(\frac{2\nu}{ks}\right)^{1/3}$ is the length
scale associated with the viscous regularization and $\lambda_{\alpha}=\frac{2\alpha}{ks}$
is the length scale associated with the Rayleigh friction. The discussion
in the following restricts to the real part of $H_{\lambda_{\nu},\lambda_{\alpha}}$
because the decay of the imaginary part is the same for all the regularizations.

Introducing the function $\tilde{H}_{\lambda_{\nu},\lambda_{\alpha}}(Y)=H_{\lambda_{\nu},\lambda_{\alpha}}(\lambda_{\nu}Y)$
and performing a reasoning very similar to the one of the previous
section, we deduce that the real part of $\tilde{H}_{\lambda_{\nu},\lambda_{\alpha}}(Y)$
is given by

\begin{equation}
\Re\left[\tilde{H}_{\lambda_{\nu},\lambda_{\alpha}}(Y)\right]=\frac{1}{\lambda_{\nu}}\left[\frac{\lambda_{\nu}^{2}(k^{2}+l^{2})+\frac{\lambda_{\alpha}}{\lambda_{\nu}}}{Y^{2}}-\frac{\left(\lambda_{\nu}^{2}(k^{2}+l^{2})+\frac{\lambda_{\alpha}}{\lambda_{\nu}}\right)^{3}+6\lambda_{\nu}^{3}l(k^{2}+l^{2})+6\lambda_{\alpha}l+2}{Y^{4}}+\textrm{O}\left(\frac{1}{Y^{6}}\right)\right]\,.\label{eq:appendix-viscosity-friction-real}
\end{equation}
We thus see that three different regimes are possible in the limit
of small Rayleigh friction ($\alpha/s\ll1$) and small viscosity ($\lambda_{\nu}l\ll1)$:
$(i)$ the viscosity is negligible compared to the Rayleigh friction:
$\lambda_{\alpha}/\lambda_{\nu}\gg1$; $(ii)$ the Rayleigh friction
is negligible compared to the viscosity $\lambda_{\alpha}/\lambda_{\nu}\ll\lambda_{\nu}^{2}(k^{2}+l^{2})\ll1$
and $(iii)$ and intermediate case $1\gg\lambda_{\alpha}/\lambda_{\nu}\gg\lambda_{\nu}^{2}(k^{2}+l^{2})$. 

\textbf{Case $(i)$}: $\lambda_{\alpha}/\lambda_{\nu}\gg1$. We see
that equation (\ref{eq:appendix-viscosity-friction-real}) reduces
to 

\begin{equation}
\Re\left[\tilde{H}_{\lambda_{\nu},\lambda_{\alpha}}(Y)\right]\underset{\frac{\lambda_{\alpha}}{\lambda_{\nu}}\gg1}{\sim}\frac{1}{\lambda_{\nu}}\frac{\lambda_{\alpha}}{\lambda_{\nu}}\left[\frac{1}{Y^{2}}-\left(\frac{\lambda_{\alpha}}{\lambda_{\nu}}\right)^{2}\frac{1}{Y^{4}}+\textrm{O}\left(\frac{1}{Y^{6}}\right)\right]\,,\label{eq:appendix-viscosity-friction-real-1}
\end{equation}
and thus

\begin{equation}
\Re\left[H_{\lambda_{\nu},\lambda_{\alpha}}(y)\right]\underset{\lambda_{\nu}\ll y\ll\lambda_{\alpha}}{\sim}-\lambda_{\alpha}^{3}\frac{1}{y^{4}}\qquad\textrm{and }\qquad\Re\left[H_{\lambda_{\nu},\lambda_{\alpha}}(y)\right]\underset{y\gg\lambda_{\alpha}}{\sim}\lambda_{\alpha}\frac{1}{y^{2}}\,.\label{eq:viscosity-real1-1}
\end{equation}
We thus obtain that, even if the viscosity is negligible with respect
to the Rayleigh friction, the behavior $1/y^{2}$ is modified by the
presence of a small viscosity in $1/y^{4}$ for a small enough $y$,
corresponding to $\lambda_{\nu}\ll y\ll\lambda_{\alpha}$.

\textbf{Case $(ii)$}: $\lambda_{\alpha}/\lambda_{\nu}\ll\lambda_{\nu}^{2}(k^{2}+l^{2})\ll1$.
We see that equation (\ref{eq:appendix-viscosity-friction-real})
reduces to the case we have already discussed in the previous section
where the Rayleigh friction was not present, see Eq. (\ref{eq:appendix-h-lambda-asymptotic}).
The only difference with respect to that case is that the crossover
point between the $1/y^{4}$ and $1/y^{2}$ behavior is slightly shifted
by the presence of a small Rayleigh friction.

\textbf{Case $(iii)$}: $\lambda_{\nu}^{2}(k^{2}+l^{2})\ll\lambda_{\alpha}/\lambda_{\nu}\ll1$.
With a similar reasoning as before, we obtain

\begin{equation}
\Re\left[H_{\lambda_{\nu},\lambda_{\alpha}}(y)\right]\underset{\lambda_{\nu}\ll y\ll\sqrt{\frac{2\lambda_{\nu}^{3}}{\lambda_{\alpha}}}}{\sim}-\frac{2\lambda_{\nu}^{3}}{y^{4}}\qquad\textrm{and }\qquad\Re\left[H_{\lambda_{\nu},\lambda_{\alpha}}(y)\right]\underset{y\gg\sqrt{\frac{2\lambda_{\nu}^{3}}{\lambda_{\alpha}}}}{\sim}\lambda_{\alpha}\frac{1}{y^{2}}\,.\label{eq:viscosity-real1-1-1}
\end{equation}
We thus see in this case the appearance of the new length scale $\sqrt{\frac{2\lambda_{\nu}^{3}}{\lambda_{\alpha}}}$,
corresponding to the crossover between the $1/y^{4}$ decay given
by the regularization due to the viscosity and the $1/y^{2}$ decay
given by the Rayleigh friction.

\section{Convergence of Reynolds stress divergence on stationary points}

\label{appendix:reynolds-stress-yc} We prove in this Appendix that
the divergence of the Reynolds stress $-\mathbb{E}_{U}\left[\left\langle v_{m}^{(y)}\omega_{m}\right\rangle(y)\right]$
converges even when evaluated at points $y=y_{c}$, where $y_{c}$
are stationary for the zonal flow $U$: $U'(y_{c})=0$. Even more
generically, we prove that $\mathbb{E}_{U}\left[\left\langle\omega_{m}({\bf r}_{1},t))v_{m}^{(y)}({\bf r}_{2},t))\right\rangle\right]$
converges, no matter for which values of ${\bf r}_{1}$ and ${\bf r}_{2}$.

In this case, to the best of our knowledge, no theoretical results
are available in literature. However, strong numerical evidences are
reported in \cite{Bouchet_Morita_2010PhyD} and permit to infer relations
analogous to equations (\ref{eq:orr-mechanism-voricity}), (\ref{eq:orr-mechanism-velocity-x}),
(\ref{eq:orr-mechanism-velocity-y}) and (\ref{eq:orr-mechanism-stream})
 for $y=y_{c}$. The main differences in this case are: $(i)$
the algebraic decay of the velocity and of the stream-function depends
on the parity of the initial datum; $(ii)$ the exponent of the algebraic
decay for $y=y_{c}$ is different. 

For initial conditions even with respect to $y\to-y$, the numerical
results in \cite{Bouchet_Morita_2010PhyD} can be summarized by replacing
Eq. (\ref{eq:orr-mechanism-voricity}), (\ref{eq:orr-mechanism-velocity-x}),
(\ref{eq:orr-mechanism-velocity-y}) and (\ref{eq:orr-mechanism-stream})
by 
\begin{equation}
\left|\tilde{\omega}_{k}(y_{c},t)\right|\underset{t\to\infty}{\sim}\frac{1}{t}\label{eq:appendix-stationary-omega-even}
\end{equation}
 
\begin{equation}
\left|\tilde{v}_{k}^{(x)}(y_{c},t)\right|\underset{t\to\infty}{\sim}\frac{1}{t}
\end{equation}
 
\begin{equation}
\left|\tilde{v}_{k}^{(y)}(y_{c},t)\right|\underset{t\to\infty}{\sim}\frac{1}{t^{2}}\,.\label{eq:even damping v_y}
\end{equation}
 Moreover, the asymptotic profile $\tilde{\omega}_{k}^{\infty}(y)$
appearing in Eq. (\ref{eq:orr-mechanism-voricity}), (\ref{eq:orr-mechanism-velocity-x}),
(\ref{eq:orr-mechanism-velocity-y}) and (\ref{eq:orr-mechanism-stream})
is equivalent to $(y-y_{c})^{2}$ in the vicinity of $y_{c}$. We
observe that these results are about the asymptotic behavior of the
modulus of vorticity and velocity and that the oscillating behavior
of these quantities is not known. This fact does not allow to obtain
estimates for the asymptotic behavior of derivatives of the vorticity
and of the velocity.

For initial conditions odd with respect to $y\to-y$,the numerical
results in \cite{Bouchet_Morita_2010PhyD} can be summarized by replacing
Eq. (\ref{eq:orr-mechanism-voricity}), (\ref{eq:orr-mechanism-velocity-x}),
(\ref{eq:orr-mechanism-velocity-y}) and (\ref{eq:orr-mechanism-stream})
by 
\begin{equation}
\left|\tilde{\omega}_{k}(y_{c},t)\right|\underset{t\to\infty}{\sim}\frac{1}{\sqrt{t}}
\end{equation}
 
\begin{equation}
\left|\tilde{v}_{k}^{(x)}(y_{c},t)\right|\underset{t\to\infty}{\sim}\frac{1}{t^{3/2}}
\end{equation}
 
\begin{equation}
\left|\tilde{v}_{k}^{(y)}(y_{c},t)\right|\underset{t\to\infty}{\sim}\frac{1}{t^{3/2}}\,.\label{eq:odd damping v_y}
\end{equation}
 Again, the asymptotic profile $\tilde{\omega}_{k}^{\infty}(y)$ appearing
in Eq. (\ref{eq:orr-mechanism-voricity}), (\ref{eq:orr-mechanism-velocity-x}),
(\ref{eq:orr-mechanism-velocity-y}) and (\ref{eq:orr-mechanism-stream})
vanishes as $(y-y_{c})^{2}$ in the vicinity of $y_{c}$.

Using these results, we can prove that the divergence of Reynolds
stress at $y=y_{c}$ converges. It is useful here to consider $\tilde{g}_{kl}$
defined, as in section \ref{sub:Numerical-solutions-of}, by $g_{kl}(x_{1},x_{2},y_{1},y_{2},t)=\mbox{e}^{ik(x_{1}-x_{2})}\tilde{g}_{kl}(y_{1},y_{2},t)+\mbox{C.C.}$
with
\begin{equation}
\frac{\partial\tilde{g}_{kl}}{\partial t}+L_{U,k}^{\alpha(1)}\tilde{g}_{kl}+L_{U,-k}^{\alpha(2)}\tilde{g}_{kl}=\mbox{e}^{il(y_{1}-y_{2})}\,.\label{eq:gk-alpha-1}
\end{equation}
Moreover, because the Orr mechanism works differently for initial
conditions with different parity, it is useful to decompose $\tilde{g}_{kl}$
as follow

\begin{equation}
\tilde{g}_{kl}=\tilde{g}_{kl,cc}-i\tilde{g}_{kl,cs}+i\tilde{g}_{kl,sc}+\tilde{g}_{kl,ss}
\end{equation}
 where $g_{kl,cc}$ solves the Lyapunov equation 
\begin{equation}
\frac{\partial\tilde{g}_{kl,cc}}{\partial t}+L_{U,k}^{0(1)}\tilde{g}_{kl,cc}+L_{U,-k}^{0(2)}\tilde{g}_{kl,cc}=\cos(ly_{1})\cos(ly_{2}),\label{eq:appendix-gk-cc}
\end{equation}
 and $\tilde{g}_{kl,cs}$, $\tilde{g}_{kl,sc}$, $\tilde{g}_{kl,ss}$
solve the same Lyapunov equation with the right hand side replaced,
respectively, by $\cos(ly_{1})\sin(ly_{2})$, $\sin(ly_{1})\cos(ly_{2})$
and $\sin(ly_{1})\sin(ly_{2})$. The computation needed to show the
convergence of the divergence of the Reynolds stress obtained as linear
transforms of the four functions $\tilde{g}_{kl,cc}$, $\tilde{g}_{kl,cs}$,
$\tilde{g}_{kl,sc}$ and $\tilde{g}_{kl,ss}$ are very similar. We
thus report in detail only the one corresponding to $\tilde{g}_{kl,cc}$.

With the notation $\tilde{g}_{kl,cc}=\Delta_{k}^{(2)}h_{kl,cc}$
and using Eq. (\ref{eq:lyapunov-solution-real-time}), we can write
\begin{equation}
h_{kl,cc}^{\infty}(y_{1},y_{2})=\int_{0}^{\infty}{\rm d}t_{1}\,\tilde{\omega}_{k}(y_{1},t_{1})\tilde{v}_{-k}^{(y)}(y_{2},t_{1})\,,\label{appendix:orr-gklcc}
\end{equation}
 where $\tilde{\omega}_{k}$ is the solution to the deterministic
evolution $\partial_{t}+L_{U,k}^{0}$ and $\tilde{\omega}_{k}(y_{1},0)=\cos(ly_{1})$
and $\tilde{v}_{-k}^{(y)}(y_{2},0)=\cos(ly_{2})$. The convergence
of the this integral has already been established in the case $y_{1}\neq y_{c}$
and $y_{2}\neq y_{c}$ in section \ref{subsec:reynolds-orr}.

For any value of $y_{1}$ and $y_{2}$, we see from equations (\ref{eq:orr-mechanism-voricity}),
(\ref{eq:orr-mechanism-velocity-y}), (\ref{eq:appendix-stationary-omega-even})
and (\ref{eq:even damping v_y}) that the absolute value of the integrand
of Eq. (\ref{appendix:orr-gklcc}) decays to zero at least as fast
as $1/t^{2}$. The convergence of $h_{kl,cc}$ is thus ensured for
all the values of $y_{1}$ and $y_{2}$.

We leave to the reader the very similar computation needed to show
that also the Reynolds stresses obtained from $\tilde{g}_{kl,cs}$,
$\tilde{g}_{kl,sc}$ and $\tilde{g}_{kl,ss}$ converge. The slowest
decay of the integrand of expressions like Eq. (\ref{appendix:orr-gklcc})
which is encountered is $1/t^{3/2}$, still sufficient to ensure the
convergence of $h_{kl}$.

We conclude stressing that the results presented in this appendix
rely on conclusions drawn from numerical results reported in \cite{Bouchet_Morita_2010PhyD}.
\bibliographystyle{plain}
\bibliography{All-2013-03,Fbouchet}

\begin{thebibliography}{10}

\bibitem{BakasIoannou2013SSST}
Nikolaos Bakas and Petros Ioannou.
\newblock A theory for the emergence of coherent structures in beta-plane
  turbulence.
\newblock {\em preprint arXiv:1303.6435}, 2013.

\bibitem{Berhanu2007}
M~Berhanu, R~Monchaux, S~Fauve, N~Mordant, F~P\'{e}tr\'{e}lis, A~Chiffaudel,
  F~Daviaud, B~Dubrulle, L~Mari\'{e}, F~Ravelet, M~Bourgoin, Ph~Odier, J.-F
  Pinton, and R~Volk.
\newblock {Magnetic field reversals in an experimental turbulent dynamo}.
\newblock {\em Europhysics Letters (EPL)}, 77(5):59001, March 2007.

\bibitem{Binney_Tremaine_1987_Galactic_Dynamics}
J.~{Binney} and S.~{Tremaine}.
\newblock {\em {Galactic dynamics}}.
\newblock Princeton, NJ, Princeton University Press, 1987, 747 p., 1987.

\bibitem{boffetta2012two}
Guido Boffetta and Robert~E Ecke.
\newblock Two-dimensional turbulence.
\newblock {\em Annual Review of Fluid Mechanics}, 44:427--451, 2012.

\bibitem{Bouchet:2004_PRE_StochasticProcess}
F.~{Bouchet}.
\newblock {Stochastic process of equilibrium fluctuations of a system with
  long-range interactions}.
\newblock {\em Phys. Rev. E}, 70(3):036113, September 2004.

\bibitem{Bouchet_Dauxois:2005_PRE}
F.~{Bouchet} and T.~{Dauxois}.
\newblock {Prediction of anomalous diffusion and algebraic relaxations for
  long-range interacting systems, using classical statistical mechanics}.
\newblock {\em Phys. Rev. E}, 72(4):045103, October 2005.

\bibitem{Bouchet_Morita_2010PhyD}
F.~{Bouchet} and H.~{Morita}.
\newblock {Large time behavior and asymptotic stability of the 2D Euler and
  linearized Euler equations}.
\newblock {\em Physica D Nonlinear Phenomena}, 239:948--966, June 2010.

\bibitem{Bouchet_Simonnet_2008}
F.~{Bouchet} and E.~{Simonnet}.
\newblock {Random Changes of Flow Topology in Two-Dimensional and Geophysical
  Turbulence}.
\newblock {\em Physical Review Letters}, 102(9):094504, March 2009.

\bibitem{BouchetVenaille-PhysicsReport}
F.~{Bouchet} and A.~{Venaille}.
\newblock {Statistical mechanics of two-dimensional and geophysical flows}.
\newblock {\em Physics Reports}, 515:227--295, 2012.

\bibitem{Bouchet_Dauxois_2005_AlgebraicCorrelations}
Freddy Bouchet and Thierry Dauxois.
\newblock Kinetics of anomalous transport and algebraic correlations in a
  long-range interacting system.
\newblock In {\em Journal of Physics: Conference Series}, volume~7, page~34.
  IOP Publishing, 2005.

\bibitem{brehier2012strong}
Charles-Edouard Br{\'e}hier.
\newblock Strong and weak order in averaging for spdes.
\newblock {\em Stochastic Processes and their Applications}, 2012.

\bibitem{Bricmont_Kupianen_2001_Comm_Math_Phys_Ergodicity2DNavierStokes}
J.~{Bricmont}, A.~{Kupiainen}, and R.~{Lefevere}.
\newblock {Ergodicity of the 2D Navier-Stokes Equations with Random Forcing}.
\newblock {\em Com.. Math. Phys.}, 224:65--81, 2001.

\bibitem{Campa_Dauxois_Ruffo_Revues_2009_PhR...480...57C}
A.~{Campa}, T.~{Dauxois}, and S.~{Ruffo}.
\newblock {Statistical mechanics and dynamics of solvable models with
  long-range interactions}.
\newblock {\em Phys. Rep.}, 480:57--159, 2009.

\bibitem{Case_1960_Phys_Fluids}
K.~M. {Case}.
\newblock {Stability of Inviscid Plane Couette Flow}.
\newblock {\em Physics of Fluids}, 3:143--148, 1960.

\bibitem{Chavanis_Quasilinear_2000PhRvL}
P.~H. {Chavanis}.
\newblock {Quasilinear Theory of the 2D Euler Equation}.
\newblock {\em Physical Review Letters}, 84:5512--5515, June 2000.

\bibitem{Chavanis_2001PhRvE_64_PointsVortex}
P.~H. {Chavanis}.
\newblock {Kinetic theory of point vortices: Diffusion coefficient and
  systematic drift}.
\newblock {\em Phys. Rev. E}, 64(2):026309, 2001.

\bibitem{Chavanis_houches_2002}
P.~H. {Chavanis}.
\newblock Statistical mechanis of two-dimensional vortices and stellar systems.
\newblock In T.~{Dauxois}, S.~{Ruffo}, E.~{Arimondo}, and M.~{Wilkens},
  editors, {\em Dynamics and Thermodynamics of Systems With Long Range
  Interactions}, volume 602 of {\em Lecture Notes in Physics}, pages 208--289.
  Springer-Verlag, 2002.

\bibitem{Chertkov_Connaughton_andco_2007_PRL_EnergyCondesation}
M.~{Chertkov}, C.~{Connaughton}, I.~{Kolokolov}, and V.~{Lebedev}.
\newblock {Dynamics of Energy Condensation in Two-Dimensional Turbulence}.
\newblock {\em Phys. Rev. Lett.}, 99:084501, 2007.

\bibitem{constantinou2012emergence}
Navid~C Constantinou, Petros~J Ioannou, and Brian~F Farrell.
\newblock Emergence and equilibration of jets in beta-plane turbulence.
\newblock {\em arXiv preprint arXiv:1208.5665}, 2012.

\bibitem{da2003ergodicity}
Giuseppe Da~Prato and Arnaud Debussche.
\newblock Ergodicity for the 3d stochastic navier--stokes equations.
\newblock {\em Journal de math{\'e}matiques pures et appliqu{\'e}es},
  82(8):877--947, 2003.

\bibitem{danilov2004scaling}
Sergey Danilov and David Gurarie.
\newblock Scaling, spectra and zonal jets in beta-plane turbulence.
\newblock {\em Physics of fluids}, 16:2592, 2004.

\bibitem{delsole1996quasi}
Timothy DelSole and Brian~F Farrell.
\newblock The quasi-linear equilibration of a thermally maintained,
  stochastically excited jet in a quasigeostrophic model.
\newblock {\em Journal of the atmospheric sciences}, 53(13):1781--1797, 1996.

\bibitem{Drazin_Reid_1981}
P.~G. {Drazin} and W.~H. {Reid}.
\newblock {\em {Hydrodynamic stability}}.
\newblock Cambridge university press, 2004, second edition.

\bibitem{Dritschel_McIntyre_2008JAtS}
D.~G. {Dritschel} and M.~E. {McIntyre}.
\newblock {Multiple Jets as PV Staircases: The Phillips Effect and the
  Resilience of Eddy-Transport Barriers}.
\newblock {\em Journal of Atmospheric Sciences}, 65:855, 2008.

\bibitem{Dubin_ONeil_1988_PhysRevLett_Kinetic_Point_Vortex}
D.~H.~E. Dubin and T.~M. O\char39{}Neil.
\newblock Two-dimensional guiding-center transport of a pure electron plasma.
\newblock {\em Phys. Rev. Lett.}, 60(13):1286--1289, 1988.

\bibitem{Farrell_Ioannou_JAS_2007}
B.~F. {Farrell} and P.~J. {Ioannou}.
\newblock {Structure and Spacing of Jets in Barotropic Turbulence}.
\newblock {\em Journal of Atmospheric Sciences}, 64:3652, 2007.

\bibitem{Farrel_Ioannou}
Brian~F. Farrell and Petros~J. Ioannou.
\newblock Structural stability of turbulent jets.
\newblock {\em Journal of Atmospheric Sciences}, 60:2101--2118, 2003.

\bibitem{ferrario1997ergodic}
Benedetta Ferrario.
\newblock Ergodic results for stochastic navier-stokes equation.
\newblock {\em Stochastics: An International Journal of Probability and
  Stochastic Processes}, 60(3-4):271--288, 1997.

\bibitem{flandoli1995ergodicity}
Franco Flandoli and Bohdan Maslowski.
\newblock Ergodicity of the 2-d navier-stokes equation under random
  perturbations.
\newblock {\em Communications in mathematical physics}, 172(1):119--141, 1995.

\bibitem{FW84}
M.~I. {Freidlin} and A.~D. { Wentzell}.
\newblock {\em {Random Perturbations of Dynamical Systems}}.
\newblock Springer, New York, 1984.

\bibitem{galperin2010geophysical}
Boris Galperin, Semion Sukoriansky, and Nadejda Dikovskaya.
\newblock Geophysical flows with anisotropic turbulence and dispersive waves:
  flows with a $\beta$-effect.
\newblock {\em Ocean Dynamics}, 60(2):427--441, 2010.

\bibitem{galperin2001universal}
Boris Galperin, Semion Sukoriansky, and Huei-Ping Huang.
\newblock Universal n spectrum of zonal flows on giant planets.
\newblock {\em Physics of Fluids}, 13:1545, 2001.

\bibitem{Gardiner_1994_Book_Stochastic}
C.~W. {Gardiner}.
\newblock {\em {Handbook of stochastic methods for physics, chemistry and the
  natural sciences}}.
\newblock Springer Series in Synergetics, Berlin: Springer, |c1994, 2nd
  ed.~1985.~Corr.~3rd printing 1994, 1994.

\bibitem{gourcy2007large}
Mathieu Gourcy.
\newblock A large deviation principle for 2d stochastic navier--stokes
  equation.
\newblock {\em Stochastic processes and their applications}, 117(7):904--927,
  2007.

\bibitem{Hairer_Mattingly_2006ergodicity}
Martin Hairer and Jonathan~C Mattingly.
\newblock Ergodicity of the 2d navier-stokes equations with degenerate
  stochastic forcing.
\newblock {\em Annals of Mathematics}, pages 993--1032, 2006.

\bibitem{Hairer_Mattingly_2008spectral}
Martin Hairer and Jonathan~C Mattingly.
\newblock Spectral gaps in wasserstein distances and the 2d stochastic
  navier--stokes equations.
\newblock {\em The Annals of Probability}, 36(6):2050--2091, 2008.

\bibitem{jaksic2012large}
Vojkan Jaksic, Vahagn Nersesyan, Claude-Alain Pillet, and Armen Shirikyan.
\newblock Large deviations from a stationary measure for a class of dissipative
  pde's with random kicks.
\newblock {\em arXiv preprint arXiv:1212.0527}, 2012.

\bibitem{kasahara1980effect}
Akira Kasahara.
\newblock Effect of zonal flows on the free oscillations of a barotropic
  atmosphere.
\newblock {\em Journal of Atmospheric Sciences}, 37:917--929, 1980.

\bibitem{RZKhasminskii}
R~Z Khasminskii.
\newblock On an averaging principle for ito stochastic differential equations.
\newblock {\em Kybernetika}, 4:260--279, 1968.

\bibitem{Kuksin_Penrose_2005_JPhysStat_BalanceRelations}
S.~{Kuksin} and 0.~{Penrose}.
\newblock {A family of balance relations for the two-dimensional Navier-Stokes
  equations with random forcing}.
\newblock {\em J. Stat. Phys.}, 118(3-4):437--449, 2005.

\bibitem{Kuksin_2004_JStatPhys_EulerianLimit}
S.~B. {Kuksin}.
\newblock {The eulerian limit for 2D statistical hydrodynamics}.
\newblock {\em J. Stat. Phys.}, 115:469--492, 2004.

\bibitem{kuksin2001ergodicity}
Sergei Kuksin and Armen Shirikyan.
\newblock Ergodicity for the randomly forced 2d navier--stokes equations.
\newblock {\em Mathematical Physics, Analysis and Geometry}, 4(2):147--195,
  2001.

\bibitem{kuksin2008khasminskii}
Sergei~B Kuksin and Andrey~L Piatnitski.
\newblock Khasminskii--whitham averaging for randomly perturbed kdv equation.
\newblock {\em Journal de math{\'e}matiques pures et appliqu{\'e}es},
  89(4):400--428, 2008.

\bibitem{kuksin2012mathematics}
Sergej~B Kuksin and Armen Shirikyan.
\newblock {\em Mathematics of two-dimensional turbulence}.
\newblock Cambridge University Press Cambridge, 2012.

\bibitem{Landau_Lifshitz_1996_Book}
L.~D. {Landau} and E.~M. {Lifshitz}.
\newblock {\em {Statistical Physics. Vol. 5 of the Course of Theoretical
  Physics}}.
\newblock Pergamon Press, 1980.

\bibitem{loxley2013bistability}
PN~Loxley and BT~Nadiga.
\newblock Bistability and hysteresis of maximum-entropy states in decaying
  two-dimensional turbulence.
\newblock {\em Physics of Fluids}, 25:015113, 2013.

\bibitem{Maassen2003}
S.~R. Maassen, H.~J.~H. Clercx, and G.~J.~F. {Van Heijst}.
\newblock {Self-organization of decaying quasi-two-dimensional turbulence in
  stratified fluid in rectangular containers}.
\newblock {\em Journal of Fluid Mechanics}, 495:19--33, November 2003.

\bibitem{Majda_Wang_Book_Geophysique_Stat}
A.~J. {Majda} and X.~{Wang}.
\newblock {\em {Nonlinear Dynamics and Statistical Theories for Basic
  Geophysical Flows}}.
\newblock Cambridge University Press, 2006.

\bibitem{Majda_Wang_Bombardement_2006_Comm}
A.~J. {Majda} and X.~{Wang}.
\newblock {The emergence of large-scale coherent structure under small-scale
  random bombardments}.
\newblock {\em Comm. Pure App. Maths}, 59(4):467--500, 2006.

\bibitem{Marston-APS-2011Phy}
B.~{Marston}.
\newblock {Looking for new problems to solve? Consider the climate}.
\newblock {\em Physcs Online Journal}, 4:20, March 2011.

\bibitem{Marston-2010-Chaos}
J.~B. {Marston}.
\newblock {Statistics of the general circulation from cumulant expansions}.
\newblock {\em Chaos}, 20(4):041107, December 2010.

\bibitem{Marston_Conover_Schneider_JAS2008}
J.~B. {Marston}, E.~{Conover}, and T.~{Schneider}.
\newblock {Statistics of an Unstable Barotropic Jet from a Cumulant Expansion}.
\newblock {\em Journal of Atmospheric Sciences}, 65:1955, 2008.

\bibitem{mattingly1999elementary}
JC~Mattingly and Ya~G Sinai.
\newblock An elementary proof of the existence and uniqueness theorem for the
  navier--stokes equations.
\newblock {\em Communications in Contemporary Mathematics}, 1(04):497--516,
  1999.

\bibitem{Miller:1990_PRL_Meca_Stat}
J.~Miller.
\newblock Statistical mechanics of euler equations in two dimensions.
\newblock {\em Phys. Rev. Lett.}, 65(17):2137--2140, 1990.

\bibitem{Mouhot_Villani:2009}
C.~Mouhot and C.~Villani.
\newblock On {L}andau damping.
\newblock {\em Acta Mathematica}, 207:29--201, 2011.

\bibitem{NardiniGuptaBouchet-2012-JSMTE}
C.~{Nardini}, S.~{Gupta}, S.~{Ruffo}, T.~{Dauxois}, and F.~{Bouchet}.
\newblock {Kinetic theory for non-equilibrium stationary states in long-range
  interacting systems}.
\newblock {\em Journal of Statistical Mechanics: Theory and Experiment},
  1:L01002, January 2012.

\bibitem{Nardini_Gupta_Ruffo_Dauxois_Bouchet_2012_kinetic}
Cesare Nardini, Shamik Gupta, Stefano Ruffo, Thierry Dauxois, and Freddy
  Bouchet.
\newblock Kinetic theory of nonequilibrium stochastic long-range systems: phase
  transition and bistability.
\newblock {\em Journal of Statistical Mechanics: Theory and Experiment},
  2012(12):P12010, 2012.

\bibitem{Nazarenko_PhysicsLetterA_2000}
S.~{Nazarenko}.
\newblock {On exact solutions for near-wall turbulence theory}.
\newblock {\em Physics Letters A}, 264:444--448, 2000.

\bibitem{nazarenko1999wkb}
S~Nazarenko, NK-R Kevlahan, and B~Dubrulle.
\newblock Wkb theory for rapid distortion of inhomogeneous turbulence.
\newblock {\em Journal of Fluid Mechanics}, 390(1):325--348, 1999.

\bibitem{nazarenko2000nonlinear}
S~Nazarenko, NK-R Kevlahan, and B~Dubrulle.
\newblock Nonlinear rdt theory of near-wall turbulence.
\newblock {\em Physica D: Nonlinear Phenomena}, 139(1):158--176, 2000.

\bibitem{Nicholson_1991}
D.~{Nicholson}.
\newblock {\em {Introduction to plasma theory}}.
\newblock {Wiley, New-York}, 1983.

\bibitem{GormanSchneider-QL-GCM}
Paul~A. O'Gorman and Tapio Schneider.
\newblock Recovery of atmospheric flow statistics in a general circulation
  model without nonlinear eddy-eddy interactions.
\newblock {\em Geophysical Research Letters}, 34(22):n/a--n/a, 2007.

\bibitem{ParkerKrommes2013SSST}
Jeffrey~B. Parker and John~A. Krommes.
\newblock Zonal flow as pattern formation: Merging jets and the ultimate jet
  length scale.
\newblock {\em preprint arXiv:1301.5059}, 2013.

\bibitem{PedloskyBook}
J.~{Pedlosky}.
\newblock {\em {Geophysical fluid dynamics}}.
\newblock 1982.

\bibitem{Pope_2000_Livre}
Stephen~B Pope.
\newblock {\em Turbulent flows}.
\newblock Cambridge university press, 2000.

\bibitem{Ravelet_Marie_Chiffaudel_Daviaud_PRL2004}
Florent Ravelet, Louis Mari\'e, Arnaud Chiffaudel, and Francois Daviaud.
\newblock Multistability and memory effect in a highly turbulent flow:
  Experimental evidence for a global bifurcation.
\newblock {\em Phys. Rev. Lett.}, 93(16):164501, 2004.

\bibitem{Robert:1990_CRAS}
R.~{Robert}.
\newblock {Etats d'\'equilibre statistique pour l'\'ecoulement bidimensionnel
  d'un fluide parfait}.
\newblock {\em C. R. Acad. Sci.}, 1:311:575--578, 1990.

\bibitem{Robert:1991_JSP_Meca_Stat}
R.~{Robert}.
\newblock {A maximum-entropy principle for two-dimensional perfect fluid
  dynamics}.
\newblock {\em J. Stat. Phys.}, 65:531--553, 1991.

\bibitem{Schmeits2001}
M.~J. Schmeits and H.~A. Dijkstra.
\newblock Bimodal behavior of the kuroshio and the gulf stream.
\newblock {\em J.~Phys. Oceanogr.}, 31:3435--56, 2001.

\bibitem{shirikyan2004exponential}
Armen Shirikyan.
\newblock Exponential mixing for 2d navier-stokes equations perturbed by an
  unbounded noise.
\newblock {\em Journal of Mathematical Fluid Mechanics}, 6(2):169--193, 2004.

\bibitem{Sommeria1986}
J.~Sommeria.
\newblock {Experimental study of the two-dimensional inverse energy cascade in
  a square box}.
\newblock {\em Journal of Fluid Mechanics}, 170:139--68, 1986.

\bibitem{Srinivasan-Young-2011-JAS}
K.~{Srinivasan} and W.~R. {Young}.
\newblock {Zonostrophic Instability}.
\newblock {\em Journal of the atmospheric sciences}, 69(5):1633--1656, 2011.

\bibitem{Sritharana_Sundarb_2006}
S.~S. {Sritharana} and P.~{Sundarb}.
\newblock {Large deviations for the two-dimensional Navier-Stokes equations
  with multiplicative noise}.
\newblock {\em Stochastic Processes and their Applications}, 116:1636--1659,
  2006.

\bibitem{tobias2013direct}
SM~Tobias and JB~Marston.
\newblock Direct statistical simulation of out-of-equilibrium jets.
\newblock {\em Physical Review Letters}, 110(10):104502, 2013.

\bibitem{VallisBook}
G.~K. {Vallis}.
\newblock {\em {Atmospheric and Oceanic Fluid Dynamics}}.
\newblock 2006.

\bibitem{Weeks_Tian_etc_Swinney_Ghil_Science_1997}
E.~R. {Weeks}, Y.~{Tian}, J.~S. {Urbach}, K.~{Ide}, H.~L. {Swinney}, and
  M.~{Ghil}.
\newblock {Transitions Between Blocked and Zonal Flows in a Rotating Annulus}.
\newblock {\em Science}, 278:1598, 1997.

\bibitem{Weinam_Mattingly_2001_Comm_Pure_Appl_Math_Ergodicity_NS}
E.~{Weinan} and J.~C. {Mattingly}.
\newblock {Ergodicity for the Navier-Stokes equation with degenerate random
  forcing: Finite-dimensional approximation}.
\newblock {\em Comm. Pure Appl. Math.}, 54:1386--1402, 2001.

\bibitem{Yamaguchi_Bouchet_Dauxois_2007_JSMTE_Anomalous_Diffusion}
Y.~Y. {Yamaguchi}, F.~{Bouchet}, and T.~{Dauxois}.
\newblock {Algebraic correlation functions and anomalous diffusion in the
  Hamiltonian mean field model}.
\newblock {\em J. Stat. Mech.}, 1:20, January 2007.

\bibitem{Yin_Montgomery_Clercx_2003PhFluids}
Z.~{Yin}, D.~C. {Montgomery}, and H.~J.~H. {Clercx}.
\newblock {Alternative statistical-mechanical descriptions of decaying
  two-dimensional turbulence in terms of ``patches'' and ``points''}.
\newblock {\em Phys. Fluids}, 15:1937--1953, 2003.

\end{thebibliography}

\end{document}